\documentclass[preprint,12pt,3p]{elsarticle}

\usepackage{amsmath}
\usepackage{multirow}
\usepackage{soul}
\usepackage{booktabs}
\usepackage{url}
\usepackage[hidelinks,breaklinks=true]{hyperref}
\newcommand{\multicomment}[1]{}
\usepackage{enumitem}
\usepackage{siunitx}
\usepackage{tabularx}
\usepackage{subcaption}
\usepackage{amssymb}
\usepackage{bbding}
\usepackage[table]{xcolor}
\definecolor{mygreen}{HTML}{1a9850}
\definecolor{myred}{HTML}{fb8072}

\usepackage[normalem]{ulem}

\usepackage{bm}

\journal{Computers in Human Behavior}

\begin{document}

\begin{frontmatter}

\title{Dark Personality Traits and Online Toxicity:\\Linking Self-Reports to Reddit Activity}

\def\thefootnote{*}\footnotetext{Equal contributions.}
\author[iit,unipi-inf]{Aldo Cerulli$^*$}
\author[iit,unipi-inf]{Benedetta Tessa$^*$}
\author[unifg]{Giuseppe La Selva}
\author[unisr]{Oronzo Mazzeo}
\author[iit,unipi-ing]{Lorenzo Cima}
\author[unifg]{Lucia Monacis}
\author[iit]{Stefano Cresci}

\affiliation[iit]{organization={Institute of Informatics and Telematics, National Research Council},addressline={ \\ Via G. Moruzzi 1}, 
 postcode={56124}, 
 city={Pisa},
 country={Italy}}

\affiliation[unipi-inf]{organization={Dept. of Computer Science, University of Pisa},country={Italy}}

\affiliation[unipi-ing]{organization={Dept. of Information Engineering, University of Pisa},country={Italy}}

\affiliation[unifg]{organization={Dept. of Humanistic Studies, University of Foggia},country={Italy}}

\affiliation[unisr]{organization={Dept. for the Promotion of Human Science and Quality of Life, San Raffaele Telematic University},country={Italy}}

\begin{abstract}
Dark personality traits have long been associated with antisocial and toxic online behaviors, yet their relationship with observable online activity remains unclear. We investigate the association between validated dark personality measures, self-reported experiences of online incivility, and linguistic and behavioral features extracted from real-world user activity. To this end, we developed a Web application that securely links responses to validated psychological questionnaires collected via Amazon Mechanical Turk with participants’ Reddit activity. This yielded a dataset of nearly 57K comments (2.2M tokens) from 114 users, represented through a broad set of linguistic and behavioral features. Our analyses reveal a clear distinction between self-reported and observed behavior. Dark personality traits show consistent associations with self-reported engagement in uncivil interactions. However, no validated dark personality dimension significantly predicts text-derived toxicity or linguistic features. In contrast, self-reported experiences of engaging in or being targeted by toxic behavior are robustly reflected in users’ language, exhibiting consistent associations with measures of negativity, moral framing, and emotional intensity. Taken together, these findings highlight a gap between stable personality traits and their manifestation in surface-level linguistic signals. While computational features effectively capture behavioral engagement in online incivility, they do not provide reliable proxies for underlying personality constructs within the present framework. Our results underscore the importance of grounding computational approaches in validated psychological measures and point to the need for richer, context-aware representations to better understand the relationship between personality and online behavior.
\\\\
\textcolor{red}{Article published in \textit{Computers in Human Behavior}. DOI: \href{https://doi.org/10.1016/j.chb.2026.109085}{10.1016/j.chb.2026.109085}. Please, cite the published version.}
\end{abstract}


\end{frontmatter}

\section{Introduction}
Online social platforms have transformed how individuals communicate, exchange information, and form communities~\cite{dwivedi2018social}. Yet these same environments have become fertile ground for a range of societal challenges~\cite{gillespie2020content}. Among the most pressing ones are the unchecked spread of fake content and misinformation~\cite{groh2022deepfake}, the amplification of ideological polarization, and the pervasive presence of hateful and toxic speech \cite{giorgi2025human,alvisi2026toxicity}. The latter issue is particularly consequential as hostile online interactions are closely associated with cyberbullying, psychological harm, and in some cases, escalation into offline violence~\cite{graf2022did}. These risks have intensified public, academic, and policy interest in understanding the drivers of harmful online behavior and in developing effective strategies to mitigate its impact~\cite{giorgi2025human,cima2025contextualized}.

A large body of research has demonstrated that online behavior is not only shaped by platform design or situational context, but is also influenced by individual personality traits~\cite{kurek2019did}. Personality affects how users interact with others, the language they choose, and the norms they adopt in digital spaces~\cite{bogolyubova2018dark}. This relationship also holds for various forms of online \textit{mis}behavior, as individuals with certain personality dispositions are more likely to engage in harmful, antisocial, or disruptive online conduct. Several studies have linked traits from the so-called \textit{Dark Tetrad}---Machiavellianism, narcissism, psychopathy, and everyday sadism---to behaviors such as online trolling, cyberbullying, and the production or endorsement of hateful and toxic content \cite{alavi2023dark,kurek2019did}. For example, individuals high in sadistic tendencies have been found to derive enjoyment from provoking or distressing others, while elevated psychopathy scores are associated with impulsive and hostile online interactions~\cite{kurek2019did}. Similarly, narcissism has been connected to aggressive responses when personal status is threatened \cite{mahadevan2024conceptualizing}, and Machiavellianism to strategic manipulation and antagonistic discourse in digital environments \cite{calic2023dark}. These findings underscore that toxic and aggressive online behaviors have been associated with underlying psychological dispositions. As harmful content threatens the safety and inclusivity of online communities and the well-being of their members, deepening the understanding of how dark personality traits shape the production and spread of toxic speech is relevant and timely. A nuanced grasp of these dynamics can inform both psychological theory and the development of ethically sound, evidence-based moderation strategies to curb online harm~\cite{cresci2022personalized,biselli2025mitigating}.
Recent advances suggest promising avenues for addressing this gap, highlighting the potential of integrating personality assessment with computational approaches to better characterize toxic behavior.
Yet, despite growing interest in this area, relatively few studies have directly examined the interplay between dark personality traits and the production of online toxicity. Moreover, there is little evidence that dark personalities can be reliably inferred from online discussions~\cite{borghi2025deceptive,bogolyubova2018dark}.

\paragraph{Research Focus} Taken together, the literature highlights the relevance of dark personality traits for understanding online toxicity. However, it also reveals important gaps, including limited direct links between validated psychometric measures and observed online behavior, and uncertainty about whether linguistic features can reliably capture dark traits. To address these limitations, we present a combined study of validated psychological assessments with large-scale behavioral data from Reddit. In particular, we assess whether individuals with higher dark traits differ in how they perceive, experience, or engage with toxic content, whether as producers, targets, or observers. Moreover, we investigate if linguistic features---either hand-crafted or computed automatically at scale---correlate with these traits, exploring a potential bridge between psychological traits and online data.
These considerations form the basis of the following research questions, which we address through a set of confirmatory analyses:
\begin{itemize}[noitemsep, topsep=3pt]
    \item \textit{Do dark traits correlate with self-reported experiences of online incivility?} While prior work has linked dark traits to antisocial tendencies, there is limited empirical evidence on how these traits relate to individuals’ reported experiences of online incivility. Dark personality traits differ in how individuals process, relate, and respond to social information. Consequently, examining their association with self-reported experiences of online incivility can help clarify whether these personalities are linked to differences in how individuals report experiencing and engaging in such interactions. 
    \item \textit{Do dark traits correspond to actual production of toxic content?} A key open question is whether validated personality measures are associated with observable online behavior. Personality theories posit that stable traits may correspond to consistent patterns of behavior across contexts. Since dark traits are characterized by manipulativeness, hostility, and reduced empathy, they may play a role in communicative behavior, including the production of toxic content on social media.
    \item \textit{Can hand-crafted text-derived formulas accurately estimate dark traits?} While many computational approaches rely on data-driven models, recent work has proposed interpretable, theory-informed formulas to estimate dark traits directly from linguistic features. However, the extent to which such hand-crafted proxies align with validated psychometric measures remains largely unexplored. Language is a primary medium of social interaction online, and personality differences are often associated with stable linguistic patterns. Since dark traits influence how individuals communicate, these differences may be detectable in structured features of online text, providing a potential link between users' personality and language use.
\end{itemize}

To answer these research questions, we recruited a sample of active Reddit users via crowdsourcing, and asked them to complete a comprehensive socio-psychological questionnaire of dark personality traits and online trolling tendencies. Throughout this article, we refer to the resulting trait measures---namely the dark-trait subscales and the trolling scale---as ``dimensions.'' The same users also consented to share their Reddit activity data for this study. In addition to answering the above questions, we also run a set of exploratory analyses to examine broader behavioral patterns. Unlike prior work, which has primarily relied on self-reported measures or aggregate behavioral proxies, our study leverages consented, user-level Reddit data to explore how specific dark traits relate to multiple aspects of online behavior. By combining validated psychometric measures with computational analyses of large-scale behavioral data, our work bridges personality psychology and computational social science, contributing to an interdisciplinary understanding of online misbehavior. Beyond its theoretical contributions, this study provides preliminary evidence on the extent to which dark personality traits correlate with online behavioral and linguistic patterns. These findings may inform future research on personality-aware computational models and contribute to discussions about ethically grounded approaches to content moderation.
 \section{Related Work}
\label{sec:rel-work}

\subsection{Dark Personalities and Online Misbehavior}
Extensive research has investigated the interplay between personality traits and digital footprints, seeking to elucidate the psychological mechanisms driving behaviors in online environments. Most of this literature has focused on bright (e.g., Big Five) traits, using them to interpret behavioral patterns and inform applications in contexts ranging from digital marketing~\cite{winter2021effects} to e-learning systems~\cite{koay2023students}. In contrast, comparatively little attention has been devoted to dark personality traits, despite their proven association with antisocial online conduct~\cite{alavi2023dark,kurek2019did}. Available evidence on this relationship has been systematically synthesized in a relatively recent review~\cite{moor2019systematic}. Based on 24 articles published between 2014 and 2019, the review emphasized the role of both the Dark Triad~\cite{paulus2002dark} and Tetrad---which adds sadism as proposed in~\cite{chabrol2009contributions,buckels2013behavioral}---in shaping a variety of subversive behaviors across digital platforms. These include aggressive acts such as trolling~\cite{buckels2018internet,lopes2017you} and cyberbullying~\cite{gibb2014who,goodboy2015personality}, intrusive practices like cyberstalking~\cite{smoker2017predicting}, maladaptive usage patterns such as problematic social media use~\cite{kircaburun2018dark,kircaburun2019analyzing} and social media addiction~\cite{demircioglu2021effects}, as well as other forms of technology-facilitated deviance. Such behaviors are studied in different ways, ranging from analyses of interaction patterns like posting frequency, visited pages, and clicked links, to linguistic studies of user-generated content such as status updates, posts, and comments. The latter approach has been widely adopted, as it enables the collection of multiple facets of user mental constructs and expressions, encompassing linguistic, topical~\cite{preotiuc2016studying,moskvichev2018using}, and psycho-emotional~\cite{sumner2012predicting} features---or a combination of these~\cite{bogolyubova2018dark}. Beyond the studies reviewed in~\cite{moor2019systematic}, more recent works confirmed the same trends by presenting results based on the analysis of an extended set of factors, including the time spent online and life satisfaction~\cite{alavi2025relationships}.

In summary, the majority of existing studies identified psychopathy as the trait that most strongly and consistently correlates with the aforementioned misbehaviors, followed by Machiavellianism and sadism. Narcissism, instead, showed the weakest and least consistent associations. Overall, these findings call for additional research to produce and leverage actionable results for promoting safer and more inclusive social platforms. Notably, the utility of this line of work has already surfaced for practically relevant tasks related to the detection of antisocial activities, including cyberbullying~\cite{balakrishnan2019cyberbullying} and fake product reviews~\cite{borghi2025deceptive}.

\subsection{Computational Personality Prediction}
Personality assessment has traditionally relied on self-reported questionnaires~\cite{paulhus2007self,montag2022we}. Although this approach provides accurate and consistent data, it is effortful and time-consuming. In addition, the need for human intervention and cooperation may introduce bias into the results~\cite{chraibi2020automatic}. To overcome these limitations, researchers have started exploring computational techniques to automatically infer personality traits  directly from online behavioral traces. This pursuit has led to the emergence of~\textit{computational personality prediction}, a branch of the research framework of \textit{computational personality}~\cite{levy2021personality,yang2022computational} that builds on the assumption that personality is an ``enduring configuration of characteristics''~\cite{apa2007} that are reflected in online behavior.

Over the past fifteen years, the field has experienced rapid and sustained growth, driven by the ever more pervasive usage of social media and the rising amount of personality-informed applications in high-impact domains like e-commerce~\cite{saw2022designing} and cybersecurity~\cite{tardelli2020characterizing}. This expansion has given rise to a complex and multifaceted body of literature, encompassing diverse methodological approaches, technologies, and data sources, as documented by a 2020 mapping study~\cite{chraibi2020automatic}. Personality prediction is typically approached as a supervised machine learning task, leveraging trait scores from self-assessed questionnaires as the ground truth for evaluating model performance. Interestingly, classification~\cite{utami2021personality} seems to be preferred over regression~\cite{karanatsiou2022tweets}, even though it simplifies the problem by producing discretized representations of continuous traits (e.g., low~\textit{vs.} high). Among conventional predictive methods, Support Vector Machines dominated until 2017, when deep learning emerged as the prevailing~\cite{majumder2017deep,zhao2022deep}. Beyond supervised approaches, a smaller body of work has explored unsupervised learning methods, including clustering~\cite{mushtaq2020PredictingMP} and ensemble modeling techniques that combine multiple algorithms to improve accuracy~\cite{ramezani2022automatic}.

Applications for personality inference draw on several data modalities of user-generated content, including multimedia~\cite{hickman2022automated,lukac2024speech}, biometric data~\cite{xu2022review,maliki2020personality}, and interaction markers~\cite{chen2023eye,meidenbauer2023mouse}. Among these, text is by far the most widely used modality, reflecting the central role of language in allowing individuals to describe themselves, others, and the world around them \cite{tausczik2010psychological}. Its prominence has lately led researchers to explore the potential of pre-trained Large Language Models (LLMs) for text-based automatic personality recognition, prompted by their outstanding results across multiple natural language processing tasks. Among encoder-based models, BERT has been the most extensively studied, with task-specific fine-tuning attempts, such as Personality BERT for personality classification \cite{jain2022personality}. More recent work has turned to generative LLMs explored in both zero-shot prediction \cite{ganesan2023systematic,treves2025viki} and conversational settings \cite{peters2024large}. Findings are mixed: LLMs appear to capture meaningful aspects of personality, with prediction accuracy varying across studies and traits. However, according to \cite{popa2025effective}, current general-purpose state-of-the-art LLMs---both proprietary and open-source---exhibit limited capability in extracting personality traits from text samples, particularly in uncontrolled scenarios, such as when using corpora not specifically collected for personality assessment. Overall, inferring personality from text remains a complex and challenging task, which calls for further research and experimentation. Moreover, increased attention should be devoted to dark traits, as almost all existing literature focuses on bright ones.

 \section{Methodology}
\label{sec:dataset}

\subsection{Participants and Procedures}

\multicomment{
\paragraph{Contributions} Our main contributions are as follows:
\begin{itemize}[noitemsep, topsep=3pt]
    \item We develop a dedicated Web application that integrates validated psychological questionnaires with participants’ real-world online activity data, enabling direct linkage of trait profiles to actual online behavior.
    \item We extract and analyze an extensive set of 219 linguistic and behavioral features from 56,892 Reddit comments, encompassing toxicity, emotional tone, moral framing, irony, and computational text-derived bright and dark trait estimates.
    \item We systematically compare self-reported constructs with online behavioral data, revealing that they primarily influence the production, rather than perception, of toxic interactions. Furthermore, we show that sadistic and psychopathic dispositions are most strongly associated with overtly toxic language, whereas other dark traits, such as entitlement and narcissism, manifest more subtly and often elude simple textual proxies.
    \item We demonstrate that existing hand-crafted text-based proxies for dark traits capture only partial aspects of validated psychological measures, highlighting the need for more robust computational models.
\end{itemize}
}

\label{sec:dataset-approach}

\begin{figure*}[t]
    \includegraphics[width=1\linewidth]{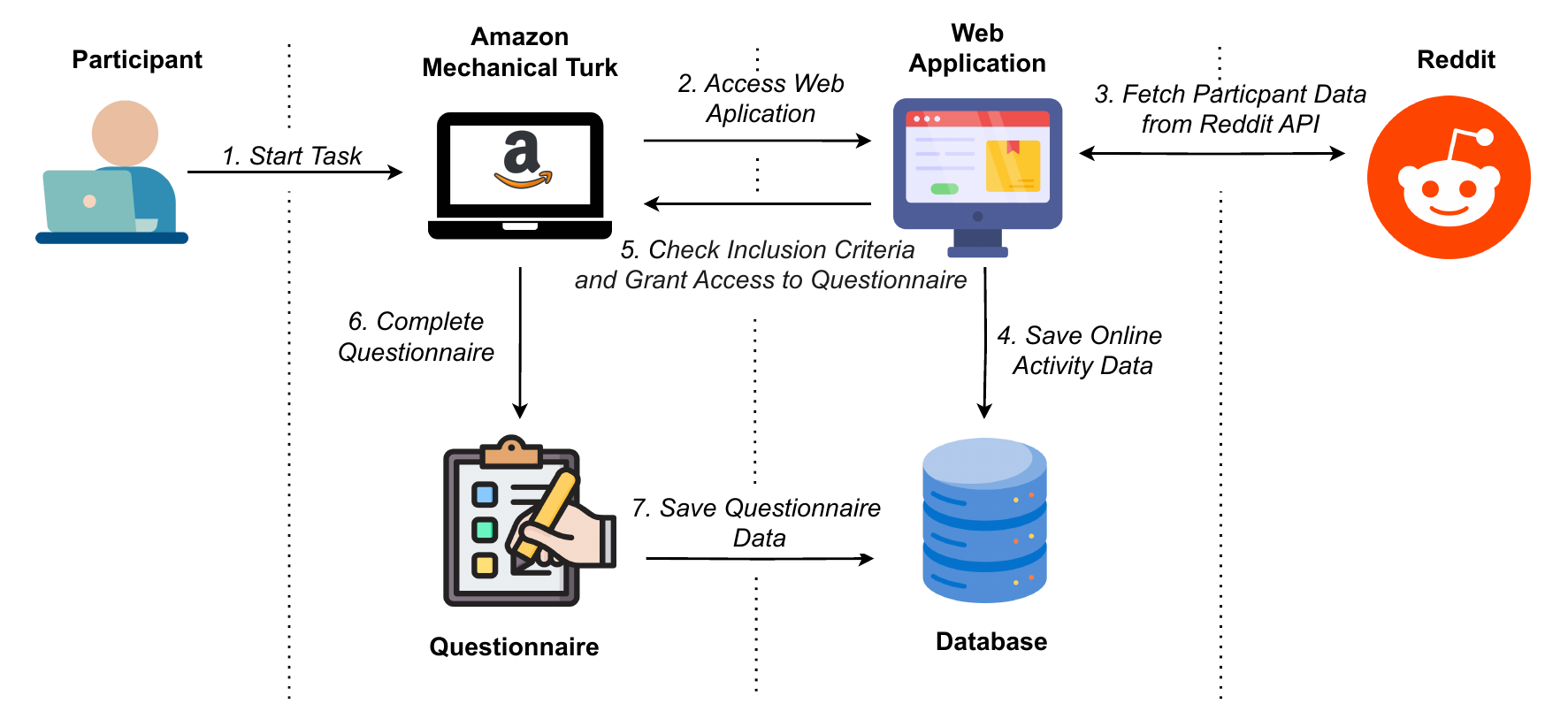}
    \centering
    \caption{Overview of our approach for linking answers to a validated psychological questionnaire with online activity data.}
    \label{fig:webapp}
\end{figure*}

Our study links rigorous psychometric measures with behavioral data derived from actual online user activity. For participant recruitment and data collection, we developed a dedicated Web application that interfaces directly with Reddit’s Application Programming Interface (API), enabling the automated collection of participant data under informed consent. The application was integrated with a crowdsourcing task deployed on Amazon Mechanical Turk, through which we recruited participants and administered a comprehensive validated psychological questionnaire. Through this combined approach, we obtained both individual-level personality assessments and the corresponding digital traces of Reddit activity. Our solution is visually presented in Figure~\ref{fig:webapp} and further described in the following. 

Participation in the crowdsourcing task was restricted to users located in the US, mirroring the prevalent user-base of Reddit~\cite{luong2022evaluating}. Before being presented with the questionnaire, prospective participants were briefed on the goal of the research.\footnote{This research received ethical approval from 
CNR's IRB (protocol \#0306210).} 
Then, they were asked to complete a fast procedure through which we linked each participant to their Reddit account. This was carried out automatically and transparently to the user via the dedicated Web application.\footnote{\url{https://ci.iit.cnr.it/pianoapp}}
In practice, participants were re-routed from within Amazon Mechanical Turk to the Web application, where they were shown detailed information about the study and were allowed to provide ìtheir informed consent. Upon consenting, participants were asked to authorize the application to collect data from their Reddit account.
Reddit authentication and data collection were handled securely via the official Reddit API.\footnote{\url{https://www.reddit.com/dev/api/}} Contextually, the application also checked that each participant had sufficient Reddit activity to justify their inclusion. Below is the list of mandatory inclusion criteria for this study:
\begin{itemize}[noitemsep, topsep=3pt]
    \item a Reddit account that was at least 30 days old at the time of participation;
    \item a Reddit account that had posted at least 50 comments;
    \item a Reddit account whose comments totaled a cumulative length of at least 1,500 tokens.\footnote{In natural language processing, a ``token'' is any unit of text resulting from tokenization, which may be a word, sub-word, multi-word expression, punctuation mark, number, or symbol. The exact definition of a token depends on the text processing methods applied. In this study, tokens were generated using spaCy’s default English tokenizer: \url{https://spacy.io/models/en\#en_core_web_sm}.}
\end{itemize}
These criteria ensure that our subsequent analyses are based on sufficient user data, thus supporting the validity of our research. Participants who did not provide informed consent or who did not meet all the above criteria were dropped from the study. Instead, eligible participants were redirected to Amazon Mechanical Turk and allowed to take the questionnaire. All participants who linked their account---independently on whether they were ultimately included in the study---were compensated with $0.05$ USD. In addition, participants who completed the psychological questionnaire were further awarded $3.22$ USD. Overall, 331 participants provided informed consent and linked their Reddit account, of whom 171 (52\%) met the inclusion criteria and were granted access to the questionnaire.

\subsection{Instruments}

\subsubsection{Questionnaire}
\label{sec:dataset-mturk}
The questionnaire consisted of 66 items\footnote{The full questionnaire is available at: \url{https://zenodo.org/records/17662973}.} organized in the following sections:
\begin{itemize}[noitemsep, topsep=3pt]
    \item \textit{Socio-Demographic}: 5 items covering age, gender, education, ethnicity, and political affiliation.
    \item \textit{Dark Side of Humanity Scale (DSHS)}~\cite{katz2022dark}: a 42-item self-report measure rated on a 6-point Likert scale ranging from \textit{Not at all like me} to \textit{Very much like me}. The instrument assesses four different factors: 
    \begin{itemize}[noitemsep, topsep=3pt]
        \item 18 items for Successful Psychopathy (SuccPsycho), which refers to a calculated, callous–manipulative style that combines primary psychopathy and Machiavellianism, characterized by strategic deceit, low remorse, and low impulsivity (example of item: \textit{``I am willing to be unethical if I believe it will help my plans succeed''});
        \item 9 items for Grandiose Entitlement (GrandEnt), which describes a pervasive belief in special privilege and superiority, expecting rule exemptions and deference, and a readiness to exploit others to boost an inflated self-image (example of item: \textit{``I deserve to receive special treatment''});
        \item 8 items for Sadistic Cruelty (SadCru), that underlines a tendency toward everyday sadism— taking pleasure in others’ suffering thought, speech, online behavior, or direct actions—driven by a desire for dominance or cruelty (example of item: \textit{``I enjoy watching people in pain''});
        \item 7 items for Entitlement Rage (EntRage), that refers to a dysregulated anger when perceived entitlements or recognition are denied, often expressed through quick-tempered, criticism-sensitive outbursts that reflect a vulnerable or defensive narcissistic core (example of item: \textit{``I get into a temper if I don’t get the recognition that I deserve''}).
    \end{itemize}
    Subscale scores were computed by averaging the responses to the items belonging to each factor.
    \item \textit{Cyberbully/Troll Deviancy Scale (CTDS)}~\cite{zezulka2016differentiating}: Trolling scale (Troll) containing 13 items that participants could rate from \textit{Never} to \textit{6 or more times}. Participants reported how often they had engaged in certain online behaviors over the past five years, explicitly targeting strangers or unknown users. Examples of items include \textit{``How often have you used profanity or insulting language towards a stranger online (just because)?''} and \textit{``How often have you posted rude comments or lies about a stranger that you did not know online?''}. As anticipated, we use the term “dimensions” to refer to both the dark tetrad traits and trolling behavior.
    \item \textit{Social Media Use}: 6 items focusing on social media habits. The first item asked participants to indicate which platforms they use regularly. 
    The remaining five items were meant to capture tendencies in social media use, admitting answers on a 5-point Likert scale ranging from \textit{Very Rarely} to \textit{Very Frequently}:
    \begin{itemize}[noitemsep, topsep=3pt]
        \item \textit{SM01:} On a weekly basis, how often do you consume content on social media?
        \item \textit{SM02:} On a weekly basis, how often do you create content on social media?
        \item \textit{SM03:} On a monthly basis, how often do you see uncivil comments/content on social media?
        \item \textit{SM04:} On a monthly basis, how often are you the target of uncivil comments/content on social media?
        \item \textit{SM05:} On a monthly basis, how often do you resort to using uncivil comments/content on social media?
    \end{itemize}
The last three items are particularly relevant to our study as they capture self-reported \textit{exposure to}, \textit{experience of}, and \textit{engagement in} uncivil or toxic online interactions. Further details on the development of \textit{SM01–SM05}, their inter-item correlations, and limitations are provided in~\ref{sec:app_soc_usage}.
\end{itemize}

The DSHS and CTDS multi-item scales showed satisfactory internal consistency, as reported in Appendix Table~\ref{tab:cronbach_alpha}. The questionnaire also included 10 control questions randomly shuffled between the psychological items to check for user attention. We rejected answers from participants who had excessively fast completion time (i.e., $< 6$ minutes) or $< 80\%$ correct answers to the control questions. Out of the 171 participants allowed to take the questionnaire, 114 (66\%) provided valid responses that we used for our subsequent analyses, 56 (32\%) were rejected, and one never submitted the questionnaire. Our sample size of 114 valid participants largely aligns with recent related literature~\cite{freyth2023social,yuan2022does,hossain2022you}. Appendix Figure~\mbox{\ref{fig:socio-demo-distribution}} and Figure~\mbox{\ref{fig:social_combs}} provide detailed socio-demographic and social media profiling of the final set of participants to the study.

\subsubsection{Behavioral Measures}
\label{sec:method-features}
To explore the relationship between dark personality traits, toxicity, and online behavior, we computed an extensive set of 219 features from a large dataset of comments made by each valid participant. Each of these features is computed for each participant by aggregating all of their comments. For clarity, we organize these features in the following groups. 

\paragraph{Basic Linguistic Characteristics}
Recent studies have brought to light the relationship between writing style and dark personality traits~\cite{borghi2025deceptive,bogolyubova2018dark}. Following this literature, we included a set of basic linguistic features that measure the number of comments per participant, and the mean and standard deviation of the number of sentences, tokens, and words. 

\paragraph{Toxicity}
One of the central aims of this study is to explore the relationship between dark traits and online toxicity. We therefore computed a large set of features capturing various facets of toxic and harmful language. We obtained toxicity scores from \texttt{Perspective API}---a well-known multilingual deep learning classifier~\cite{lees2022new} that is widely used for content moderation and in many studies about online toxicity~\cite{cima2025contextualized,trujillo2023one}. \texttt{Perspective API} provides the following scores---each in the $[0,1]$ range: toxicity, severe toxicity, obscenity, insult, identity attack, and threat. We submitted raw comment text to the API without additional preprocessing, relying on its character-level models designed to operate directly on noisy user-generated content~\cite{lees2022new}. From each score, we computed multiple features as the mean, maximum, minimum, standard deviation, median, 75th, and 90th percentile across each participant's comments. Finally, we computed two additional features as the frequency of comments with toxicity scores $> 0.25$ and $> 0.50$. These thresholds are motivated by the strongly left-skewed distribution of toxicity scores in social media data, where most values are close to zero~\cite{cima2024great}. The 0.25 cutoff captures deviations from baseline levels, while 0.5 identifies more severe toxicity. The latter threshold is commonly used in prior work~\cite{cima2025investigating,kumarswamy2023impact}.

\paragraph{Psycholinguistic Categories}
Linguistic Inquiry and Word Count (LIWC) is a lexicon that maps textual content to psycholinguistic categories, including cognitive processes, social concerns, and affective expression~\cite{boyd2022development}. Therefore, LIWC provides a way to quantify the underlying psychological and linguistic patterns in participants’ comments. By capturing how individuals express cognition, emotion, and social orientation through language, these measures allow examining the relationship between personality traits and the socio-linguistic characteristics of online behavior~\cite{borghi2025deceptive,bogolyubova2018dark}. For each participant, we averaged the scores returned by LIWC, which assigns a numerical continuous value to each category based on word usage, across all comments. Here, we used the latest version of the lexicon, which is LIWC-22.\footnote{\url{https://www.liwc.app/}} The numerosity of this group is particularly high, as it comprises more than half of the total (118 features, 52.7\%). For this reason, we further divided this group into three subgroups, namely \texttt{Psychological Processes} (40 features), \texttt{Lifestyle and Socio-Cultural Factors} (40 features), \texttt{Use of Language} (38 features) following the official LIWC-22 documentation.\footnote{\url{https://www.liwc.app/static/documents/LIWC-22\%20Manual\%20-\%20Development\%20and\%20Psychometrics.pdf}}

\paragraph{Emotion and Affect}
We used the \texttt{NRC Emotion Intensity Lexicon} (\texttt{NRC-EIL}) to quantify the strength of emotional expressions in user comments, as emotional intensity has been shown to play a key role in shaping online discourse and potentially amplifying toxic interactions~\cite{mohammad2013crowdsourcing}. \texttt{NRC-EIL} is a widely used lexicon that provides an intensity scores in the $[0,1]$ range for eight basic emotions: anger, anticipation, disgust, fear, joy, sadness, surprise, and trust. For each comment, we computed aggregated scores by averaging the intensities of matched words, allowing us to quantify the emotional strength of participants' online comments. Furthermore, we leveraged the \texttt{NRC Valence-Arousal-Dominance lexicon} (\texttt{NRC-VAD}) to capture affective aspects underlying user language, as variations in valence, arousal, and dominance are closely associated with personality traits and may influence the emergence of toxic or antagonistic behavior online~\cite{mohammad2018obtaining}. \texttt{NRC-VAD} maps words to three continuous affective components: valence (positivity \textit{vs.} negativity), arousal (intensity of emotion), and dominance (sense of control). Similarly to \texttt{NRC-EIL}, we computed features by averaging at the participant level the scores associated with each matched word.

\paragraph{Contextual Emotions}
In addition to the NRC-based lexicons, we also used \texttt{EmoAtlas}, a computational framework that provides a complementary perspective on emotional analysis~\cite{semeraro2025emoatlas}. While \texttt{NRC-EIL} and \texttt{NRC-VAD} are grounded in psychological theory and rely on word-level associations, \texttt{EmoAtlas} adopts a data-driven approach to infer emotional tone based on the overall context of a comment rather than individual words in isolation. It assigns standardized \textit{z}-scores for eight discrete emotions: joy, trust, fear, surprise, sadness, disgust, anger, and anticipation. By integrating resources originating from both psychology and computer science, we combine theoretically grounded emotion models with data-driven, context-sensitive computational methods. This joint use potentially allows capturing a richer, multi-faceted view of the emotional dynamics underlying participant comments.

\paragraph{Moral Foundations}
Moral foundations theory posits that human moral reasoning is structured around a set of universal categories, such as care \textit{vs.} harm, fairness \textit{vs.} cheating, loyalty \textit{vs.} betrayal, authority \textit{vs.} subversion, and purity \textit{vs.} degradation~\cite{graham2013moral}. Prior research has shown that these foundations influence how individuals evaluate social situations and communicate their values online~\cite{hopp2021extended}. Thus, understanding the moral framings used online by participants in our study can provide valuable insights into how individuals with different dark personality traits express attitudes and engage in potentially toxic discussions. To capture these framings, we employed \texttt{FrameAxis}~\cite{kwak2021frameaxis}, a method that quantifies the extent to which text leans toward specific conceptual poles derived from antonym pairs (e.g., authority \textit{vs.} respect, care \textit{vs.} harm). For example, participants high in narcissism may emphasize authority frames in discussions, while those high in psychopathy may exhibit reduced sensitivity to care. By incorporating these features, we move beyond surface-level emotion or toxicity measures and gain a deeper understanding of the moral constructs underpinning participants’ online language. For each comment, \texttt{FrameAxis} provides intensity scores for the virtue and vice of each moral axis (e.g., care.virtue and care.vice) as well as a bias score. These values were then averaged across each participant's comments, resulting in one set of moral framing features per individual.

\paragraph{Irony}
Irony is a subtle communicative strategy that can be used to express criticism or dissent by stating the opposite of what is actually meant. Furthermore, there is research showing that different dark personalities use irony in different ways~\cite{fanslau2023dark}. For this reason, we computed an irony score for each participant comment via a BERT-based deep learning model~\cite{barbieri2020tweeteval}. Then, we aggregated comment-level scores to obtain features such as the average, minimum, maximum value, and standard deviation for each participant.

\paragraph{Text-Derived Dark Triad Traits}
To further enrich our analysis and enable a deeper comparison between questionnaire-derived personality measures and computationally inferred markers, we incorporated text-based estimates of the Dark Triad traits based on the approach recently proposed by Borghi and Ratcharak~\cite{borghi2025deceptive}. In their study, the authors leveraged LIWC categories and insights from personality theory~\cite{yousaf2023dark} to define a set of hand-crafted formulas that provide Machiavellianism, narcissism, and psychopathy scores directly from textual data. They further showed that these scores are significantly correlated with fake review posting behavior~\cite{borghi2025deceptive}. Building on this work, we re-implemented the formulas by Borghi and Ratcharak and applied them to our dataset of Reddit comments, obtaining for each participant a set of text-derived scores corresponding to the three Dark Triad constructs. These features serve a dual purpose in our study: \textit{(i)} they allow us to explore the alignment between language-based personality estimates and validated questionnaire-based measures, and \textit{(ii)} they provide an additional lens to investigate how dark personality traits manifest in participants’ online discourse. 

\subsection{Data Analysis}
\label{sec:method-data}

\subsubsection{Dimension Scores and Labels}
We analyze the dimension scores derived for each participant in two complementary ways: \textit{(i)} using the continuous values directly, and \textit{(ii)} converting them into binary labels to indicate whether a participant exhibits a given dimension. To obtain the binary labels, we compute the mean score $\mu$ and standard deviation $\sigma$ for each dimension across all participants, and classify participant $i$ as exhibiting dimension $k$ if their score for that dimension exceeds the threshold $s_{i,k} > \mu_k + \sigma_k$. Similar procedures are commonly adopted to transform continuous scores into categorical representations, especially within the domain of personality classification~\cite{guleva2022personality}.  

\begin{figure*}[t]
    \centering
    \begin{subfigure}{0.85\textwidth}
        \includegraphics[width=1\linewidth]{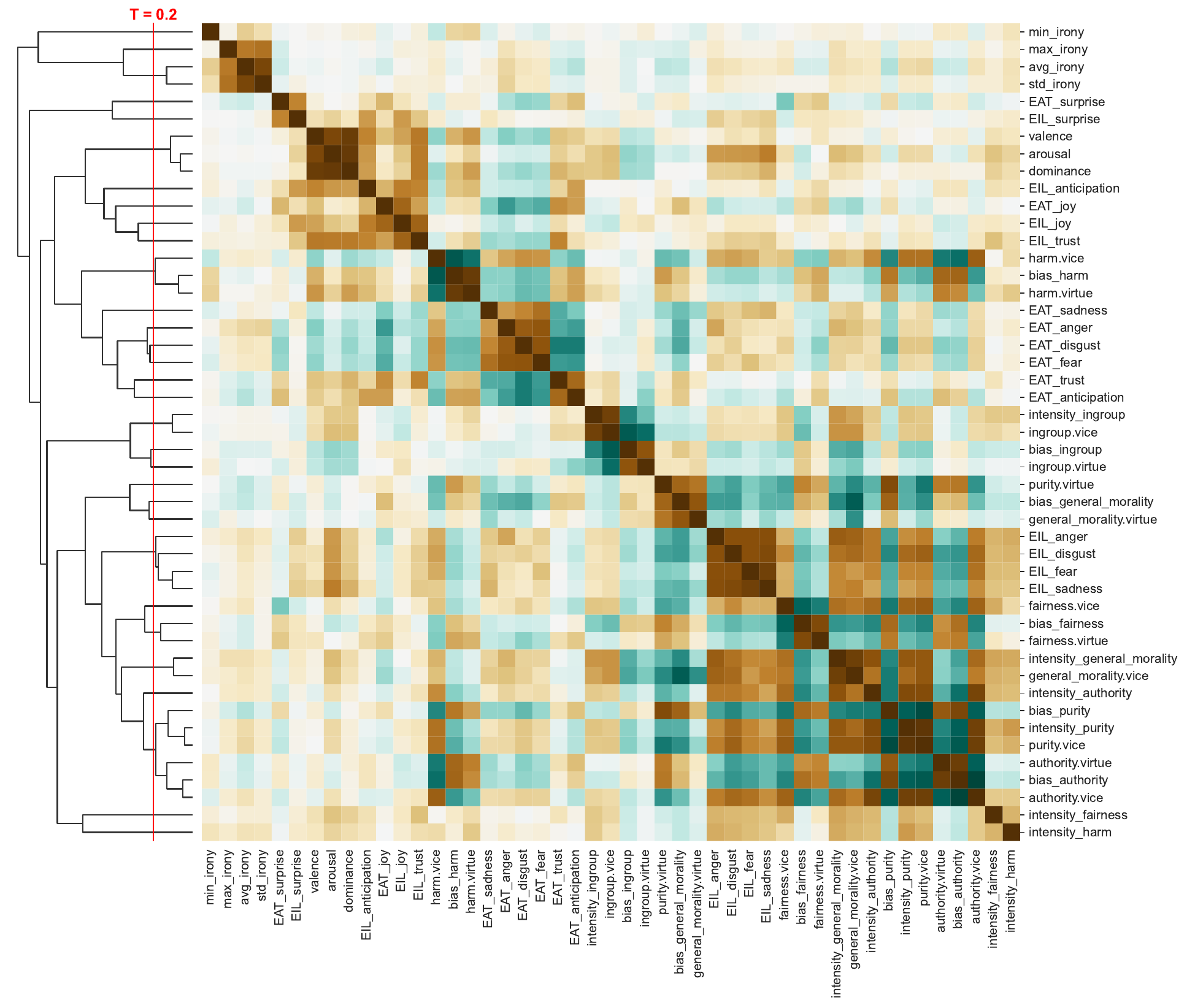}
        \centering
    \end{subfigure}
    \hspace{3pt}
    \begin{subfigure}{0.04\textwidth}
        \includegraphics[width=0.95\linewidth]{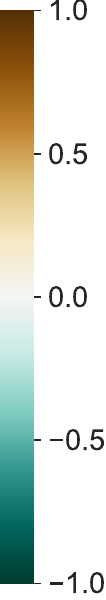}
        \vspace{1.5cm}
        \centering
    \end{subfigure}
    \caption{Pairwise Spearman correlation matrix for the features in the emotions, affect, irony, and moral foundations group, together with the hierarchical clustering dendrogram used for feature selection. Each cell represents the correlation between a pair of features, with color indicating the strength and direction of the association. The dendrogram on the left shows the hierarchical clustering structure and the vertical red line indicates the threshold used to define clusters. The block-like square patterns visible along the diagonal correspond to the clusters of strongly correlated features identified by this procedure. For each of these clusters, we select its medoid as the representative feature of the whole cluster.}
    \label{fig:dendrogram_other}
\end{figure*}

\begin{figure*}[t]
    \centering
    \begin{subfigure}{0.45\textwidth}
        \includegraphics[width=1\linewidth]{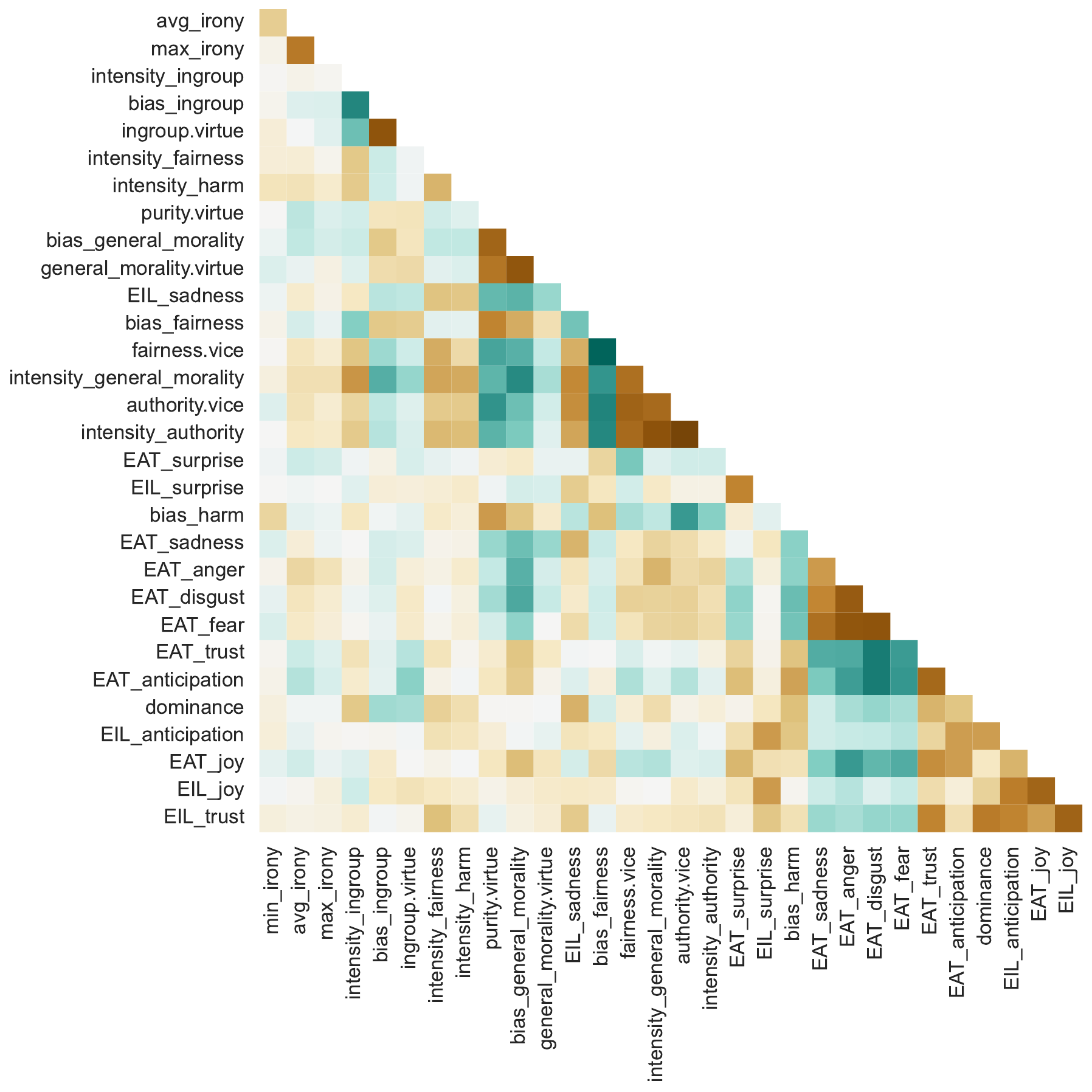}
        \centering
    \end{subfigure}
    \begin{subfigure}{0.075\textwidth}
        \includegraphics[width=0.5\linewidth]{pictures/colorbar_cropped.pdf}
        \vspace{2.85cm}
        \centering
    \end{subfigure}
    \caption{Pairwise Spearman correlation matrix for the selected features in the emotions, affect, irony, and moral foundations group after feature selection. The procedure removed 16 features (34\% reduction), yielding a more compact set with reduced redundancy while preserving the main information captured by the group.}
    \label{fig:selected_other}
\end{figure*}

\subsubsection{Feature Selection}
\label{sec:feature_selection}
To study the online behavior of the participants, we consider an initial set of 219 text-derived features, described in Section~\ref{sec:method-features}. Some of these features capture closely related aspects of behavior and are therefore strongly correlated, leading to redundancy and an inflated number of statistical tests without providing independent information. To address this issue, we perform a feature selection procedure aimed at reducing redundancy while preserving interpretability. The procedure consists of three main steps: \textit{(i)} correlation estimation, \textit{(ii)} clustering, and \textit{(iii)} representative feature selection. First, we compute pairwise Spearman correlations $\rho$ between features within each group. We then convert correlations into a distance measure $d_{ij} = 1 - |\rho_{ij}|$, so that highly correlated features are close in distance. Using this distance, we apply hierarchical clustering with average linkage. The resulting dendrogram is cut at a threshold of 0.2, corresponding to grouping features with absolute correlation $|\rho| \geq 0.8$. For each cluster identified in this way, we retain a single representative feature, defined as the feature with the highest average correlation with all others in the cluster (i.e., the medoid). Unlike dimensionality reduction techniques such as PCA, which yield latent components, this approach retains representative features that can be directly analyzed. This feature selection process yields a reduced set of 154 features (i.e., a reduction of 29\% compared to the original set), markedly decreasing redundancy and the effective number of hypotheses tested. The extent of reduction varies across feature groups. For instance, toxicity features are substantially reduced due to strong internal correlations (66\% reduction), whereas LIWC features are almost unaffected (11\% reduction), reflecting their lower degree of inter-correlation. To further exemplify the feature selection process, Figure~\ref{fig:dendrogram_other} shows the correlation matrix of the features in the emotions, affect, irony, and moral foundations group, together with the hierarchical clustering dendrogram and the threshold used to define clusters and select the representative features for this group. Instead, Figure~\ref{fig:selected_other} shows the correlation matrix of the remaining features of this group, after feature selection. As shown, some residual correlations persist among the selected features, but the overall degree of inter-correlation is substantially lower than in the original feature set. Inter-correlation and feature selection results for the remaining groups of features are reported in~\ref{sec:app_feature_selection}.

\subsubsection{Confirmatory and Exploratory Analyses}
\label{sec:analyses}
To investigate the relationship between dark traits, trolling tendencies, social media usage, and linguistic behaviors, we conduct two complementary strands of analyses: \textit{(i)} confirmatory and \textit{(ii)} exploratory. The confirmatory analyses address specific research questions formulated a priori. For each question, we examine a small and interpretable set of features---such as self-reported social media use, toxicity indicators, or text-derived estimates dark traits---and assess their correlations with dark trait scores and other relevant questionnaire answers. Given the limited number of comparisons in this stage, we control the family-wise error rate by applying the Bonferroni correction to adjust the correlation \textit{p}-values and account for multiple hypotheses testing. The exploratory analyses broaden the scope of this study to a much larger set of linguistic and behavioral features---including LIWC categories, affective lexicons, moral framing, and other fine-grained textual markers. While these features are less directly interpretable, they are valuable as potential signals for computational models of personality inference and online misbehavior~\cite{tessa2025beyond,tessa2025quantifying}. In this case, we correct the large number of correlation \textit{p}-values with the Benjamini–Hochberg procedure, which controls the false discovery rate and offers greater statistical power than more conservative alternatives in high-dimensional settings~\cite{thissen2002quick}. Across both strands, we rely on Spearman’s rank correlations $\rho$ to capture monotonic associations under the null hypothesis of no relationship ($\rho = 0$).

\subsubsection{Statistical Power and Sensitivity Analysis}
\label{sec:power}
All statistical analyses are conducted at the participant level ($N=114$), with linguistic and behavioral features aggregated per user. As such, statistical power is determined by the number of participants rather than the volume of textual data. To assess the sensitivity of our analyses given the sample size and the scope of hypothesis testing, we adopted a twofold strategy that distinguishes between confirmatory and exploratory analyses, reflecting their different inferential objectives and multiple-comparison procedures. For the confirmatory analyses, which involve relatively small families of hypotheses and Bonferroni correction, we conducted a sensitivity analysis based on the standard Fisher $z$-transformation for correlation tests. Specifically, for each family of tests, we computed the minimum population correlation required to achieve 80\% statistical power under the corresponding family-wise significance threshold and number of comparisons.

For the exploratory analyses, which involve larger families of hypotheses and the Benjamini–Hochberg correction, a single minimum detectable effect size cannot be derived analytically. We therefore performed a simulation-based sensitivity analysis tailored to each exploratory setting. For each family of tests, we generated synthetic datasets assuming a mixture of null and non-null associations, with 25\% of correlations set to a moderate effect size ($\rho=0.3$) and the remaining 75\% set to zero. For each simulated dataset, we computed \textit{p}-values using two-sided correlation tests with sample size $N=114$, applied the same Benjamini–Hochberg procedure used in the main analyses, and recorded the proportion of non-null effects successfully detected after correction. This procedure was repeated $r=1000$ times for each setting to obtain stable estimates of detection rates. The results of these analyses, both for the confirmatory and exploratory strands, are reported alongside the corresponding results sections.

To further strengthen the robustness of our findings, we employ bootstrap resampling to compute 95\% confidence intervals for all correlation estimates that preserve statistical significance after \textit{p}-values correction. Bootstrap resampling involves repeatedly drawing random samples with replacement from the observed data to approximate the variability of a statistic~\cite{davison1997bootstrap}. Results are reported in Appendix Table~\ref{tab:bootstrap}. This approach provides a non-parametric and distribution-free way to assess the stability of our results.
 \section{Results}

For each participant who gave their informed consent and met the inclusion criteria described in Section \ref{sec:dataset-approach}, we collected account-level metadata (e.g., total number of comments posted, account creation timestamp), the full text of each posted comment, and some comment metadata (e.g., creation timestamp, subreddit in which the comment was posted). 
In summary, we collected 2.2M tokens (mean per comment $\mu = 42.34$, $\sigma = 47.30$), 152,741 sentences ($\mu = 2.91$, $\sigma = 2.50$), 56,892 comments (mean per participant $\mu = 499.05$, $\sigma = 450.42$) posted by the 114 valid participants. Despite a relatively low number of participants, this dataset greatly exceeds that of related studies in terms of available textual data~\cite{bogolyubova2018dark}.

\begin{figure}[th!]
    \centering
    \begin{subfigure}{0.49\textwidth}
        \includegraphics[width=\linewidth]{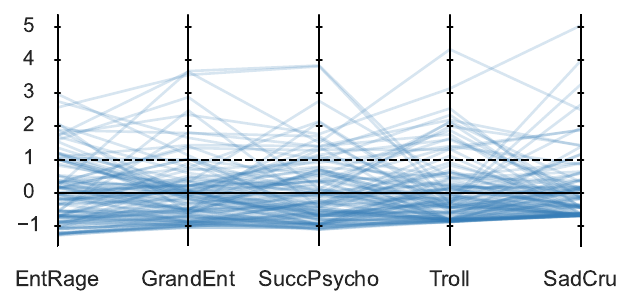}
        \caption{All participants ($N = 114$).}
        \label{fig:results-dist-all_norm}
    \end{subfigure}
    \hfill
    \begin{subfigure}{0.49\textwidth}
        \includegraphics[width=\linewidth]{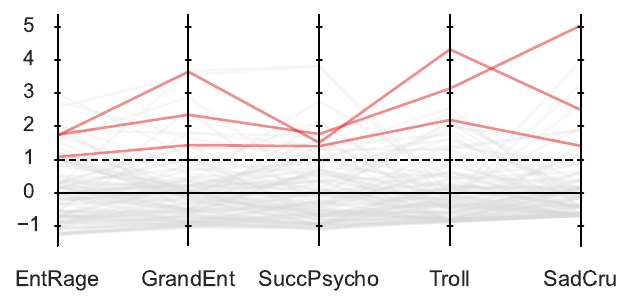}
        \caption{Participants exhibiting all traits ($N = 3$).}
        \label{fig:results-dist-all-dark_norm}
    \end{subfigure}
    \begin{subfigure}{0.49\textwidth}
        \includegraphics[width=\linewidth]{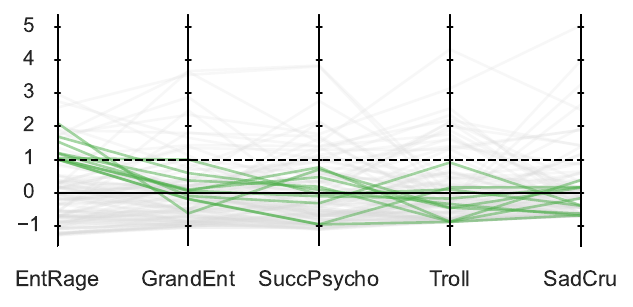}
        \caption{Participants exclusively exhibiting EntRage ($N = 10$).}
        \label{fig:results-dist-ent-rage_norm}
    \end{subfigure}
    \hfill
    \begin{subfigure}{0.49\textwidth}
        \includegraphics[width=\linewidth]{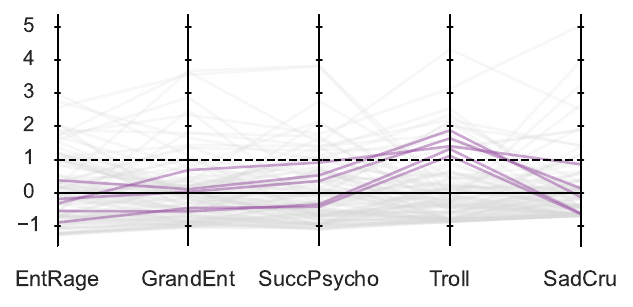}
        \caption{Participants exclusively exhibiting Trolling ($N = 5$).}
        \label{fig:results-dist-trolling_norm}
    \end{subfigure}
    \begin{subfigure}{0.49\textwidth}
        \includegraphics[width=\linewidth]{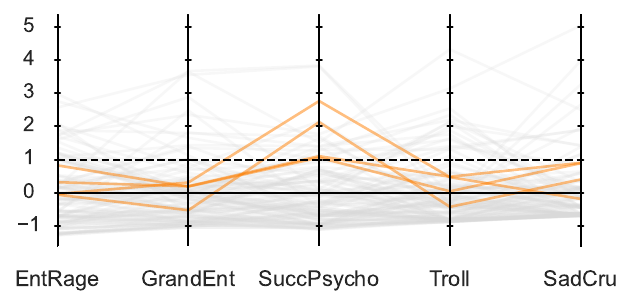}
        \caption{Participants exclusively exhibiting SuccPsycho ($N = 4$).}
        \label{fig:results-dist-succ-psycho_norm}
    \end{subfigure}
    \hfill
    \begin{subfigure}{0.49\textwidth}
        \includegraphics[width=\linewidth]{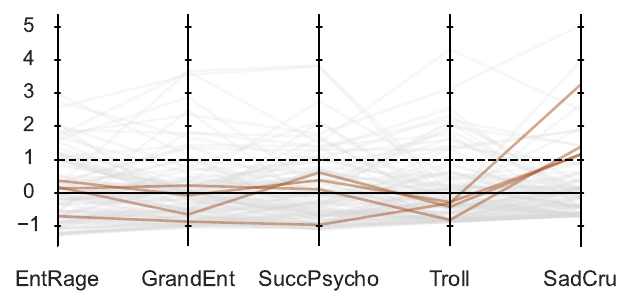}
        \caption{Participants exclusively exhibiting SadCru ($N = 4$).}
        \label{fig:results-dist-sad-cru_norm}
    \end{subfigure}
    \caption{Distribution of self-reported trait and trolling behavior scores. Scores are standardized via z-transformation. Each vertical axis represents one of the measured dimensions, and each line shows a participant’s scores across all traits and trolling behavior. The solid and dashed horizontal lines display the mean and the threshold of the dimensions, respectively. Dimensions are ordered from left to right by decreasing median score. Figure~\ref{fig:results-dist-all_norm} shows all participants, while the remaining ones highlight subsets of participants exhibiting certain dimensions.}\label{fig:parCoords_norm}
\end{figure}

\subsection{Dimension Distributions and Sample Characteristics}
We provide an overview of the distribution of our dimensions within our sample and examine how they vary across some socio-demographic groups, establishing the descriptive context for the subsequent analyses.

\subsubsection{Prevalence and Co-Occurrence of Dimensions}
Figure~\ref{fig:parCoords_norm} shows the distributions of the scores assigned to each participant based on their responses to the questionnaire. To enable meaningful comparisons across subscales, we standardized the scores using a z-transformation. Specifically, each score was centered by subtracting the mean $\mu$ and scaled by the standard deviation $\sigma$, such that $z = \frac{s - \mu}{\sigma}$, resulting in distributions having $\mu=0$ and  $\sigma=1$. The distribution of the raw (non-standardized) scores is instead reported in~\ref{sec:app_original_scores}. In Figure~\ref{fig:parCoords_norm}, each vertical axis corresponds to a dimension and each line corresponds to the scores of a participant across the dimensions. The horizontal solid line and the dashed line indicate the mean $\mu$ and the $\mu + \sigma$ threshold, respectively. Dimensions are ordered from left to right by decreasing median score, as per Table~\ref{tab:binarization_thresholds}. Figure~\ref{fig:results-dist-all_norm} shows all participant scores. Instead, Figure~\ref{fig:results-dist-all-dark_norm} highlights participants whose scores exceed the threshold for all dimensions ($N = 3$), while Figures~\ref{fig:results-dist-ent-rage_norm} to~\ref{fig:results-dist-sad-cru_norm} highlight participants whose scores exceed the threshold exclusively for EntRage ($N = 10$), Trolling ($N = 5$), SuccPsycho ($N = 4$), and SadCru ($N = 4$), respectively. The distributions shown in Figure~\ref{fig:parCoords_norm} vary considerably across dimensions. Scores for EntRage, GrandEnt, and SuccPsycho are overall higher and more evenly spread across the scale, indicating greater variability in participant responses. In contrast, scores for Trolling and SadCru are markedly lower and more homogeneous, with most participants clustering near the low end and only a few exhibiting large values. At the individual level, Figure~\ref{fig:parCoords_norm} reveals distinct behavioral patterns. A small subset of participants consistently score highly across multiple dimensions, as evidenced by relatively horizontal lines positioned above the bulk of other trajectories. Conversely, the presence of pronounced spikes surfaces cases where some participants score highly on a single dimension but remain low on other ones. Despite these phenomena, the majority of participants display uniformly low scores across dimensions, reflected by the dense concentration of flat lines near the lower range of the scales.

\begin{table}
\centering
\small
\setlength{\tabcolsep}{5.3pt}
\begin{tabular}{lccccccccc}
    \toprule
    & & \multicolumn{8}{c}{\textbf{summary statistics}} \\
    \cmidrule{3-10}
    \textbf{dimension} & \textbf{participants} & \textit{min} & \textit{25\textsuperscript{th}} & \textit{median} & \textit{75\textsuperscript{th}} & \textit{max} & \textit{IQR} & \textit{skewness} & \textit{kurtosis} \\
    \bottomrule
    EntRage & 24 (21\%) & -1.24 & -0.71	& -0.19 & 0.47 & 2.95 & 1.18 & 0.87 & 0.12 \\
    GrandEnt & 13 (11\%) & -1.00 & -0.74 & -0.20 & 0.19 & 3.66 & 0.93 & 1.76 & 3.36 \\
    SuccPsycho & 19 (17\%) & -1.07 & -0.76 & -0.28 & 0.52 & 3.84 & 1.28 & 1.43 & 2.37 \\
    Troll & 17 (15\%) & -0.84 & -0.84 & -0.31 & 0.34 & 4.28 & 1.17 & 1.67 & 2.84 \\
    SadCru & 13 (11\%) & -0.66 & -0.66 & -0.41 & 0.13 & 5.04 & 0.79 & 2.51 & 7.44 \\
    \bottomrule
    \end{tabular}
\caption{Summary statistics of standardized scores per dimension, and participants who exhibit those dimensions. Dimensions are ordered by decreasing median score.}
\label{tab:binarization_thresholds}
\end{table}
 
\begin{figure*}[t]
    \includegraphics[width=0.9\textwidth]{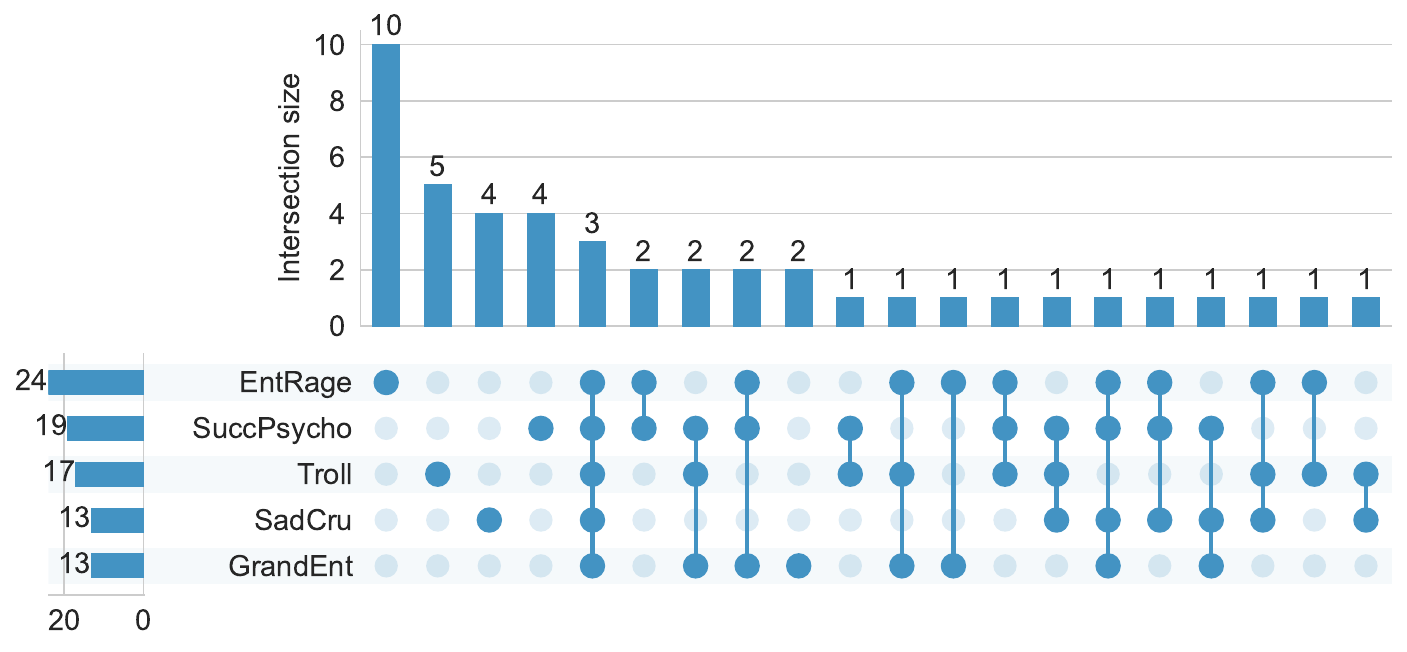}
    \centering
    \caption{Distribution of the combinations of dimensions exhibited by participants in our study. Dimensions and their combinations are shown from top to bottom and from left to right in decreasing order of frequency.}
    \label{fig:traits_combs}
\end{figure*}

Table~\ref{tab:binarization_thresholds} reports the number and percentage of participants exhibiting each dimension. Based on this binary assignment, 45 (39.5\%) participants exhibit at least one dimension, while the remaining 69 (60.5\%) exhibit none. Additionally, the table summarizes the distributional properties of each dimension, including measures of central tendency, dispersion, and shape. Figure~\ref{fig:traits_combs} displays the frequency of each dimension when exhibited individually, as well as that of all possible co-occurrences---that is, when one or more participants simultaneously exhibit more than one dimension. Dots indicate the dimensions considered in a given combination, while the height of the corresponding bar represents the number of participants who scored above threshold for all dimensions of that combination. We verified the robustness of the results reported in Table~\ref{tab:binarization_thresholds} and Figure~\ref{fig:traits_combs} by repeating the analysis under alternative threshold values, finding that the relative ranking and distribution of trait combinations remained highly consistent across specifications.
Figure~\ref{fig:traits_combs} reveals that the most frequent combinations involve participants exhibiting a single dimension in isolation, with intersection sizes ranging from $N=10$ participants for EntRage, down to $N=4$ for SuccPsycho. These cases are highlighted in Figures~\ref{fig:results-dist-ent-rage_norm} to~\ref{fig:results-dist-sad-cru_norm}. This trend holds across all dimensions except GrandEnt, which appears individually in only two cases, suggesting that this dimension is more commonly expressed alongside others. Three participants exhibit all five dimensions simultaneously, representing a small but distinct subgroup characterized by broadly elevated scores, as highlighted in Figure~\ref{fig:results-dist-all-dark_norm}. Beyond these cases, all other co-occurrences involve fewer than three participants, indicating that high scores across multiple dimensions are generally rare within our sample. These results align with previous research showing that dark personality traits are interrelated yet largely independent constructs~\cite{paulus2002dark,delisi2021overlapping}.

\subsubsection{Socio-Demographic Factors}
To explore potential links between our dimensions and socio-demographic factors, here we analyze differences in participant scores across \textit{(i)} gender, \textit{(ii)} educational background, and \textit{(iii)} political orientation. Each of these socio-demographic questions offered more than two response options, but some categories were seldom selected, while others could be meaningfully combined with minimal loss of information, as shown in Appendix Figure~\ref{fig:socio-demo-distribution}. To simplify the comparison between participant subgroups, we therefore grouped responses to each of the considered socio-demographic questions into two categories. Answers for gender included females ($N=59$ participants), males ($N=50$), and non-binary individuals ($N=5$). The latter were dropped to due to their low statistical relevance. For educational background, participants reported one of three categories: high school or less ($N=17$), some college ($N=28$), college graduate or more ($N=69$). The first two categories were merged into a single non-graduate group, while the third was retained as graduate. Finally, for political orientation, participants indicated Democratic ($N=56$), Lean Democratic ($N=34$), Lean Republican ($N=15$), Republican ($N=9$). We grouped these into two larger categories: Democrat ($N=90$) and Republican ($N=24$). For each socio-demographic attribute we then applied a Mann–Whitney \textit{U} test to compare dimension scores between the two groups.

\begin{table}
\centering
\small
\setlength{\tabcolsep}{4pt}
\begin{tabular}{lcccrp{5pt}ccrlp{5pt}ccrl}
    \toprule
    && \multicolumn{3}{c}{\textbf{gender}} & & \multicolumn{4}{c}{\textbf{education}} & & \multicolumn{4}{c}{\textbf{political affiliation}} \\
    \cline{3-5} \cline{7-10} \cline{12-15}
    \textbf{dimension} && \textit{female} & \textit{male} & \multicolumn{1}{c}{\textit{U}} & & \textit{non-grad.} & \textit{grad.} & \multicolumn{1}{c}{\textit{U}} && & \textit{Dem.} & \textit{Rep.} & \multicolumn{1}{c}{\textit{U}} & \\
    \bottomrule
    SuccPsycho && 1.801 & 1.832 & -0.295 & & 1.864 & 1.795 & -0.009 & & & 1.810 & 1.870 & -0.587 & \\
    GrandEnt && 1.887 & 1.799 & 0.569 & & 1.735 &  1.905 & -1.699 & * & & 1.736 & 2.218 & -2.193 & ** \\
    SadCru && 1.305 & 1.327 & 0.128 & & 1.294 & 1.333 & -0.316 & & & 1.305 & 1.364 & 0.070 & \\
    EntRage && 2.069 & 1.937 & 0.769 & & 1.994 & 2.023 & -0.177 & & & 2.018 & 1.990 & 0.118 & \\
    Troll && 1.495 & 1.514 & -1.116 & & 1.505 & 1.480 & -0.128 & & & 1.394 & 1.850 & -2.332 & ** \\
    \bottomrule
    \multicolumn{15}{l}{\footnotesize *: $p < 0.1$, **: $p < 0.05$}
    \end{tabular}
\caption{Differences in dimension scores between participants belonging to different socio-demographic groups. For each dimension and socio-demographic attribute, we report mean scores of the two groups, computed on raw, non-standardized dimension scores, and the results of a Mann–Whitney \textit{U} test. No difference remains significant after Bonferroni correction.}
\label{tab:traits_persData}
\end{table}
 
Table~\ref{tab:traits_persData} reports the results of this analysis. After applying Bonferroni correction, none of the tested differences remain significant. Before correction, some differences reach conventional significance thresholds, particularly for GrandEnt and Troll scores across political affiliation and education. However, these effects do not survive correction for multiple comparisons and should therefore be interpreted with caution. Overall, these results suggest that socio-demographic factors show limited and non-robust associations with the measured dimensions in this sample.

Finally, to further contextualize our sample, we compared its DSHS dimension scores with those reported in the original validation study by Katz et al.~\cite{katz2022dark}. The comparison is presented in~\ref{sec:dshs-validation}. We found that while our participants exhibit slightly lower levels of dark trait endorsement---particularly males, the differences remain within one standard deviation of the reference distribution, indicating that our sample is not markedly atypical.

\subsection{Confirmatory Analyses}
\label{sec:results-confirmatory}
This section presents the results of the confirmatory analyses that address our research questions. The sensitivity analysis described in Section~\ref{sec:power} indicates that, given the sample size and Bonferroni correction, these confirmatory analyses have 80\% power to detect correlations of approximately $\rho \approx 0.31$--$0.34$, depending on the number of tests. Thus, these analyses are primarily sensitive to moderate effect sizes, while smaller associations may not be reliably detected. Uncorrected associations are reported for completeness but are not interpreted as robust findings in light of multiple-comparison adjustments.

\begin{figure*}[t]
    \includegraphics[width=\textwidth]{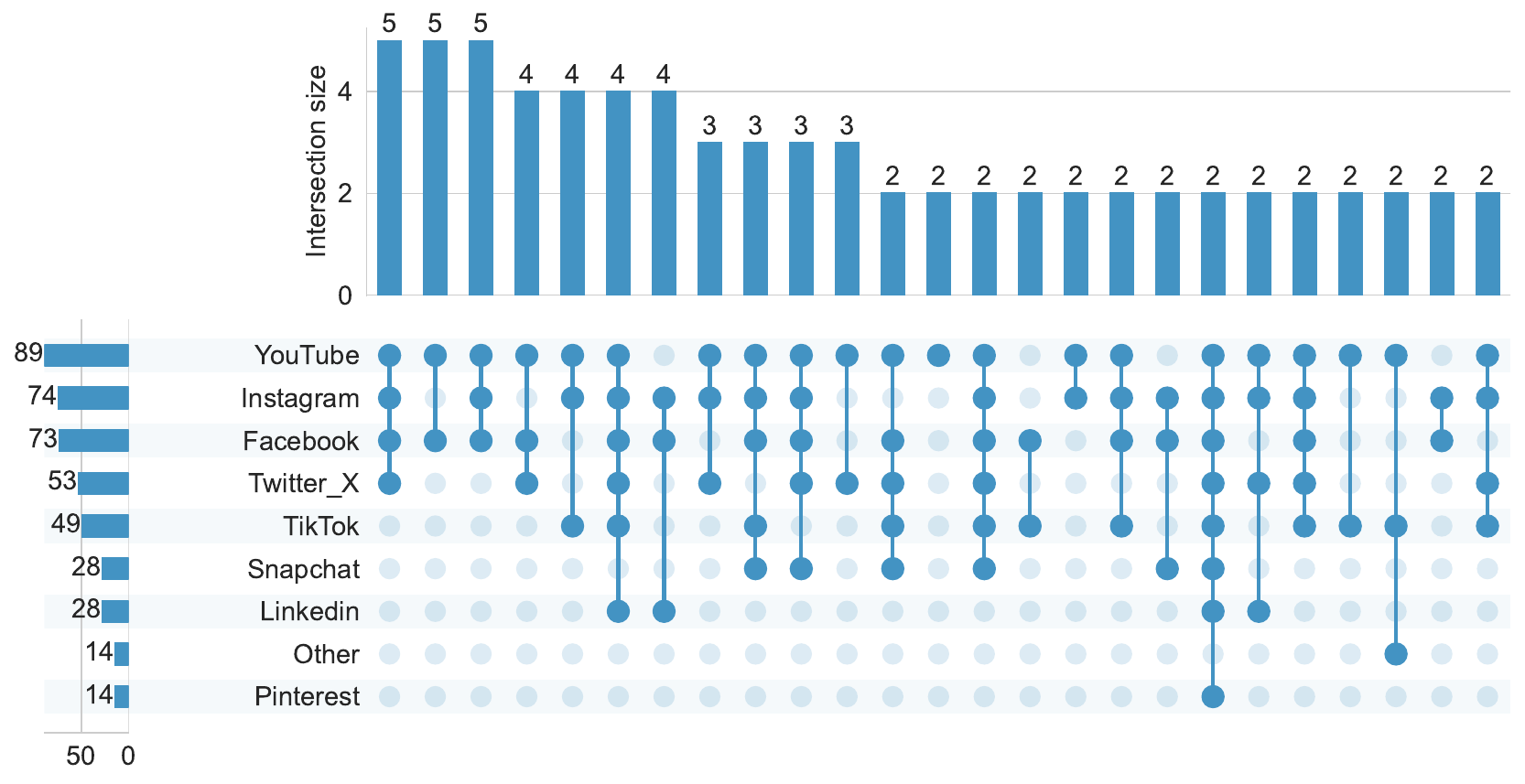}
    \centering
    \caption{Distribution of the combinations of platforms used, in addition to Reddit, by participants in our study. Platforms and their combinations are shown from top to bottom and from left to right in decreasing order of frequency. Only combinations with 2+ occurrences are shown.}
    \label{fig:social_combs}
\end{figure*}

\subsubsection{Self-Reported Social Media Use}
Figure~\ref{fig:social_combs} shows the combinations of platforms mostly used by the 114 participants to our study.\footnote{We note that, in addition to the platforms explicitly shown in Figure~\ref{fig:social_combs}, all participants also regularly used Reddit, as that was a strict inclusion criterion for this study.} As shown in figure, most participants used YouTube, Instagram, Facebook, and X, which is overall consistent with the socio-demographic characteristics of our sample (e.g., the age distribution, as per panel \textit{A} of Appendix Figure~\ref{fig:socio-demo-distribution})~\cite{gottfried2024americans}. Conversely, platforms centered around short-form multimedia content and more popular among younger users---such as TikTok and Snapchat---were used by a comparatively smaller number of participants. Similarly, both the professional networking platform LinkedIn and Pinterest showed lower adoption rates within our sample.

\begin{table}[t]
\centering
\small
\begin{tabular}{lcrlccrlrl}
    \toprule
    && \multicolumn{8}{c}{\textbf{self-reported social media use}} \\
    \cmidrule{3-10}
    \textbf{dimension} && \multicolumn{2}{c}{\textit{SM01}} & \textit{SM02} & \textit{SM03} & \multicolumn{2}{c}{\textit{SM04}} & \multicolumn{2}{c}{\textit{SM05}}\\
    \midrule
    SuccPsycho && 0.162 & * & -0.018 & -0.025 & 0.112 & & 0.336 & \uline{***} \\
    GrandEnt && -0.035 && -0.062 & -0.062 & -0.005 && 0.160 & * \\
    SadCru && -0.002 &&  { 0.046} & { 0.005} & 0.149 && 0.225 & ** \\
    EntRage && 0.069 && -0.021 & { 0.095} & 0.168 & * & 0.262 & *** \\
    Troll && 0.116 && -0.076 & { 0.078} & 0.063 && 0.365 & \uline{***} \\
    \bottomrule
    \multicolumn{10}{l}{\footnotesize *: $p < 0.1$, **: $p < 0.05$, ***: $p < 0.01$}
\end{tabular}
\caption{Spearman rank correlation coefficients between dimension scores and social media habits. \textit{SM01}--\textit{SM05} correspond to the questions from the \textit{Social Media Use} section of the questionnaire. Asterisks denote statistical significance of the uncorrected correlation \textit{p}-values. Underlined asterisks denote statistical significance after Bonferroni correction.}
\label{tab:traits_socUsage2}
\end{table}

\begin{figure*}[t]
    \centering
    \begin{subfigure}{0.49\textwidth}
        \includegraphics[width=\linewidth]{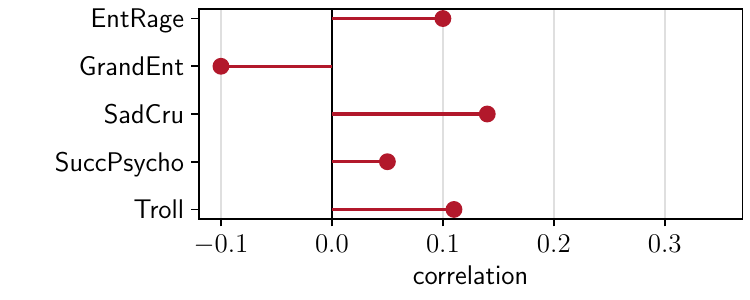}
        \centering
    \caption{Dimension scores and median toxicity.}
    \label{fig:selected_toxicity-traits}
    \end{subfigure}
    \begin{subfigure}{0.49\textwidth}
        \includegraphics[width=\linewidth]{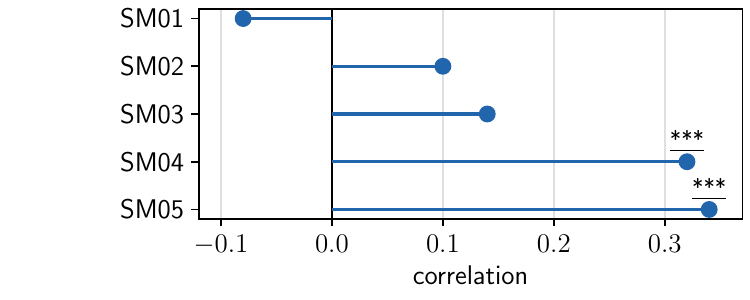}
        \centering
    \caption{Social media habits and median toxicity.}
    \label{fig:selected_toxicity-sm-habits}
    \end{subfigure}
    \caption{Spearman rank correlation coefficients between median toxicity, dimension scores, and social media habits. SM01–SM05 correspond to the questions from the \textit{Social Media Use} section of the questionnaire. Asterisks denote statistical significance of the uncorrected correlation \textit{p}-values. Underlined asterisks denote statistical significance after Bonferroni correction. ***: $p < 0.01$}
    \label{fig:selected_toxicity}
\end{figure*}

To address the first research question---\textit{Do dark traits correlate with self-reported experiences of online incivility?}---we compute Spearman rank correlations between participants' dimensions and their ordinal answers to the five items on social media use. Results are reported in Table~\ref{tab:traits_socUsage2}, along with
the statistical significance of both uncorrected and corrected (underlined) \textit{p}-values. For questions \textit{SM01} to \textit{SM04}, most correlations are weak and not statistically significant. Instead, all investigated dimensions have significant positive correlations with \textit{SM05}, which concerns the production of uncivil comments. Trolling and SuccPsycho exhibit moderate and highly significant correlations, even after correction for multiple hypotheses testing, followed by EntRage, SadCru, and GrandEnt. These results suggest that while our dimensions are only marginally related to how often individuals consume or are targeted by uncivil content online, they are more strongly associated with the production of such content. Furthermore, they show that the strength of this relationship varies between the different dimensions.

\subsubsection{Engagement in Toxic Speech}
Next, we examine rank correlations between participant scores in dark traits and trolling behavior and a text-derived measure of toxicity computed from their Reddit comments. This analysis directly addresses the question: \textit{Do dark traits correspond to actual production of toxic content?} As shown in Figure~\ref{fig:selected_toxicity-traits}, no dimension exhibits a statistically significant correlation with median toxicity after Bonferroni correction. While a few weak positive associations are observed at the uncorrected level (e.g., for SadCru, Trolling, and EntRage), none of these survive correction. Thus, in this sample, we found no evidence of moderate or larger associations between validated dark personality measures and the production of toxic content, as captured by our text-derived toxicity indicators. Smaller associations may nevertheless exist, but could not be reliably detected under the present design, feature set, and sample size. This null result represents the primary outcome of this analysis. It indicates that, despite theoretical expectations linking dark traits to antisocial or harmful behaviors, such relationships are not robustly associated with observable toxicity patterns in participants’ Reddit activity, at least under the present measurement and analytical framework.

In contrast, correlations involving self-reported social media behaviors reveal a different pattern. Figure~\ref{fig:selected_toxicity-sm-habits} shows no significant associations for general consumption (\textit{SM01}) or production (\textit{SM02}) of online content, and for exposure to toxic content (\textit{SM03}). However, both being the target of toxic interactions (\textit{SM04}) and admitting to producing them (\textit{SM05}) show moderate and statistically significant correlations with median toxicity, even after correction. These findings suggest that participants display a degree of self-awareness regarding their own online behavior, as self-reported production of uncivil content aligns with observed toxicity. Interestingly, the comparable strength of associations for \textit{SM04} indicates that individuals who report being targeted by toxic behavior are also more likely to produce it. This symmetry suggests a form of perpetrator–victim reciprocity, aligning with prior research showing that toxic exchanges online often develop through reciprocal and escalating interaction patterns, in which exposure to hostility increases the likelihood of responding in kind~\cite{cheng2017anyone,erreygers2018positive}.

Overall, the absence of significant associations between DSHS dimensions and text-derived toxicity after correction should not necessarily be interpreted as evidence that dark traits are unrelated to harmful online behavior. Rather, it may indicate that these traits operate as distal dispositions whose behavioral expression depends on situational and contextual factors. Recent longitudinal evidence~\cite{wang2026dark} suggests that dark personality traits predict cyber aggression primarily indirectly, through toxic online disinhibition, rather than through a direct effect. In this regard, several studies~\cite{kurek2019did,wu2023individuals} have demonstrated that individuals high in antisocial traits are more likely to construe online environments as unregulated and anonymous, which in turn heightens their propensity to engage in aggressive behaviors in digital contexts. In addition, a recent meta-analysis~\cite{barlett2024meta} has shown that the majority of predictors and outcomes of cyber aggression remain significantly associated with cyber aggression perpetration and victimization even after controlling for traditional offline misbehavior, supporting the view that cyber aggression reflects partially distinct processes rather than merely an online extension of offline aggression.

\subsubsection{Text-Derived Dark Triad Traits}
In~\cite{borghi2025deceptive}, Borghi and Ratcharak proposed three hand-crafted formulas to compute text-derived estimations of psychopathy, Machiavellianism, and narcissism, grounded in an extensive body of prior research on the psycholinguistic markers of Dark Triad traits~\cite{preotiuc2016studying,sumner2012predicting,bogolyubova2018dark,holtzman2019linguistic}. Here, we compute rank correlations between the scores produced by these formulas and those derived from our questionnaires.

This analysis contributes to answering the following question: \textit{Can hand-crafted text-derived formulas accurately estimate dark traits?} 

After correction, Table~\ref{tab:traits_darkTriad} reports no statistically significant correlations between our scores and those obtained from Borghi and Ratcharak’s formulas for psychopathy and narcissism. Instead, at the uncorrected level, their estimates of Machiavellianism showed positive correlations with EntRage, SuccPsycho, and SadCru. Notably, the two approaches used to measure dark traits present conceptual differences, rather than measurement limitations. In particular, Successful Psychopathy in DSHS represents a unique factor combining Machiavellianism and primary (low-impulsivity) psychopathy, whereas the formulas in \cite{borghi2025deceptive} capture more antisocial characteristics consistent with secondary (high-impulsivity) psychopathy. In addition, while the Dark Triad conceptualizes narcissism as a single dimension, the DSHS distinguishes between two separate facets: Grandiose Entitlement, which reflects a grandiose narcissistic style characterized by beliefs of superiority, deservingness, and special treatment, and Entitlement Rage, which captures a more vulnerable and reactive form of narcissism marked by anger and resentment when such expectations are frustrated. Importantly, the DSHS also incorporates the factor of Sadistic Cruelty, referring to the dispositional tendency to derive pleasure from observing or inflicting physical or psychological pain and humiliation on others. Consequently, the lack of correlation between text-derived and self-report dark traits should be interpreted cautiously, as it may stem from differences in construct definitions, measurement strategies, or a combination of both. Given the conceptual difference discussed above, the two measures are not directly comparable. This limited convergence highlights the challenges of developing reliable textual proxies for the more complex dark personality framework of the DSHS, and points to the need for further refinement in this area.

\begin{table}
\centering
\small
\begin{tabular}{lccrlc}
\toprule
 && \multicolumn{4}{c}{\textbf{text-derived Dark Triad traits}}\\
 \cmidrule{3-6}
 \textbf{dimension} && \textit{Psychopathy} & \multicolumn{2}{c}{\textit{Machiavellianism}} & \textit{Narcissism} \\
 \midrule
 SuccPsycho && -0.026 & {\qquad\;\;0.182} & * & -0.103 \\
 GrandEnt && -0.057 & -0.014 & & -0.043 \\
 SadCru && -0.120 & 0.174 & * & -0.060 \\
 EntRage && { 0.052} & 0.196 & ** & { 0.007} \\
 Troll && { 0.020} & 0.069 & & -0.097 \\
 \bottomrule
 \multicolumn{6}{l}{\footnotesize *: $p < 0.1$, **: $p < 0.05$}
\end{tabular}
\caption{Spearman rank correlation coefficients between dimension scores and text-derived Dark Triad scores based on the approach from~\cite{borghi2025deceptive}. Asterisks denote statistical significance of the uncorrected correlation \textit{p}-values. No correlation is significant after Bonferroni correction.}
\label{tab:traits_darkTriad}
\end{table}
 
\multicomment{
\subsubsection{Text-Derived Bright Traits}
Finally, we assess rank correlations between our dimension scores and text-derived ones for bright (i.e., Big Five) traits, thus addressing the question: \textit{What is the relationship between bright and dark traits?} Results are presented in Table~\ref{tab:traits_bigFive}. The majority of correlations are non-significant. Yet, two patterns emerge at the uncorrected level. Conscientiousness shows weak positive correlations with both GrandEnt and SadCru, and extraversion is negatively correlated with Trolling. However, none of these associations remain significant after applying the Bonferroni correction. While these results should therefore be interpreted cautiously, they tentatively suggest that individuals with higher self-discipline and goal orientation can still display entitlement or callousness in specific contexts. More notably, the negative correlation between extraversion and trolling aligns with psychological theory in that individuals who are more sociable and oriented toward positive interactions are less inclined to engage in disruptive or antagonistic online behavior~\cite{leng2020bridging,xu2024meta}.}

\subsection{Exploratory Analyses}
\label{sec:results-exploratory}
Here, we present findings of the exploratory analyses designed to uncover possible additional behavioral patterns. The simulation-based sensitivity analysis described in Section~\ref{sec:power} shows that, under the Benjamini–Hochberg correction and the observed testing scope, approximately 79--80\% of true correlations of magnitude $\rho=0.3$ are recovered. This indicates that, despite the large number of hypotheses, the exploratory analyses are adequately powered to detect moderate associations, while smaller effects may be less consistently identified. Some uncorrected associations are reported for completeness but are not interpreted as robust findings given the multiple testing context.

\begin{figure*}[t]
    \centering
    \begin{subfigure}{1\textwidth}
        \includegraphics[height=0.3\textheight]{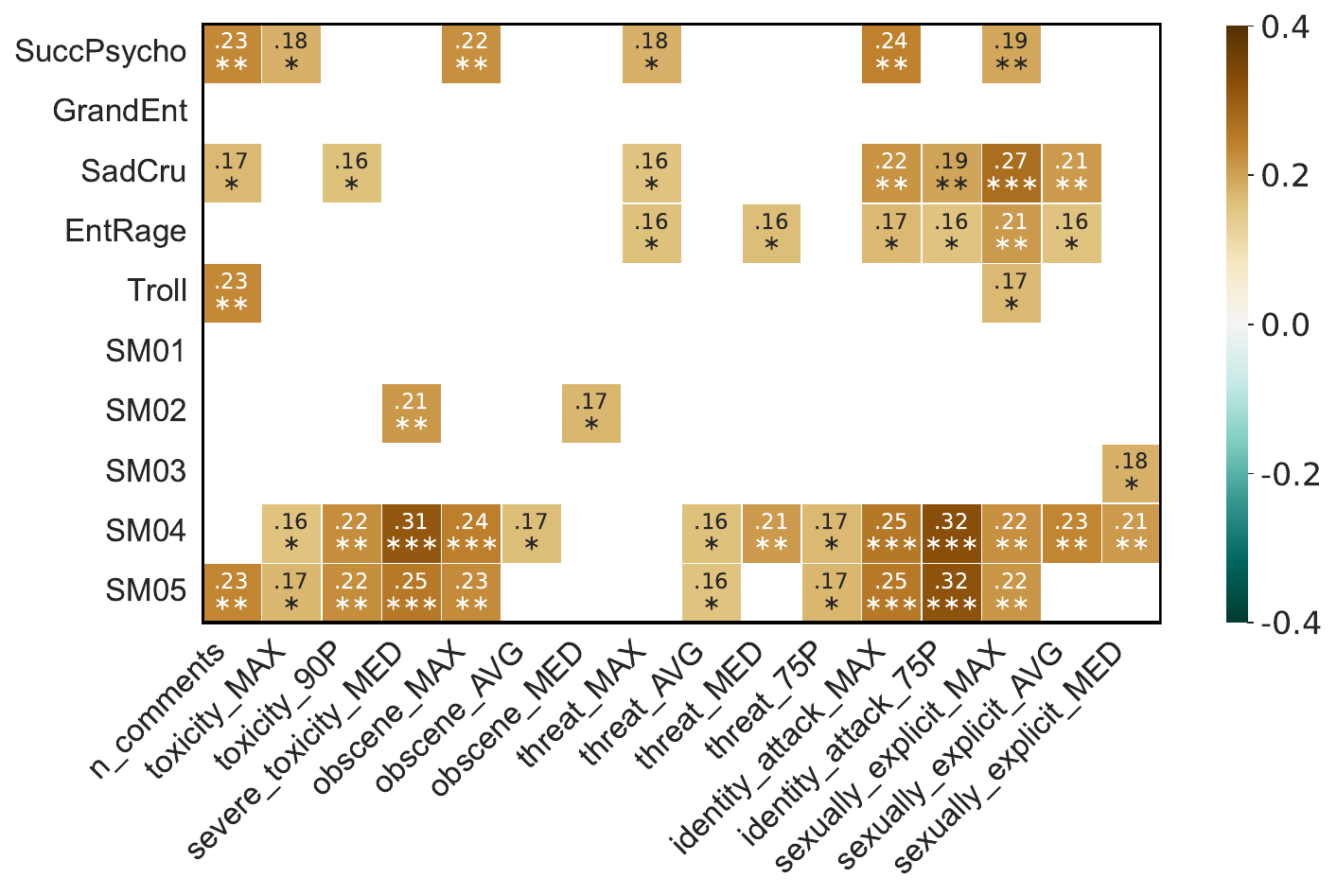}
        \centering
        \caption{Correlations significant at uncorrected $p < 0.1$.}
        \label{fig:hm_traits_tox_uncorrected}
    \end{subfigure}
    \begin{subfigure}{1\textwidth}
        \includegraphics[width=0.16\linewidth]{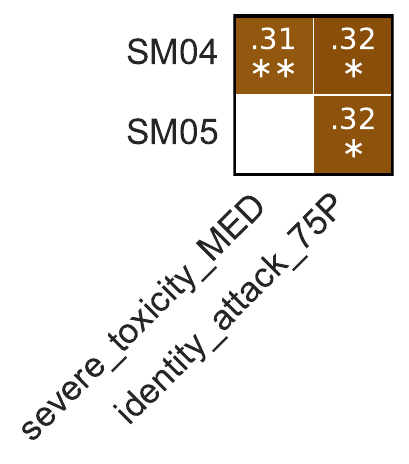}
        \centering
        \caption{Correlations significant at corrected $p < 0.1$. Corrections are obtained with the Benjamini–Hochberg procedure.}
        \label{fig:hm_traits_tox_corrected}
    \end{subfigure}
    \caption{Spearman rank correlation coefficients between dimension scores and social media habits (\textit{y} axis), and nuanced toxicity features (\textit{x} axis). Cell colors indicate the strength and direction of the correlations. Cell texts report correlation coefficients and their statistical significance. \textit{SM01}--\textit{SM05} correspond to the questions from the \textit{Social Media Use} section of the questionnaire. Asterisks denote statistical significance of the correlations: *: $p < 0.1$, **: $p < 0.05$, ***: $p < 0.01$.}
    \label{fig:hm_traits_tox}
\end{figure*}

\subsubsection{Fine-Grained Toxicity Subtypes}
Online harmful speech can manifest in multiple distinct forms beyond overall toxicity~\cite{banko2020unified}. To capture this nuance, we leverage fine-grained indicators provided by Google’s \texttt{Perspective API}---including severe toxicity, obscene language, insults, threats, identity attacks, and sexually explicit content---and compute representative features for each participant based on summary statistics of these indicators. Figure~\ref{fig:hm_traits_tox} reports correlations between these features, dimension scores, and social media use variables, both before and after applying the Benjamini–Hochberg correction.

\begin{figure*}[htbp]
    \centering
    \begin{subfigure}{1\textwidth}
        \includegraphics[height=0.24\textheight]{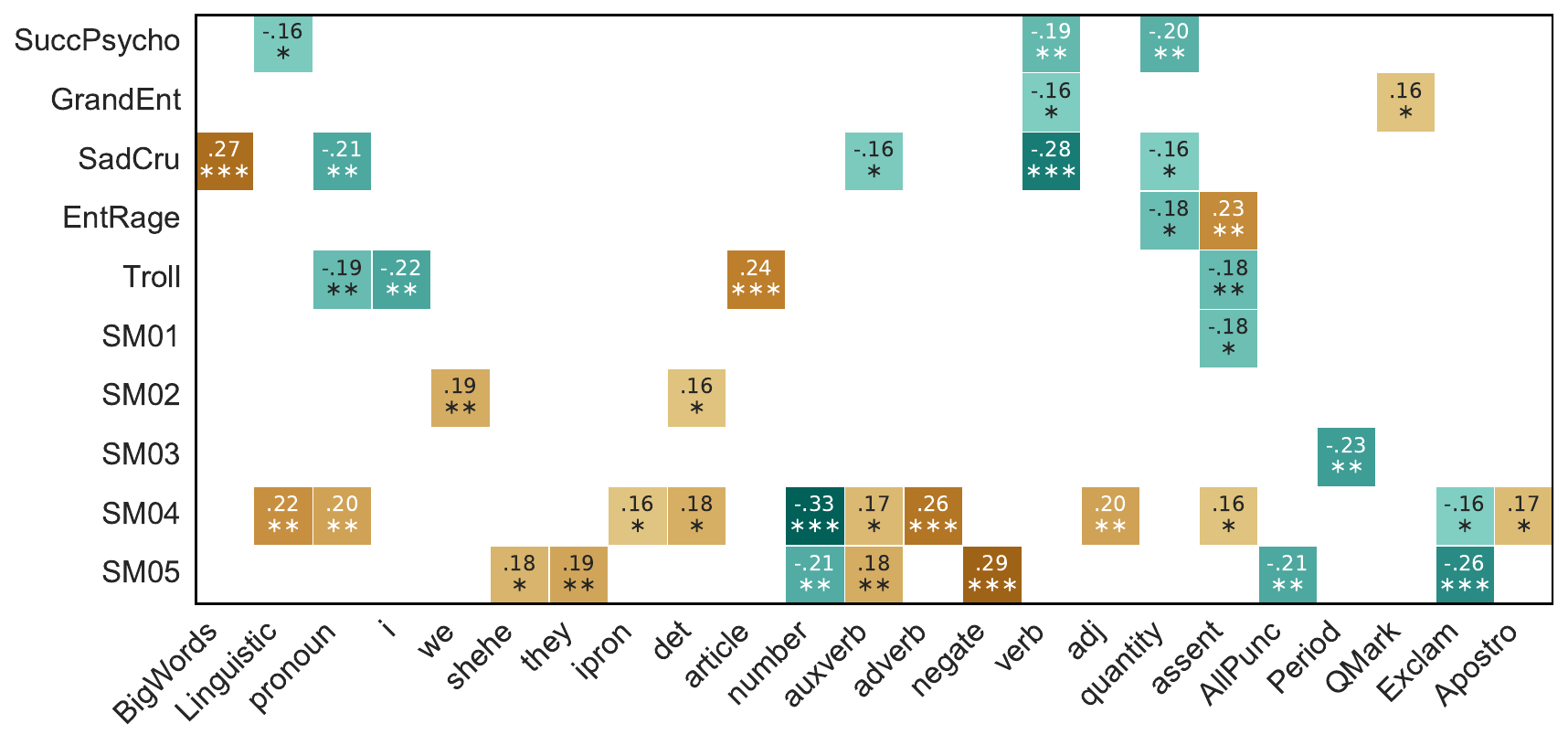}
        \centering
        \vspace{-5.5pt}
        \caption{Use of language. Correlations significant at uncorrected $p < 0.1$.}
        \label{fig:hm_liwc_ling}
    \end{subfigure}
    \begin{subfigure}{1\textwidth}
        \includegraphics[height=0.24\textheight]{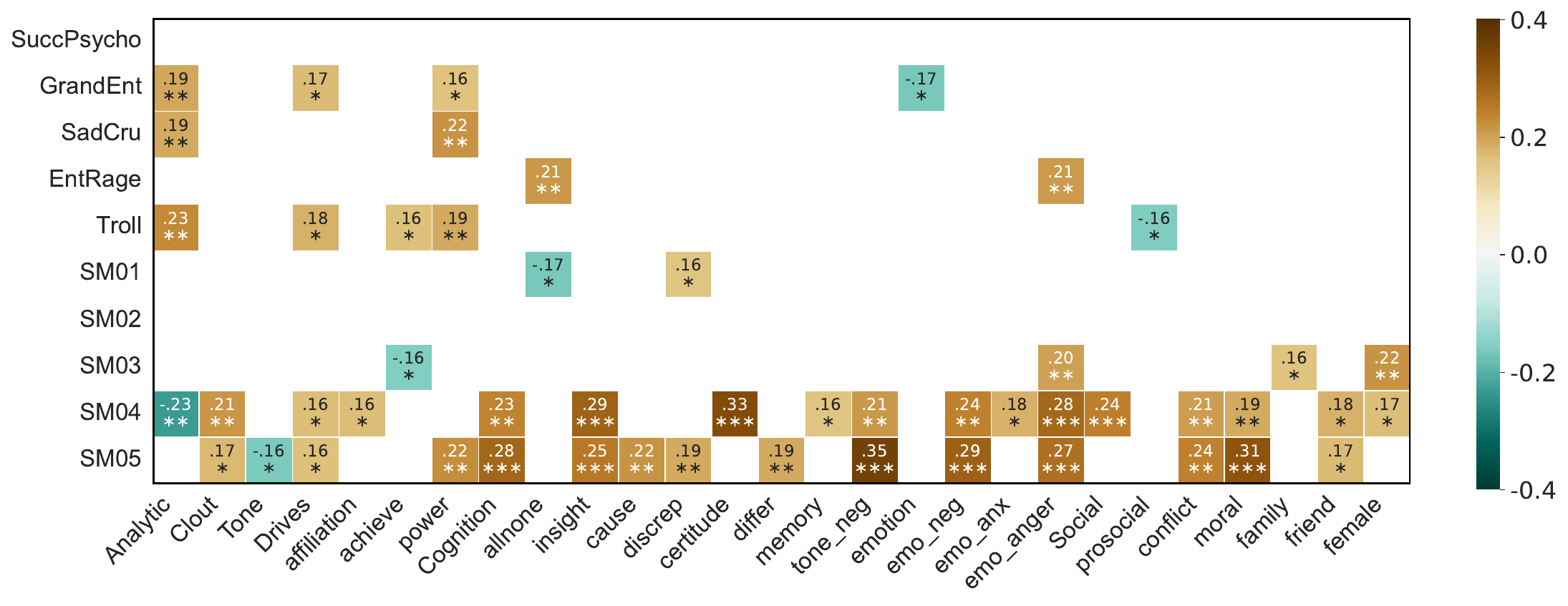}
        \centering
        \vspace{-5.5pt}
        \caption{Psychological processes. Correlations significant at uncorrected $p < 0.1$.}
        \label{fig:hm_liwc_psycho}
    \end{subfigure}
    \begin{subfigure}{1\textwidth}
        \includegraphics[height=0.24\textheight]{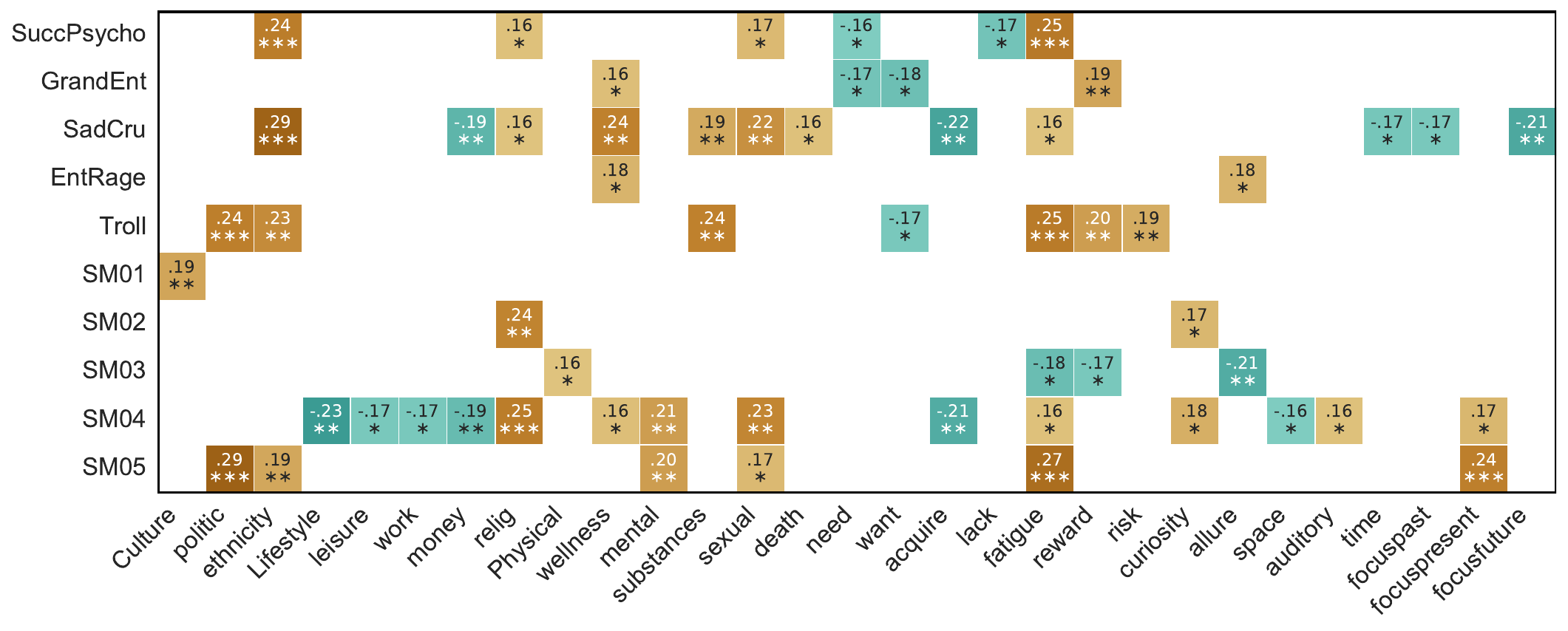}
        \centering
        \vspace{-5.5pt}
        \caption{Lifestyle and Socio-Cultural Factors. Correlations significant at uncorrected $p < 0.1$.}
        \label{fig:hm_liwc_lifestyle}
    \end{subfigure}
    \begin{subfigure}{1\textwidth}
        \includegraphics[width=0.15\linewidth]{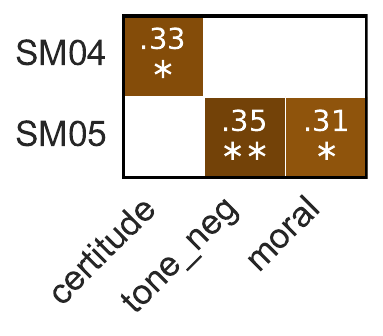}
        \centering
        \vspace{-5.5pt}
        \caption{Correlations significant at corrected \textit{p} < 0.1. Corrections are obtained with the Benjamini–Hochberg procedure.}
        \label{fig:hm_liwc_corrected}
    \end{subfigure}
    \caption{Spearman rank correlation coefficients between dimension scores and social media habits (\textit{y} axis), and LIWC features (\textit{x} axis). Only dimensions and social media habits questions with at least one significant correlation are shown. Cell colors indicate the strength and direction of the correlations. Cell texts report correlation coefficients and their statistical significance. \textit{SM01}--\textit{SM05} correspond to the questions from the \textit{Social Media Use} section of the questionnaire. Asterisks denote statistical significance of the correlations: *: $p < 0.1$, **: $p < 0.05$, ***: $p < 0.01$.}
    \label{fig:hm_liwc}
\end{figure*}

At the uncorrected level, several weak associations emerge between dark personality dimensions and specific toxicity subtypes. In particular, SadCru and, to a lesser extent, SuccPsycho show positive correlations with indicators related to identity attacks and sexually explicit content, broadly consistent with their theoretical links to antagonistic and transgressive tendencies~\cite{paulus2002dark,chabrol2009contributions,buckels2013behavioral}. However, none of these associations remain statistically significant after correction for multiple comparisons. In contrast, the only correlations that survive correction involve self-reported social media behaviors, specifically being the target of (\textit{SM04}) or engaging in (\textit{SM05}) toxic interactions. These variables show consistent and moderate associations with two toxicity subtypes, namely indicators of \textit{severe toxicity} and \textit{identity attacks}. More specifically, a feature based on \textit{severe toxicity} correlates positively with \textit{SM04}, while a feature capturing \textit{identity attacks} shows positive correlations with both \textit{SM04} and \textit{SM05}.

Taken together, these results indicate that, in this sample, validated dark personality dimensions do not significantly predict fine-grained linguistic manifestations of toxic behavior after correction. Instead, the robust signals are driven by self-reported behavioral experiences, which show consistent alignment with observed toxicity patterns.

\subsubsection{Linguistic Inquiry and Word Count}
The analysis of the LIWC subgroups presented in Figure~\ref{fig:hm_liwc} highlights clear differences in how linguistic patterns relate to dark personality traits, trolling behavior, and self-reported social media use. At the uncorrected level, several associations emerge, particularly for features capturing psychological processes such as negative tone, moral language, and emotional expression. In this setting, both being the target of uncivil interactions (\textit{SM04}) and engaging in such behavior (\textit{SM05}) show multiple moderate correlations with these categories. In contrast, the associations involving dark personality dimensions are fewer and generally weaker. Traits such as SadCru and Trolling account for most of the uncorrected correlations, especially within lifestyle and socio-cultural categories, while SuccPsycho, EntRage, and GrandEnt exhibit only sparse and modest associations. However, none of these relationships remain statistically significant after correction. Overall, only three correlations survive correction, all involving self-reported social media behaviors. \textit{Certitude} shows a positive association with \textit{SM04}, while both \textit{tone\_neg} and \textit{moral} are positively correlated with \textit{SM05}. No LIWC feature shows a statistically significant association with any dark personality dimension after correction.

These results indicate that, in this sample, validated dark personality traits do not exhibit statistically significant associations with linguistic patterns captured by LIWC features once multiple testing is accounted for. Instead, the robust associations are limited to self-reported experiences of online incivility, suggesting that linguistic markers such as certainty, negativity, and moral framing primarily reflect behavioral engagement in uncivil interactions rather than underlying personality traits.

\begin{figure*}[th!]
    \centering
    \begin{subfigure}{1\textwidth}
        \includegraphics[height=0.35\textheight]{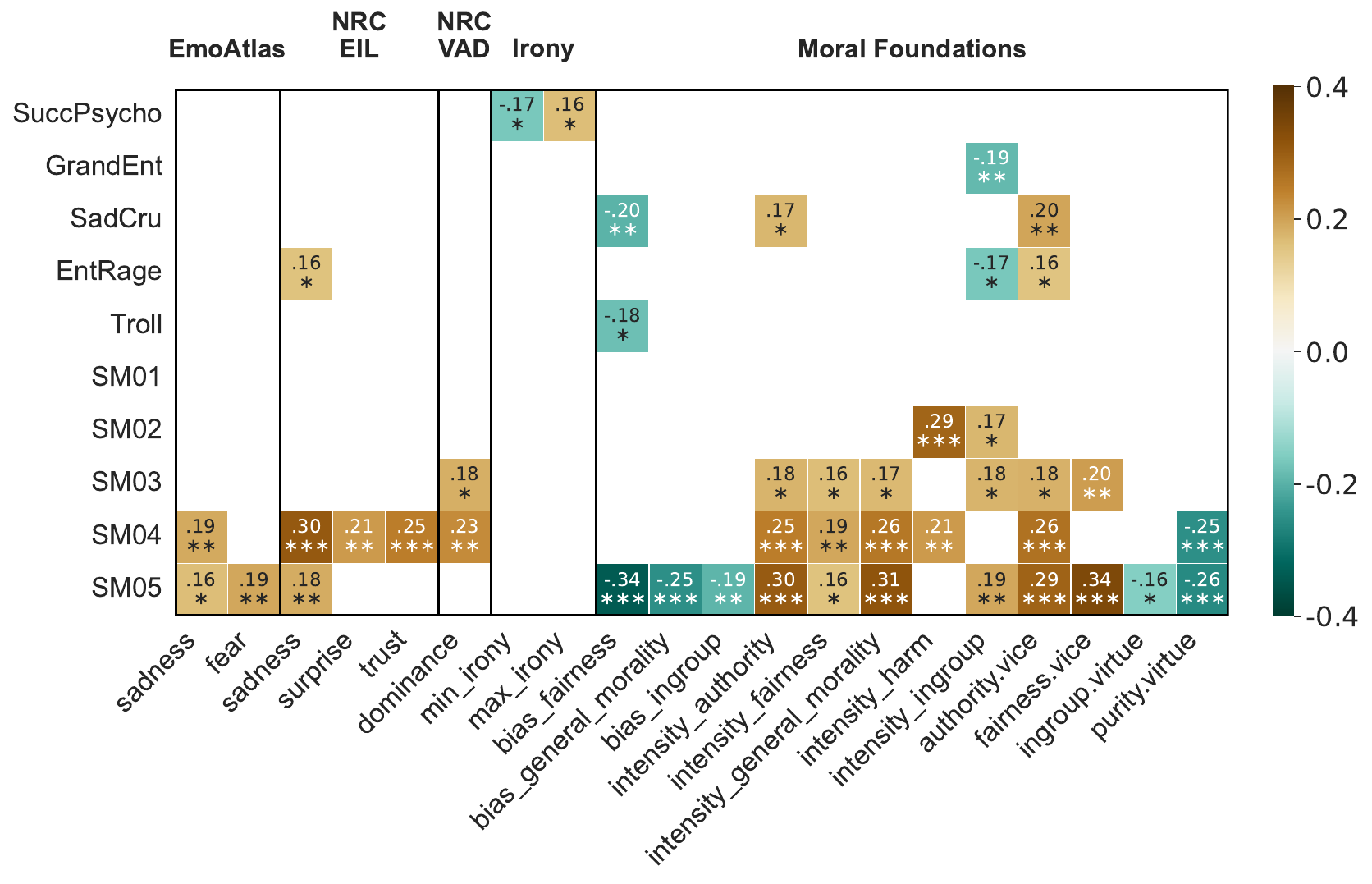}
        \centering
        \caption{Correlations significant at uncorrected $p < 0.1$.}
        \label{fig:hm_rem_uncorrected}
    \end{subfigure}
    \begin{subfigure}{1\textwidth}
        \includegraphics[width=0.28\linewidth]{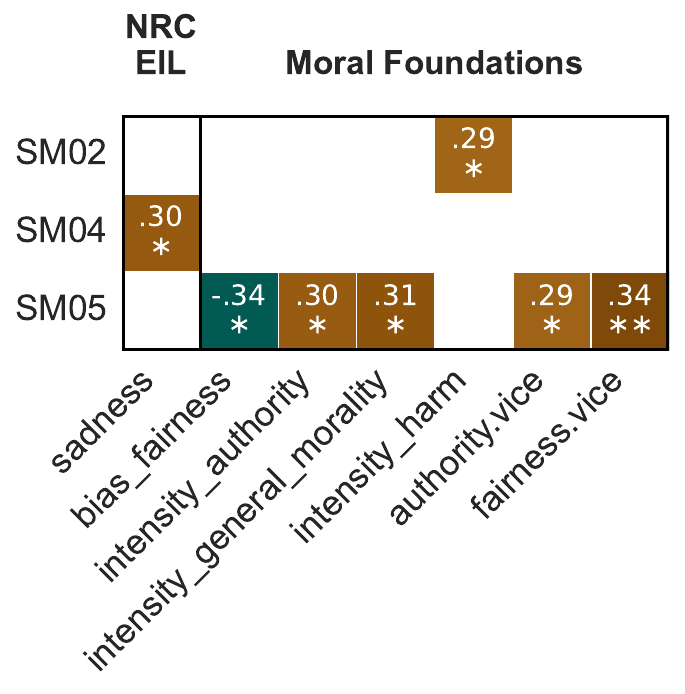}
        \centering
        \caption{Correlations significant at corrected $p < 0.1$. Corrections are obtained with the Benjamini–Hochberg procedure.}
        \label{fig:hm_rem_corrected}
    \end{subfigure}
    \caption{Spearman rank correlation coefficients between dimension scores and social media habits (\textit{y} axis), and features capturing emotions, affect, and moral foundations (\textit{x} axis). Cell colors indicate the strength and direction of the correlations. Cell texts report correlation coefficients and their statistical significance. \textit{SM01}--\textit{SM05} correspond to the questions from the \textit{Social Media Use} section of the questionnaire. Asterisks denote statistical significance of the correlations: *: $p < 0.1$, **: $p < 0.05$, ***: $p < 0.01$.}
    \label{fig:hm_rem}
\end{figure*}

\subsubsection{Emotions, Affect, Irony, Moral Foundations}
Here we examine associations between dimension scores and self-reported social media behaviors on the one hand, and features derived from \texttt{EmoAtlas}, \texttt{NRC-EIL}, \texttt{NRC-VAD}, irony, and moral foundations on the other. Figure~\ref{fig:hm_rem} reports correlations both before and after applying the Benjamini–Hochberg correction. At the uncorrected level, a broad set of associations emerges, largely driven by self-reported social media behaviors rather than personality dimensions. Both producing (\textit{SM05}) and being targeted by (\textit{SM04}) toxic content correlate with numerous features related to emotional intensity and moral framing. In contrast, dark personality dimensions exhibit only sparse and generally weak associations.

After correction, the pattern becomes substantially more selective. Similarly to toxicity and LIWC, no statistically significant correlations remain between any feature and the dark personality dimensions. Instead, the surviving associations involve only self-reported behaviors. Specifically, \textit{SM05} (producing toxic content) shows a negative correlation with \textit{bias\_fairness}, and positive correlations with \textit{intensity\_authority}, \textit{intensity\_general\_morality}, \textit{authority.vice}, and \textit{fairness.vice}. In addition, \textit{SM04} (being targeted by toxic content) is positively correlated with \textit{sadness}, and \textit{SM02} (general content production) is positively associated with \textit{intensity\_harm}.

Therefore, these results indicate that, in this sample, validated dark personality dimensions do not significantly predict emotional, affective, or moral linguistic features once multiple testing is taken into account. Instead, the robust associations are confined to self-reported experiences of online incivility, which are reflected in patterns of emotional expression and moral framing. While some uncorrected associations involving dark traits may suggest theoretically plausible tendencies, they do not provide reliable evidence of stable relationships under the present analytical conditions.

 \section{Discussion}
\label{sec:discussion}
By bridging validated psychometric assessment with extensive Reddit activity data, our study explores both the promise and the limits of computational text analysis for understanding and responding to online toxic behavior.

\subsection{Insights into Personality, Perceptions, and Online Conduct}
Confirmatory analyses based on interpretable features yield four main findings that directly address core research questions about the interplay between personality traits, self-reported behaviors, and actual online activity.

\paragraph{Do dark traits correlate with self-reported experiences of online incivility?}
Participants’ self-reports indicate that dark personality traits and trolling behavior are weakly associated with general patterns of social media use, such as content consumption or exposure to incivility. In contrast, all dimensions show positive correlations with the self-reported production of uncivil comments (\textit{SM05}). These associations are strongest and robust for Trolling and SuccPsycho, while they are non-significant after correction for EntRage, SadCru, and GrandEnt. These results reflect relationships between self-reported personality traits and self-reported behavioral tendencies. As such, they indicate that individuals scoring higher on dark dispositions are more likely to report engaging in uncivil interactions.

\paragraph{Do dark traits correspond to actual production of toxic content?} 
The analysis of text-derived toxicity features indicates that validated dark personality dimensions do not significantly predict the production of toxic content in our sample. After correction for multiple comparisons, no DSHS dimension shows a statistically significant association with any toxicity measure derived from participants’ Reddit activity. This result provides a clear answer to the research question: under the present measurement and analytical framework, dark personality traits do not exhibit robust links to observable toxic behavior. By contrast, self-reported engagement in uncivil interactions shows a different pattern. Participants who report frequently producing (\textit{SM05}) or being targeted by (\textit{SM04}) toxic content display consistently higher levels of toxicity in their comments. These findings indicate a strong alignment between self-reported experiences of online incivility and observable linguistic behavior. Importantly, however, this alignment reflects consistency between behavioral self-reports and behavioral traces, rather than a direct link between personality traits and behavior. These results highlight a lack of robust associations between validated personality measures and surface-level behavioral indicators.

\paragraph{Can hand-crafted text-derived formulas accurately estimate dark traits?} The comparison between Borghi and Ratcharak’s hand-crafted textual proxies for Dark Triad traits~\cite{borghi2025deceptive} and our validated Dark Tetrad scores shows limited convergence between the two measurement approaches in this sample. After correction, no statistically significant associations are observed between the text-derived estimates and the questionnaire-based dimensions. While a small number of weak associations emerge at the uncorrected level---primarily involving Machiavellianism estimates and dimensions such as EntRage, SuccPsycho, and SadCru---these effects are not stable and do not support a robust correspondence between the two approaches. In contrast, proxies for psychopathy and narcissism exhibit no meaningful relationship with their DSHS counterparts. More broadly, these results highlight the challenges of mapping complex and multidimensional personality constructs onto simple rule-based linguistic representations. At the same time, the observed lack of alignment should be interpreted with caution, as the two approaches rely on only partially overlapping conceptualizations of dark personality traits (Dark Triad vs Dark Tetrad). Taken together, these findings suggest that the relationship between text-derived personality proxies and validated psychometric measures remains limited and sensitive to both construct definitions and methodological choices, motivating the need for more robust and better-validated approaches to text-based personality inference.

\subsection{Interpreting the Absence of Trait–Behavior Associations}
While our analyses do not reveal statistically significant associations between validated dark personality dimensions and linguistic or behavioral indicators after correction, these null findings can be interpreted in light of existing theoretical frameworks in personality psychology and cyberpsychology. Prior work suggests that different dark traits are associated with distinct behavioral tendencies. For example, sadistic and psychopathic dispositions have been linked to impulsivity, reduced empathy, and a propensity for direct hostility, whereas traits such as narcissism and Machiavellianism are more often associated with status-seeking, reputational management, and strategic or indirect forms of antagonism~\cite{paulus2002dark,buckels2013behavioral,patrick2009triarchic,morf2001unraveling,jones2016nature}. However, our results indicate that such differences are not associated with robust, detectable patterns in surface-level linguistic features or toxicity scores within the present analytical framework. It is important to interpret these null findings cautiously. Because our confirmatory analyses were powered primarily to detect moderate and larger effects, we cannot rule out the existence of smaller personality–behavior associations that the present design and feature set were underpowered to detect. One possible interpretation is that the behavioral expression of dark traits in online environments is more context-dependent and less directly observable than previously assumed. In particular, traits associated with overt hostility may be expected to produce strong lexical signals, while more strategic or instrumental traits may be associated with subtler communicative behaviors, such as selective engagement, rhetorical framing, or interaction dynamics, which are not captured by shallow, text-based features~\cite{elsherief2021latent}. From this perspective, the absence of significant associations in our study does not necessarily contradict theoretical expectations, but rather highlights the limitations of current feature sets in capturing the full range of behavioral expressions associated with dark personality traits. This interpretation is consistent with our broader findings, which show that linguistic features are effective in capturing overt behavioral engagement in toxic interactions, but not stable personality dispositions.

To address this limitation, future work should explore richer and more context-aware representations of online behavior. In particular, experimentation with feature sets complementary to those used in this work---such as those aimed at surfacing persuasion strategies or conversational tactics~\cite{barron2019proppy,razuvayevskaya2024comparison}---may improve the ability to detect linguistic signals associated with traits such as narcissism and Machiavellianism. Additionally, the observed associations between moral-framing features and self-reported engagement in toxic interactions suggest that moral rhetoric may play a role in structuring hostile exchanges, further supporting the need for feature representations that capture discourse-level and interactional dynamics. Overall, these considerations point to the existence of diverse and potentially indirect pathways through which dark traits may operate online, and highlight the need for more nuanced methodological approaches rather than relying on shallow, one-size-fits-all linguistic signatures.

\subsection{Feature Families and Trait Estimators}
Our findings help clarify what different families of computational features can---and cannot---measure regarding online misbehavior. Tools such as LIWC and moral foundation lexicons are most sensitive to overt markers of emotional stance and moral positioning, making them effective for detecting heightened negativity, absolutist framing, and moralizing rhetoric that often accompany toxic exchanges. Similarly, \texttt{NRC-EIL} highlights variations in the intensity of expressed affect. Perspective API toxicity scores remain useful for quantifying nuances in hostile language, although their subjectivity and known cultural biases must be acknowledged~\cite{nogara2025toxic}. Overall, these approaches provide valuable signal when behavior is explicit and affect-laden. However, our results also show that these feature families are limited in their ability to capture latent personality dispositions. The strong associations observed for \textit{SM04} (being targeted by toxic content) and \textit{SM05} (producing toxic content) serve as a behavioral baseline, indicating that while linguistic features reliably capture engagement in uncivil interactions, they do not extend to robust inference of underlying personality traits. In this light, surface lexicon methods and simple features appear well suited for detecting signal-level toxicity, but their use as proxies for psychological constructs should be approached with caution. The mixed results obtained when re-implementing Borghi and Ratcharak's~\cite{borghi2025deceptive} hand-crafted formulas further support this interpretation. Rather than providing reliable estimates of personality traits, such approaches should be regarded as preliminary approximations that capture behavioral signals without fully representing the underlying constructs. Advancing this line of research therefore requires richer, more context-aware models validated against robust ground-truth data, rather than relying on shallow lexical features alone.

\subsection{Revisiting LIWC–Personality Associations}
Our findings regarding LIWC features differ from several prior studies that report significant associations between linguistic markers and dark personality traits~\cite{sumner2012predicting,preotiuc2016studying}. Earlier work has identified correlations between Dark Triad traits and LIWC categories capturing negative emotion, hostility, and social orientation. By contrast, in our analyses no LIWC feature exhibits a statistically significant association with validated dark personality dimensions after correcting for multiple comparisons. At face value, this discrepancy may suggest a lack of convergence between our results and the existing literature. However, a closer examination reveals that these differences can be largely explained by methodological and conceptual factors.

A first important distinction concerns the statistical treatment of multiple comparisons. Prior studies reported LIWC associations based on uncorrected significance thresholds, despite testing a large number of markers~\cite{sumner2012predicting,preotiuc2016studying}. In contrast, our analysis applies the Benjamini–Hochberg procedure to control the false discovery rate across all tested features. This difference is consequential given the small magnitude of the effects involved. Indeed, Holtzman et al., who explicitly address this issue, show that LIWC associations with narcissism are generally very small (often $|\rho| < 0.10$) and that nearly all of them disappear under strict multiple-comparison correction, even in large samples~\cite{holtzman2019linguistic}. From this perspective, the absence of significant LIWC–trait associations in our study is consistent with the expectation that such effects, if present, are subtle and difficult to detect robustly under conservative statistical control. Moreover, differences in the operationalization of personality constructs also play a role. Much of the prior literature focuses on the Dark Triad using brief instruments such as the SD3 or the Dirty Dozen~\cite{sumner2012predicting,preotiuc2016studying}, whereas our study employs the DSHS framework, which captures a broader and partly distinct set of dimensions, including constructs such as sadistic cruelty and entitlement rage. While some overlap exists---for example, Successful Psychopathy shares features with psychopathy and Machiavellianism---the constructs are not equivalent, and direct correspondence between linguistic markers identified in prior work and those expected here should not be assumed. Finally, the scale and analytical scope of the studies differ substantially. Prior work relies on larger samples and explores a wide range of linguistic features without dimensionality reduction, increasing the likelihood of detecting small effects. In contrast, our analysis combines a more modest sample size with a feature selection procedure designed to reduce redundancy and the effective number of hypotheses tested. While this approach improves interpretability and statistical robustness, it also makes the analysis less sensitive to weak and diffuse associations.

Taken together, these considerations suggest a coherent interpretation of the broader literature. LIWC features may capture certain aspects of dark personality traits, but the corresponding signals are generally small, context-dependent, and sensitive to analytical choices such as feature selection and multiple-testing correction. As a result, findings based on uncorrected analyses or large feature sets may overstate the robustness of these associations. In this light, our results do not contradict prior work but rather provide a more conservative estimate of the strength and stability of LIWC–trait relationships. More broadly, they indicate that while lexicon-based features are effective at capturing overt behavioral and affective signals, their ability to serve as reliable proxies for complex personality constructs remains limited. Consistent with this interpretation, recent work shows that LIWC-derived emotional features can robustly predict incivility in large-scale social media corpora, particularly through expressions of anger and, to a lesser extent, sadness and anxiety~\cite{stevens2022emotions}. However, these studies focus on situational emotional processes rather than stable personality traits, reinforcing the view that LIWC features are well suited to capturing transient affective states associated with uncivil behavior, while remaining limited as indicators of underlying personality dispositions.

\subsection{Availability of Online Activity Data}
The need for robust ground-truth data naturally leads to the broader challenge of data availability in computational social science and online behavior research~\cite{assenmacher2022benchmarking}. Here, our analyses were limited to Reddit, since it currently represents one of the few major platforms offering openly accessible data. While this choice enabled us to link validated psychological questionnaires with online activity data, it also imposed dire constraints, as our recruitment through Amazon Mechanical Turk revealed only limited overlap with the active Reddit user base. This resulted in a reduced sample size despite considerable effort. At the same time however, our framework demonstrates a flexible path forward. As illustrated in Figure~\ref{fig:webapp}, our study design and data collection approach is platform-agnostic, as extending it to other platforms would only involve modifying the data acquisition and storage components (i.e., steps 3 and 4 in figure)---a relatively limited effort. In principle, this would allow for richer and more diverse datasets that combine personality assessments with authentic online behavior across multiple social platforms. Yet, progress in this direction may be hindered by the realities of the current ``post-API'' era~\cite{tromble2021have}. Many platforms have restricted or closed programmatic access, constraining the ability of researchers to study online behavior at scale. This shift has far-reaching consequences, as it limits transparency and hinders empirical work across a large number of tasks and domains. Against this backdrop, the availability of ethically collected, high-quality datasets that integrate psychometric and behavioral data becomes all the more critical. We therefore envision future research building upon the approach introduced here to extend data collection across platforms wherever feasible, thereby mitigating current bottlenecks in data access and paving the way for more effective downstream research.

\subsection{Limitations and Future Work}
This study faces several limitations. First, the modest sample size constrains both statistical power and the generalizability of our findings. While our sensitivity and power analyses indicate that the study is adequately powered to detect moderate associations, smaller effect sizes may not be reliably identified given the sample size and the scope of multiple testing, and should therefore be interpreted with appropriate caution. Second, recruiting participants through Amazon Mechanical Turk and restricting behavioral data to Reddit may introduce sample biases and further limit applicability to other populations or platforms with different affordances, habits, and cultures. Third, our methodological choices carry inherent limitations. Dark traits and trolling tendencies were measured through self-reports, which are susceptible to self-presentation bias, though this remains standard practice in personality research. Likewise, automated tools used to estimate toxicity and other linguistic features can be error-prone and introduce uncertainties or systematic biases. A further limitation concerns the comparison between text-derived estimates of Dark Triad traits and the DSHS dimensions. These constructs are not directly equivalent, as the DSHS captures a broader and partly distinct set of traits. As a result, the lack of alignment observed between text-derived formulas and questionnaire-based measures may reflect differences in construct definitions, rather than solely limitations of the text-based estimation approach. This mismatch constrains the interpretability of the comparison and highlights the need for computational tools specifically designed to capture Dark Tetrad constructs. Aspects of social media use (\textit{SM01}–\textit{SM05}) were assessed using single-item measures rather than validated multi-item scales. As a consequence, such assessments might suffer from reduced content validity, limited sensitivity, and the inability to evaluate internal consistency reliability. Moreover, although our feature set was extensive and yielded multiple significant correlations, it was necessarily selective. Other relevant aspects of online communication---such as the use of persuasion strategies, subreddits participation, or broader interaction dynamics---remain unexplored. We also acknowledge the heuristic nature of the threshold used to classify participants along each dimension. While our choice does not rely on established clinical or normative cutoffs, a sensitivity analysis in which the threshold was systematically varied shows that the resulting distribution of trait combinations remains largely stable across reasonable threshold values. Finally, even if our feature selection procedure substantially reduced redundancy among features, some residual correlations inevitably remain. This residual dependence may affect the assumptions underlying multiple-comparison correction procedures such as Benjamini–Hochberg, potentially leading to imperfect control of false discovery rates.

Future work could address several of the limitations outlined above. The approach that we employed in this study to recruit participants and collect high-quality data could be extended to platforms beyond Reddit with limited effort, enabling larger sample sizes, greater statistical power, improved generalizability, and cross-platform comparisons. Our results also indicate that while the features we employed capture signals of overt toxicity and behavioral engagement in uncivil interactions, they are less effective for identifying underlying personality traits. Future studies should therefore explore richer and more context-aware representations of online behavior, possibly including features that model persuasion strategies, conversational dynamics, or interaction patterns, in order to capture more nuanced behavioral correlates of personality. More broadly, rather than directly enabling automated inference of dark traits, our findings help delineate the current limits of such approaches. In this sense, they provide a foundation for future work aimed at developing computational models that more reliably connect behavioral traces to validated psychometric constructs, ideally through integration with stronger ground-truth data and more expressive modeling techniques. Finally, while the present results do not support direct personality inference for moderation purposes, they contribute to a better understanding of the relationship between behavioral signals and self-reported experiences of incivility, which may inform the design of more targeted and context-sensitive moderation strategies.
 \section{Conclusions}
\label{sec:conclusions}
We examined the relationship between dark personality traits, online toxicity, and the socio-linguistic characteristics of online user activity data. To this end, we developed a dedicated Web application that enabled the secure integration of a validated psychological questionnaire administered via Amazon Mechanical Turk with real-world Reddit activity data. This design allowed us to link individual-level trait profiles to behavioral and linguistic features derived from participants’ online comments. Our analyses yield three main findings. First, dark personality traits show consistent associations with self-reported engagement in uncivil interactions, particularly for trolling and psychopathy-related dimensions. Second, these associations do not extend to observed behavior: after correction for multiple comparisons, no validated dark personality dimension significantly predicts text-derived toxicity or linguistic features. Third, self-reported experiences of online incivility---both as targets and as perpetrators---are robustly reflected in linguistic patterns, indicating a strong alignment between behavioral self-reports and observable online activity. These results highlight a gap between validated personality measures and their manifestation in surface-level behavioral and linguistic signals. While computational features effectively capture overt behavioral engagement in toxic interactions, they do not provide reliable proxies for underlying personality traits within the present framework. This suggests that the relationship between personality and online behavior may be more indirect, context-dependent, or mediated by factors not captured by current feature sets.

Looking ahead, future research should expand this framework to larger and more diverse datasets and explore richer, more context-aware representations of online behavior. In particular, integrating features that capture conversational dynamics, persuasion strategies, or interaction patterns may help uncover subtler links between personality and behavior. More broadly, our findings help delineate the current limits of text-based personality inference and provide a foundation for developing more robust and theoretically grounded approaches to studying online toxic behavior.

\bibliographystyle{elsarticle-num}
\bibliography{references}

@article{borghi2025deceptive,
  title={Deceptive minds in digital spaces: the influence of the dark triad on posting fake online reviews},
  author={Borghi, Matteo and Ratcharak, Phatcharasiri},
  journal={Psychology \& Marketing},
  year={2025},
  publisher={Wiley Online Library}
}

@article{dwivedi2018social,
  title={Social media: The good, the bad, and the ugly},
  author={Dwivedi, Yogesh K and Kelly, Gerald and Janssen, Marijn and Rana, Nripendra P and Slade, Emma L and Clement, Marc},
  journal={Information Systems Frontiers},
  volume={20},
  number={3},
  pages={419--423},
  year={2018},
  publisher={Springer}
}

@article{bogolyubova2018dark,
  title={Dark personalities on Facebook: Harmful online behaviors and language},
  author={Bogolyubova, Olga and Panicheva, Polina and Tikhonov, Roman and Ivanov, Viktor and Ledovaya, Yanina},
  journal={Computers in human Behavior},
  volume={78},
  pages={151--159},
  year={2018},
  publisher={Elsevier}
}

@article{tessa2025beyond,
  title={Beyond trial-and-error: Predicting user abandonment after a moderation intervention},
  author={Tessa, Benedetta and Cima, Lorenzo and Trujillo, Amaury and Avvenuti, Marco and Cresci, Stefano},
  journal={Engineering Applications of Artificial Intelligence},
  volume={162},
  pages={112375},
  year={2025},
  publisher={Elsevier}
}

@article{gillespie2020content,
  title={Content moderation, AI, and the question of scale},
  author={Gillespie, Tarleton},
  journal={Big Data \& Society},
  volume={7},
  number={2},
  year={2020},
  publisher={SAGE Publications Sage UK: London, England}
}

@inproceedings{giorgi2025human,
 author = {Giorgi, Tommaso and Cima, Lorenzo and Fagni, Tiziano and Avvenuti, Marco and Cresci, Stefano},
 booktitle = {The 19th International AAAI Conference on Web and Social Media (ICWSM'25)},
 organization = {AAAI},
 pages = {653--670},
 title = {{Human and LLM biases in hate speech annotations: A socio-demographic analysis of annotators and targets}},
 year = {2025}
}

@article{alvisi2026toxicity,
  title={From Toxicity to Conformity: Adaptive user behavior to social norms in {Telegram} communities},
  author={Alvisi, Lorenzo and Popa, Victoria and Cola, Guglielmo and Tardelli, Serena and Tesconi, Maurizio},
  journal={Scientific Reports},
  year={2026},
  publisher={Nature Publishing Group UK London}
}

@article{groh2022deepfake,
  title={Deepfake detection by human crowds, machines, and machine-informed crowds},
  author={Groh, Matthew and Epstein, Ziv and Firestone, Chaz and Picard, Rosalind},
  journal={Proceedings of the National Academy of Sciences},
  volume={119},
  number={1},
  pages={e2110013119},
  year={2022},
  publisher={National Academy of Sciences}
}

@article{alavi2023dark,
  title={Dark tetrad of personality, cyberbullying, and cybertrolling among young adults},
  author={Alavi, Masoumeh and Latif, Adibah Abdul and Ramayah, T and Tan, Jia Yue},
  journal={Current Psychology},
  volume={42},
  number={32},
  year={2023},
  publisher={Springer}
}

@article{kurek2019did,
  title={‘I did it for the LULZ’: How the dark personality predicts online disinhibition and aggressive online behavior in adolescence},
  author={Kurek, Anna and Jose, Paul E and Stuart, Jaimee},
  journal={Computers in Human Behavior},
  volume={98},
  pages={31--40},
  year={2019},
  publisher={Elsevier}
}

@inproceedings{barbieri2020tweeteval,
  title={{TweetEval: Unified benchmark and comparative evaluation for tweet classification}},
  author={Barbieri, Francesco and Camacho-Collados, Jose and Anke, Luis Espinosa and Neves, Leonardo},
  booktitle={Findings of the Association for Computational Linguistics: EMNLP 2020},
  pages={1644--1650},
  year={2020}
}

@article{fanslau2023dark,
  title={Dark triad predictors of irony and sarcasm use: An investigation in a {Polish} sample},
  author={Fanslau, Agnieszka and Ka{\l}owski, Piotr and Olech, Micha{\l} and Rowicka, Magdalena and Branowska, Katarzyna and Olechowska, Anna and Zarazi{\'n}ska, Anna and Siemieniuk, Aleksandra and Banasik-Jemielniak, Natalia},
  journal={Personality and Individual Differences},
  volume={214},
  year={2023}
}

@article{semeraro2025emoatlas,
  title={EmoAtlas: An emotional network analyzer of texts that merges psychological lexicons, artificial intelligence, and network science},
  author={Semeraro, Alfonso and Vilella, Salvatore and Improta, Riccardo and De Duro, Edoardo Sebastiano and Mohammad, Saif M and Ruffo, Giancarlo and Stella, Massimo},
  journal={Behavior Research Methods},
  volume={57},
  number={2},
  pages={77},
  year={2025},
  publisher={Springer}
}

@article{boyd2022development,
  title={The development and psychometric properties of LIWC-22},
  author={Boyd, Ryan L and Ashokkumar, Ashwini and Seraj, Sarah and Pennebaker, James W},
  journal={Austin, TX: University of Texas at Austin},
  volume={10},
  pages={1--47},
  year={2022}
}

@article{kwak2021frameaxis,
  title={{FrameAxis}: Characterizing microframe bias and intensity with word embedding},
  author={Kwak, Haewoon and An, Jisun and Jing, Elise and Ahn, Yong-Yeol},
  journal={PeerJ Computer Science},
  volume={7},
  year={2021}
}

@inproceedings{mohammad2018obtaining,
  title={Obtaining reliable human ratings of valence, arousal, and dominance for 20,000 English words},
  author={Mohammad, Saif},
  booktitle={Proceedings of the 56th annual meeting of the association for computational linguistics (volume 1: Long papers)},
  pages={174--184},
  year={2018}
}

@article{mohammad2013crowdsourcing,
  title={Crowdsourcing a word--emotion association lexicon},
  author={Mohammad, Saif M and Turney, Peter D},
  journal={Computational intelligence},
  volume={29},
  number={3},
  pages={436--465},
  year={2013},
  publisher={Wiley Online Library}
}

@inproceedings{cima2025contextualized,
 author = {Cima, Lorenzo and Miaschi, Alessio and Trujillo, Amaury and Avvenuti, Marco and Dell'Orletta, Felice and Cresci, Stefano},
 booktitle = {The 34th ACM Web Conference (WWW'25)},
 organization = {ACM},
 pages = {5022--5033},
 title = {{Contextualized counterspeech: Strategies for adaptation, personalization, and evaluation}},
 year = {2025}
}

@article{luong2022evaluating,
  title={Evaluating Reddit as a crowdsourcing platform for psychology research projects},
  author={Luong, Raymond and Lomanowska, Anna M},
  journal={Teaching of Psychology},
  volume={49},
  number={4},
  year={2022},
  publisher={Sage Publications Sage CA: Los Angeles, CA}
}

@article{calic2023dark,
  title={The dark side of Machiavellian rhetoric: Signaling in reward-based crowdfunding performance},
  author={Calic, Goran and Arseneault, Rene and Ghasemaghaei, Maryam},
  journal={Journal of Business Ethics},
  volume={182},
  number={3},
  pages={875--896},
  year={2023},
  publisher={Springer}
}

@inproceedings{trujillo2023one,
 author = {Trujillo, Amaury and Cresci, Stefano},
 booktitle = {The 15th International ACM Web Science Conference (WebSci'23)},
 organization = {ACM},
 pages = {55--64},
 title = {{One of many: Assessing user-level effects of moderation interventions on r/The\_Donald}},
 year = {2023}
}

@article{freyth2023social,
  title={Social media use and personality: Beyond self-reports and trait-level assessments},
  author={Freyth, Lennart and Batinic, Bernad and Jonason, Peter K},
  journal={Personality and Individual Differences},
  volume={202},
  pages={111960},
  year={2023},
  publisher={Elsevier}
}

@article{yuan2022does,
  title={Does Facebook activity reveal your dark side? Using online language features to understand an individual’s dark triad and needs},
  author={Yuan, Cuixin and Hong, Ying and Wu, Junjie},
  journal={Behaviour \& Information Technology},
  volume={41},
  number={2},
  pages={292--306},
  year={2022},
  publisher={Taylor \& Francis}
}

@article{hossain2022you,
  title={Are you a cyberbully on social media? Exploring the personality traits using a fuzzy-set configurational approach},
  author={Hossain, Mohammad Alamgir and Quaddus, Mohammed and Warren, Matthew and Akter, Shahriar and Pappas, Ilias},
  journal={International Journal of Information Management},
  volume={66},
  pages={102537},
  year={2022},
  publisher={Elsevier}
}

@article{koay2023students,
    author = {Koay, Kian Yeik and Poon, Wai Ching},
    title = {Students' cyberslacking behaviour in e-learning environments: the role of the Big Five personality traits and situational factors},
    journal = {Journal of Applied Research in Higher Education},
    volume = {15},
    number = {2},
    pages = {521-536},
    year = {2023},
    month = {04},
}

@article{moor2019systematic,
title = {A systematic literature review of the relationship between dark personality traits and antisocial online behaviours},
journal = {Personality and Individual Differences},
volume = {144},
pages = {40-55},
year = {2019},
author = {Lily Moor and Joel R. Anderson}
}

@article{paulus2002dark,
title = {The Dark Triad of personality: Narcissism, Machiavellianism and psychopathy},
journal = {Journal of Research in Personality},
volume = {36},
number = {6},
pages = {556–563},
year = {2002},
author = {Delroy Paulhus and Kevin M. Williams}
}

@article{buckels2013behavioral,
author = {Buckels, Erin and Jones, Daniel and Paulhus, Delroy},
year = {2013},
month = {09},
title = {Behavioral Confirmation of Everyday Sadism},
volume = {24},
journal = {Psychological science},
}

@article{chabrol2009contributions,
title = {Contributions of psychopathic, narcissistic, Machiavellian, and sadistic personality traits to juvenile delinquency},
journal = {Personality and Individual Differences},
volume = {47},
number = {7},
pages = {734-739},
year = {2009},
author = {Henri Chabrol and Nikki {Van Leeuwen} and Rachel Rodgers and Natalène Séjourné}
}

@article{buckels2018internet,
author = {Buckels, Erin and Trapnell, Paul and Andjelovic, Tamara and Paulhus, Delroy},
year = {2018},
month = {04},
title = {Internet Trolling and Everyday Sadism: Parallel Effects on Pain Perception and Moral Judgment},
volume = {87},
journal = {Journal of Personality}
}

@article{lopes2017you,
title = {Who do you troll and Why: An investigation into the relationship between the Dark Triad Personalities and online trolling behaviours towards popular and less popular Facebook profiles},
journal = {Computers in Human Behavior},
volume = {77},
pages = {69-76},
year = {2017},
author = {Barbara Lopes and Hui Yu}
}

@article{gibb2014who,
title = {Who does that anyway? Predictors and personality correlates of cyberbullying in college},
journal = {Computers in Human Behavior},
volume = {38},
pages = {8-16},
year = {2014},
author = {Zebbedia G. Gibb and Paul G. Devereux}
}

@article{goodboy2015personality,
title = {The personality profile of a cyberbully: Examining the Dark Triad},
journal = {Computers in Human Behavior},
volume = {49},
pages = {1-4},
year = {2015},
author = {Alan K. Goodboy and Matthew M. Martin},
}

@article{smoker2017predicting,
title = {Predicting perpetration of intimate partner cyberstalking: Gender and the Dark Tetrad},
journal = {Computers in Human Behavior},
volume = {72},
pages = {390-396},
year = {2017},
author = {Melissa Smoker and Evita March}
}

@article{kircaburun2019analyzing,
  title={Analyzing the links between problematic social media use, dark triad traits, and self-esteem},
  author={Kircaburun, Kagan and Demetrovics, Zsolt and Tosunta{\c{s}}, {\c{S}}ule Bet{\"u}l},
  journal={International Journal of Mental Health and Addiction},
  volume={17},
  number={6},
  pages={1496--1507},
  year={2019},
  publisher={Springer}
}

@article{kircaburun2018dark,
title = {The Dark Tetrad traits and problematic social media use: The mediating role of cyberbullying and cyberstalking},
journal = {Personality and Individual Differences},
volume = {135},
pages = {264-269},
year = {2018},
author = {Kagan Kircaburun and Peter K. Jonason and Mark D. Griffiths}
}

@article{demircioglu2021effects,
author = {Demircioğlu, Zeynep and Göncü Köse, Asli},
year = {2021},
month = {01},
title = {Effects of attachment styles, dark triad, rejection sensitivity, and relationship satisfaction on social media addiction: A mediated model},
volume = {40},
journal = {Current Psychology}
}

@inproceedings{sumner2012predicting,
author = {Sumner, Chris and Byers, Alison and Boochever, Rachel and Sumner, Chris and Byers, Alison and Boochever, Rachel and Park, Gregory},
year = {2012},
title = {Predicting Dark Triad Personality Traits from {Twitter} Usage and a Linguistic Analysis of Tweets},
booktitle = {The 11th International Conference on Machine Learning and Applications (ICMLA'12)}
}

@inproceedings{preotiuc2016studying,
  title={Studying the Dark Triad of personality through {Twitter} behavior},
  author={Preotiuc-Pietro, Daniel and Carpenter, Jordan and Giorgi, Salvatore and Ungar, Lyle},
  booktitle={The 25th ACM International Conference on Information and Knowledge Management (CIKM'16)},
  pages={761--770},
  year={2016}
}

@inproceedings{moskvichev2018using,
author = {Moskvichev, Arseny and Dubova, Marina and Menshov, Sergey and Filchenkov, Andrey},
year = {2018},
month = {09},
pages = {16-26},
title = {Using Linguistic Activity in Social Networks to Predict and Interpret Dark Psychological Traits},
booktitle = {Communications in Computer and Information Science}
}

@article{alavi2025relationships,
author = {Alavi, Masoumeh and Garg, Anchal and Wanigatunga, Niroshya},
year = {2025},
month = {05},
title = {The relationships between Dark Tetrad traits and adolescent cyberbullying and cybertrolling with online time and life satisfaction as moderators},
volume = {5},
journal = {Discover Psychology}
}

@article{balakrishnan2019cyberbullying,
title = {Cyberbullying detection on twitter using Big Five and Dark Triad features},
journal = {Personality and Individual Differences},
volume = {141},
pages = {252-257},
year = {2019},
author = {Vimala Balakrishnan and Shahzaib Khan and Terence Fernandez and Hamid R. Arabnia}
}

@inproceedings{chraibi2020automatic,
  author       = {Khaoula Chraibi and
                  Ilham Chaker and
                  Azeddine Zahi},
  title        = {Automatic Personality Prediction: {A} Systematic Mapping Study},
  booktitle    = {2020 {IEEE} Symposium Series on Computational Intelligence, {SSCI}
                  2020, Canberra, Australia, December 1-4, 2020},
  pages        = {2053--2060},
  publisher    = {{IEEE}},
  year         = {2020}
}

@article{levy2021personality,
author = {Phan, Le Vy and Rauthmann, John F.},
title = {Personality computing: New frontiers in personality assessment},
journal = {Social and Personality Psychology Compass},
volume = {15},
number = {7},
year = {2021}
}

@book{apa2007,
    title = {{APA} {Dictionary} of {Psychology}},
    editor = {VandenBos, Gary R.},
    year = {2007},
    publisher = {American Psychological Association},
    pages = {782}
}

@inproceedings{lees2022new,
  title={A new generation of {Perspective API}: Efficient multilingual character-level transformers},
  author={Lees, Alyssa and Tran, Vinh Q and Tay, Yi and Sorensen, Jeffrey and Gupta, Jai and Metzler, Donald and Vasserman, Lucy},
  booktitle={The 28th ACM SIGKDD Conference on Knowledge Discovery and Data Mining (KDD)},
  year={2022}
}

@incollection{graham2013moral,
  title={Moral foundations theory: The pragmatic validity of moral pluralism},
  author={Graham, Jesse and Haidt, Jonathan and Koleva, Sena and Motyl, Matt and Iyer, Ravi and Wojcik, Sean P and Ditto, Peter H},
  booktitle={Advances in experimental social psychology},
  volume={47},
  pages={55--130},
  year={2013},
  publisher={Elsevier}
}

@article{hopp2021extended,
  title={{The extended Moral Foundations Dictionary (eMFD): Development and applications of a crowd-sourced approach to extracting moral intuitions from text}},
  author={Hopp, Frederic R and Fisher, Jacob T and Cornell, Devin and Huskey, Richard and Weber, Ren{\'e}},
  journal={Behavior Research Methods},
  volume={53},
  number={1},
  year={2021}
}

@article{yousaf2023dark,
  title={Dark personalities and online reviews: A textual content analysis of review generation, consumption and distribution.},
  author={Yousaf, Salman and Kim, Jong Min},
  journal={Tourism Management},
  volume={98},
  year={2023}
}

@article{ramezani2022automatic,
  author       = {Majid Ramezani and
                  Mohammad{-}Reza Feizi{-}Derakhshi and
                  Mohammad Ali Balafar and
                  Meysam Asgari{-}Chenaghlu and
                  Ali{-}Reza Feizi{-}Derakhshi and
                  Narjes Nikzad{-}Khasmakhi and
                  Mehrdad Ranjbar{-}Khadivi and
                  Zoleikha Jahanbakhsh{-}Nagadeh and
                  Elnaz Zafarani{-}Moattar and
                  Taymaz Akan},
  title        = {Automatic personality prediction: an enhanced method using ensemble
                  modeling},
  journal      = {Neural Computing and Applications},
  volume       = {34},
  number       = {21},
  pages        = {18369--18389},
  year         = {2022}
}

@book{davison1997bootstrap,
  title={Bootstrap methods and their application},
  author={Davison, Anthony Christopher and Hinkley, David Victor},
  year={1997},
  publisher={Cambridge University Press}
}

@article{tausczik2010psychological,
author = {Tausczik, Yla and Pennebaker, James},
year = {2010},
month = {03},
pages = {24-54},
title = {The Psychological Meaning of Words: LIWC and Computerized Text Analysis Methods},
volume = {29},
journal = {Journal of Language and Social Psychology}
}

@article{tessa2025quantifying,
  title={Quantifying Feature Importance for Online Content Moderation},
  author={Tessa, Benedetta and Moreo, Alejandro and Cresci, Stefano and Fagni, Tiziano and Sebastiani, Fabrizio},
  journal={arXiv preprint arXiv:2510.19882},
  year={2025}
}

@inproceedings{popa2025effective,
  author       = {Victoria Popa and
                  Guglielmo Cola and
                  Caterina Senette and
                  Maurizio Tesconi},
  title        = {How effective are Large Language Models (LLMs) at inferring people's
                  personality based on texts they authored?},
  booktitle    = {Proceedings of the Joint National Conference on Cybersecurity {(ITASEC}
                  {\&} {SERICS} 2025)},
  year         = {2025}
}

@InProceedings{jain2022personality,
author="Jain, Dipika
and Kumar, Akshi
and Beniwal, Rohit",
title="Personality BERT: A Transformer-Based Model for Personality Detection from Textual Data",
booktitle="Proceedings of International Conference on Computing and Communication Networks",
year="2022",
publisher="Springer Nature Singapore",
pages="515--522"
}

@article{hickman2022automated,
author = {Hickman, Louis and Bosch, Nigel and Ng, Vincent and Saef, Rachel and Tay, Louis and Woo, Sang Eun},
year = {2022},
month = {08},
pages = {1323-1351},
title = {Automated Video Interview Personality Assessments: Reliability, Validity, and Generalizability Investigations},
journal = {Journal of Applied Psychology},
volume = {107},
number = {8}
}

@article{lukac2024speech,
author = {Lukac, Martin},
year = {2024},
month = {12},
title = {Speech-based personality prediction using deep learning with acoustic and linguistic embeddings},
volume = {14},
journal = {Scientific Reports}
}

@inproceedings{xu2022review,
  author       = {Yan Xu and
                  Yufang Tang and
                  Ching Y. Suen},
  title        = {Review of Handwriting Analysis for Predicting Personality Traits},
  booktitle    = {Structural, Syntactic, and Statistical Pattern Recognition - Joint
                  {IAPR} International Workshops, {S+SSPR} 2022},
  volume       = {13813},
  pages        = {54--63},
  publisher    = {Springer},
  year         = {2022}
}

@article{maliki2020personality,
author = {Maliki, Irfan and Sidik, M},
year = {2020},
month = {08},
title = {Personality Prediction System Based on Signatures Using Machine Learning},
volume = {879},
journal = {IOP Conference Series: Materials Science and Engineering}
}

@article{chen2023eye,
  author       = {Li Chen and
                  Wanling Cai and
                  Dongning Yan and
                  Shlomo Berkovsky},
  title        = {Eye-tracking-based personality prediction with recommendation interfaces},
  journal      = {User Model. User Adapt. Interact.},
  volume       = {33},
  number       = {1},
  pages        = {121--157},
  year         = {2023}
}

@article{meidenbauer2023mouse,
author = {Meidenbauer, Kimberly L. and Niu, Tianyue and Choe, Kyoung Whan and Stier, Andrew J. and Berman, Marc G.},
title = {Mouse movements reflect personality traits and task attentiveness in online experiments},
journal = {Journal of Personality},
volume = {91},
number = {2},
pages = {413-425},
year = {2023}
}

@article{karanatsiou2022tweets,
  author       = {Dimitra Karanatsiou and
                  Pavlos Sermpezis and
                  Dritjon Gruda and
                  Konstantinos Kafetsios and
                  Ilias Dimitriadis and
                  Athena Vakali},
  title        = {My Tweets Bring All the Traits to the Yard: Predicting Personality
                  and Relational Traits in Online Social Networks},
  journal      = {{ACM} Trans. Web},
  volume       = {16},
  number       = {2},
  pages        = {10:1--10:26},
  year         = {2022}
}

@inproceedings{mushtaq2020PredictingMP,
  title={Predicting MBTI Personality type with K-means Clustering and Gradient Boosting},
  author={Zeeshan Mushtaq and Sagar Ashraf and Nosheen Sabahat},
  booktitle={2020 IEEE 23rd International Multitopic Conference (INMIC)},
  year={2020},
  pages={1-5},
  url={https://api.semanticscholar.org/CorpusID:231684233}
}

@inproceedings{treves2025viki,
  author       = {Ben Treves and
                  Emiliano De Cristofaro and
                  Yue Dong and
                  Michalis Faloutsos},
  title        = {{VIKI:} Systematic Cross-Platform Profile Inference of Tech Users},
  booktitle    = {Proceedings of the 17th {ACM} Web Science Conference 2025, Websci
                  2025, New Brunswick, NJ, USA, May 20-24, 2025},
  pages        = {32--41},
  publisher    = {{ACM}},
  year         = {2025}
}

@inproceedings{ganesan2023systematic,
  author       = {Adithya V. Ganesan and
                  Yash Kumar Lal and
                  August H{\aa}kan Nilsson and
                  H. Andrew Schwartz},
  title        = {Systematic Evaluation of {GPT-3} for Zero-Shot Personality Estimation},
  booktitle    = {Proceedings of the 13th Workshop on Computational Approaches to Subjectivity,
                  Sentiment, {\&} Social Media Analysis, WASSA@ACL 2023, Toronto,
                  Canada, July 14, 2023},
  pages        = {390--400},
  publisher    = {Association for Computational Linguistics},
  year         = {2023}
}

@article{peters2024large,
  author       = {Heinrich Peters and
                  Moran Cerf and
                  Sandra C. Matz},
  title        = {Large Language Models Can Infer Personality from Free-Form User Interactions},
  journal      = {CoRR},
  volume       = {abs/2405.13052},
  year         = {2024},
}

@inproceedings{tardelli2020characterizing,
  author       = {Serena Tardelli and
                  Marco Avvenuti and
                  Maurizio Tesconi and
                  Stefano Cresci},
  title        = {Characterizing Social Bots Spreading Financial Disinformation},
  booktitle    = {Social Computing and Social Media. Design, Ethics, User Behavior,
                  and Social Network Analysis - 12th International Conference, {SCSM} 2020},
  volume       = {12194},
  pages        = {376--392},
  publisher    = {Springer},
  year         = {2020}
}

@article{saw2022designing,
  title={Designing for trust on E-commerce websites using two of the big five personality traits},
  author={Saw, Chian Chyi and Inthiran, Anushia},
  journal={Journal of Theoretical and Applied Electronic Commerce Research},
  volume={17},
  number={2},
  pages={375--393},
  year={2022},
  publisher={MDPI}
}

@article{utami2021personality,
  title={Personality classification of facebook users according to big five personality using SVM (support vector machine) method},
  author={Utami, Ninda Anggoro and Maharani, Warih and Atastina, Imelda},
  journal={Procedia Computer Science},
  volume={179},
  pages={177--184},
  year={2021},
  publisher={Elsevier}
}

@article{mahadevan2024conceptualizing,
  title={Conceptualizing grandiose and vulnerable narcissism as alternative status-seeking strategies: Insights from hierometer theory},
  author={Mahadevan, Nikhila},
  journal={Social and Personality Psychology Compass},
  volume={18},
  number={6},
  year={2024},
  publisher={Wiley Online Library}
}

@article{biselli2025mitigating,
  title={Mitigating Misinformation Sharing on Social Media through Personalised Nudging},
  author={Biselli, Tom and Hartwig, Katrin and Reuter, Christian},
  journal={Proceedings of the ACM on Human-Computer Interaction},
  volume={9},
  number={2},
  pages={1--44},
  year={2025},
  publisher={ACM New York, NY, USA}
}

@inproceedings{cresci2022personalized,
 author = {Cresci, Stefano and Trujillo, Amaury and Fagni, Tiziano},
 booktitle = {The 33rd ACM Conference on Hypertext and Social Media (HT'22)},
 organization = {ACM},
 pages = {248--251},
 title = {{Personalized interventions for online moderation}},
 year = {2022}
}

@article{thissen2002quick,
  title={Quick and easy implementation of the {Benjamini-Hochberg} procedure for controlling the false positive rate in multiple comparisons},
  author={Thissen, David and Steinberg, Lynne and Kuang, Daniel},
  journal={Journal of Educational and Behavioral Statistics},
  volume={27},
  number={1},
  year={2002}
}

@article{barron2019proppy,
  title={Proppy: Organizing the news based on their propagandistic content},
  author={Barr{\'o}n-Cedeno, Alberto and Jaradat, Israa and Da San Martino, Giovanni and Nakov, Preslav},
  journal={Information Processing \& Management},
  volume={56},
  number={5},
  year={2019}
}

@article{razuvayevskaya2024comparison,
  title={Comparison between parameter-efficient techniques and full fine-tuning: A case study on multilingual news article classification},
  author={Razuvayevskaya, Olesya and Wu, Ben and Leite, Jo{\~a}o A and Heppell, Freddy and Srba, Ivan and Scarton, Carolina and Bontcheva, Kalina and Song, Xingyi},
  journal={PLoS One},
  volume={19},
  number={5},
  year={2024}
}

@inproceedings{nogara2025toxic,
  title={Toxic Bias: Perspective API misreads German as more toxic},
  author={Nogara, Gianluca and Pierri, Francesco and Cresci, Stefano and Luceri, Luca and T{\"o}rnberg, Petter and Giordano, Silvia},
  booktitle={Proceedings of the international AAAI conference on web and social media},
  volume={19},
  pages={1346--1357},
  year={2025}
}

@article{assenmacher2022benchmarking,
  title={Benchmarking crisis in social media analytics: A solution for the data-sharing problem},
  author={Assenmacher, Dennis and Weber, Derek and Preuss, Mike and Calero Valdez, Andre and Bradshaw, Alison and Ross, Bj{\"o}rn and Cresci, Stefano and Trautmann, Heike and Neumann, Frank and Grimme, Christian},
  journal={Social Science Computer Review},
  volume={40},
  number={6},
  year={2022}
}

@article{tromble2021have,
  title={Where have all the data gone? A critical reflection on academic digital research in the {post-API} age},
  author={Tromble, Rebekah},
  journal={Social Media + Society},
  volume={7},
  number={1},
  year={2021}
}

@article{graf2022did,
  title={Why did you do that? Differential types of aggression in offline and in cyberbullying},
  author={Graf, Daniel and Yanagida, Takuya and Runions, Kevin and Spiel, Christiane},
  journal={Computers in Human Behavior},
  volume={128},
  pages={107107},
  year={2022},
  publisher={Elsevier}
}

@article{winter2021effects,
  title={The effects of trait-based personalization in social media advertising},
  author={Winter, Stephan and Maslowska, Ewa and Vos, Anne L},
  journal={Computers in Human Behavior},
  volume={114},
  pages={106525},
  year={2021},
  publisher={Elsevier}
}

@article{paulhus2007self,
  title={The self-report method},
  author={Paulhus, Delroy and Vazire, Simine and others},
  journal={Handbook of research methods in personality psychology},
  volume={1},
  number={2007},
  pages={224--239},
  year={2007},
  publisher={Guilford;}
}

@article{montag2022we,
  title={Do we still need psychological self-report questionnaires in the age of the Internet of things?},
  author={Montag, Christian and Dagum, Paul and Hall, Brian J and Elhai, Jon D},
  journal={Discover Psychology},
  volume={2},
  number={1},
  pages={1},
  year={2022},
  publisher={Springer}
}

@ARTICLE{majumder2017deep,
  author={Majumder, Navonil and Poria, Soujanya and Gelbukh, Alexander and Cambria, Erik},
  journal={IEEE Intelligent Systems}, 
  title={Deep Learning-Based Document Modeling for Personality Detection from Text}, 
  year={2017},
  volume={32},
  number={2},
  pages={74-79},
}

@article{zhao2022deep,
  title={Deep personality trait recognition: a survey},
  author={Zhao, Xiaoming and Tang, Zhiwei and Zhang, Shiqing},
  journal={Frontiers in psychology},
  volume={13},
  pages={839619},
  year={2022},
  publisher={Frontiers Media SA}
}

@inproceedings{guleva2022personality,
  title={Personality traits classification from EEG signals using EEGNet},
  author={Guleva, Veronika and Calcagno, Alessandra and Reali, Pierluigi and Bianchi, Anna Maria},
  booktitle={2022 IEEE 21st Mediterranean Electrotechnical Conference (MELECON)},
  pages={590--594},
  year={2022},
  organization={IEEE}
}

@article{delisi2021overlapping,
  title={Overlapping measures or constructs? An empirical study of the overlap between self-control, psychopathy, Machiavellianism and narcissism},
  author={DeLisi, Matt and Pechorro, Pedro and Maroco, Jo{\~a}o and Sim{\~o}es, M{\'a}rio},
  journal={Forensic science international: synergy},
  volume={3},
  pages={100141},
  year={2021},
  publisher={Elsevier}
}

@article{patrick2009triarchic,
  title={Triarchic conceptualization of psychopathy: Developmental origins of disinhibition, boldness, and meanness},
  author={Patrick, Christopher J and Fowles, Don C and Krueger, Robert F},
  journal={Development and psychopathology},
  volume={21},
  number={3},
  pages={913--938},
  year={2009},
  publisher={Cambridge University Press}
}

@article{leng2020bridging,
  title={Bridging personality and online prosocial behavior: The roles of empathy, moral identity, and social self-efficacy},
  author={Leng, Jie and Guo, Qingke and Ma, Bingqing and Zhang, Shuyue and Sun, Peng},
  journal={Frontiers in psychology},
  volume={11},
  pages={575053},
  year={2020},
  publisher={Frontiers Media SA}
}

@article{xu2024meta,
  title={A meta-analysis of the relationship between personality traits and cyberbullying},
  author={Xu, Weilin and Zhao, Baobao and Jin, Cancan},
  journal={Aggression and Violent Behavior},
  volume={79},
  pages={101992},
  year={2024},
  publisher={Elsevier}
}

@article{yang2022computational,
  title={Computational personality: a survey},
  author={Yang, Liang and Li, Shuqun and Luo, Xi and Xu, Bo and Geng, Yuanling and Zeng, Zeyuan and Zhang, Fan and Lin, Hongfei},
  journal={Soft Computing},
  volume={26},
  number={18},
  pages={9587--9605},
  year={2022},
  publisher={Springer}
}

@inproceedings{banko2020unified,
  title={A unified taxonomy of harmful content},
  author={Banko, Michele and MacKeen, Brendon and Ray, Laurie},
  booktitle={Proceedings of the fourth workshop on online abuse and harms},
  pages={125--137},
  year={2020}
}

@article{gottfried2024americans,
  title={Americans’ social media use},
  author={Gottfried, Jeffrey},
  journal={Pew Research Center},
  volume={31},
  year={2024},
  publisher={JSTOR}
}

@article{morf2001unraveling,
  title={Unraveling the paradoxes of narcissism: A dynamic self-regulatory processing model},
  author={Morf, Carolyn C and Rhodewalt, Frederick},
  journal={Psychological inquiry},
  volume={12},
  number={4},
  pages={177--196},
  year={2001},
  publisher={Taylor \& Francis}
}

@article{jones2016nature,
  title={The nature of Machiavellianism: Distinct patterns of misbehavior.},
  author={Jones, Daniel N},
  year={2016},
  publisher={American Psychological Association},
  journal = {The dark side of personality: Science and practice in social, personality, and clinical psychology},
  pages = {87--107}
}

@article{elsherief2021latent,
  title={Latent hatred: A benchmark for understanding implicit hate speech},
  author={ElSherief, Mai and Ziems, Caleb and Muchlinski, David and Anupindi, Vaishnavi and Seybolt, Jordyn and De Choudhury, Munmun and Yang, Diyi},
  journal={arXiv preprint arXiv:2109.05322},
  year={2021}
}

@article{katz2022dark,
  title={The dark side of humanity scale: a reconstruction of the dark tetrad constructs},
  author={Katz, Louise and Harvey, Caroline and Baker, Ian S and Howard, Chris},
  journal={Acta Psychologica},
  volume={222},
  pages={103461},
  year={2022},
  publisher={Elsevier}
}

@article{zezulka2016differentiating,
  title={Differentiating cyberbullies and internet trolls by personality characteristics and self-esteem},
  author={Zezulka, Lauren A and Seigfried-Spellar, Kathryn},
  year={2016},
  journal = {Journal of Digital Forensics, Security and Law},
}

@article{holtzman2019linguistic,
  title={Linguistic markers of grandiose narcissism: A LIWC analysis of 15 samples},
  author={Holtzman, Nicholas S and Tackman, Allison M and Carey, Angela L and Brucks, Melanie S and K{\"u}fner, Albrecht CP and Deters, Fenne Gro{\ss}e and Back, Mitja D and Donnellan, M Brent and Pennebaker, James W and Sherman, Ryne A and others},
  journal={Journal of Language and Social Psychology},
  volume={38},
  number={5-6},
  pages={773--786},
  year={2019},
  publisher={Sage Publications Sage CA}
}

@article{stevens2022emotions,
  title={Emotions and incivility in vaccine mandate discourse: natural language processing insights},
  author={Stevens, Hannah and Rasul, Muhammad Ehab and Oh, Yoo Jung and others},
  journal={Jmir Infodemiology},
  volume={2},
  number={2},
  pages={e37635},
  year={2022},
  publisher={JMIR Publications Inc.}
}

@inproceedings{cheng2017anyone,
  title={Anyone can become a troll: Causes of trolling behavior in online discussions},
  author={Cheng, Justin and Bernstein, Michael and Danescu-Niculescu-Mizil, Cristian and Leskovec, Jure},
  booktitle={Proceedings of the 2017 ACM conference on computer supported cooperative work and social computing},
  pages={1217--1230},
  year={2017}
}

@article{erreygers2018positive,
  title={Positive or negative spirals of online behavior? Exploring reciprocal associations between being the actor and the recipient of prosocial and antisocial behavior online},
  author={Erreygers, Sara and Vandebosch, Heidi and Vranjes, Ivana and Baillien, Elfi and De Witte, Hans},
  journal={New Media \& Society},
  volume={20},
  number={9},
  pages={3437--3456},
  year={2018},
  publisher={Sage publications Sage UK}
}

@article{kumarswamy2023impact,
  title={Impact of Stricter Content Moderation on Parler's Users' Discourse},
  author={Kumarswamy, Nihal and Singhal, Mohit and Nilizadeh, Shirin},
  journal={arXiv preprint arXiv:2310.08844},
  year={2023}
}

@article{cima2025investigating,
  title={Investigating the heterogeneous effects of a massive content moderation intervention via Difference-in-Differences},
  author={Cima, Lorenzo and Tessa, Benedetta and Trujillo, Amaury and Cresci, Stefano and Avvenuti, Marco},
  journal={Online Social Networks and Media},
  volume={48},
  pages={100320},
  year={2025},
  publisher={Elsevier}
}

@inproceedings{cima2024great,
 author = {Cima, Lorenzo and Trujillo, Amaury and Avvenuti, Marco and Cresci, Stefano},
 booktitle = {The 16th ACM Web Science Conference Companion (WebSci'24)},
 organization = {ACM},
 pages = {85--93},
 title = {{The Great Ban: Efficacy and unintended consequences of a massive deplatforming operation on Reddit}},
 year = {2024}
}

@article{barlett2024meta,
  title={Meta-analyses of the predictors and outcomes of cyberbullying perpetration and victimization while controlling for traditional bullying perpetration and victimization},
  author={Barlett, Christopher P and Kowalski, Robin M and Wilson, Annie M},
  journal={Aggression and violent behavior},
  volume={74},
  pages={101886},
  year={2024},
  publisher={Elsevier}
}

@article{wang2026dark,
  title={The dark triad and cyber aggression: Testing the longitudinal mediation of moral disengagement and toxic online disinhibition},
  author={Wang, Cheng-Yen and Liu, Yih-Lan and Chang, Chia-Yun},
  journal={Cyberpsychology: Journal of Psychosocial Research on Cyberspace},
  volume={20},
  number={2},
  year={2026}
}

@article{wu2023individuals,
  title={Why individuals with psychopathy and moral disengagement are more likely to engage in online trolling? The online disinhibition effect},
  author={Wu, Biyun and Xiao, Yubei and Zhou, Li and Li, Fang and Liu, Mingfan},
  journal={Journal of Psychopathology and Behavioral Assessment},
  volume={45},
  number={2},
  pages={322--332},
  year={2023},
  publisher={Springer}
}

@article{hawdon2017exposure,
  title={Exposure to online hate in four nations: A cross-national consideration},
  author={Hawdon, James and Oksanen, Atte and R{\"a}s{\"a}nen, Pekka},
  journal={Deviant behavior},
  volume={38},
  number={3},
  pages={254--266},
  year={2017},
  publisher={Taylor \& Francis}
}

@article{wachs2018associations,
  title={Associations between bystanders and perpetrators of online hate: The moderating role of toxic online disinhibition},
  author={Wachs, Sebastian and Wright, Michelle F},
  journal={International journal of environmental research and public health},
  volume={15},
  number={9},
  pages={2030},
  year={2018},
  publisher={MDPI}
}

\newpage
\begin{appendix}

\section{Self-reported Social Media Habits}
\label{sec:app_soc_usage}
\begin{figure*}[bh!]
    \includegraphics[width=0.35\linewidth]{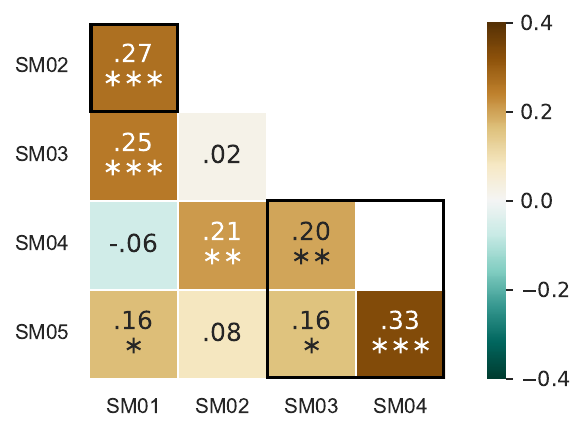}
    \centering
    \caption{Spearman rank correlation coefficients among self-reported social media habits. \textit{SM01}--\textit{SM05} correspond to the questions from the \textit{Social Media Use} section of the questionnaire. Black outlines highlight two conceptually distinct groups of items: general social media usage (SM01–SM02) and toxicity-related interactions (SM03–SM05). Asterisks denote statistical significance of the correlations: *: $p < 0.1$, **: $p < 0.05$, ***: $p < 0.01$.}
    \label{fig:hm_soc_usage}
\end{figure*}

Figure \ref{fig:hm_soc_usage} exhibits Spearman rank correlations among the five social media use items (\textit{SM01}–\textit{SM05}) included in our questionnaire. Black frames highlight correlations within two conceptual subsets of items: \textit{SM01} and \textit{SM02} were developed ad hoc for this study and capture participants' general social media activity (i.e., content consumption and creation), whereas the second subset reflects three different behaviors in uncivil interaction: passive exposure as bystander (\textit{SM03}), direct victimization (\textit{SM04}) and active engagement (\textit{SM05}). The last three items were formulated following previous research~\cite{hawdon2017exposure,wachs2018associations}. Overall, correlations tend to be low, indicating that the questions capture distinct aspects of participants’ social media behavior. The most notable exception is represented by the moderate and highly significant correlation ($\rho = 0.332$, $p < 0.01$) between being the target of uncivil content (\textit{SM04}) and producing toxic content (\textit{SM05}). Notably, aspects of social media use (SM01–SM05) were assessed using single-item measures. Future research should consider employing validated multi-item scales to address the limitations associated with single-item assessments, such as reduced content validity, limited sensitivity, and the inability to evaluate internal consistency reliability. 

\begin{table}[bh!]
\centering
\begin{tabular}{lrr}
\toprule
\textbf{Scale} & \textbf{N items} & \textbf{Cronbach's $\alpha$} \\
\midrule
DSHS\_SuccPsycho & 18  & 0.927 \\
DSHS\_GrandEnt & 9  & 0.872 \\
DSHS\_SadCru & 8  & 0.791 \\
DSHS\_EntRage & 7  & 0.800 \\
CTDS  & 13  & 0.863 \\
\bottomrule
\end{tabular}
\caption{Cronbach’s $\alpha$ coefficients and number of items for each subscale included in the questionnaire.}
\label{tab:cronbach_alpha}
\end{table}

\section{Socio-Demographic Characteristics}
\label{sec:app_personal_data}

\begin{figure*}[h]
    \begin{subfigure}[t]{0.25\textwidth}\centering
            \includegraphics[width=\textwidth]{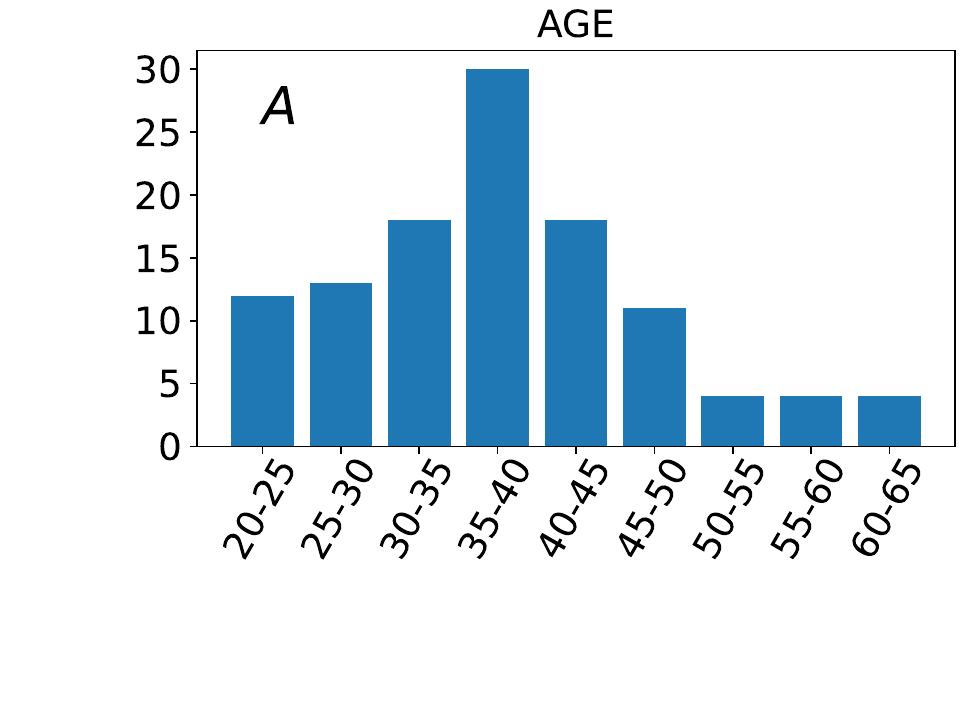}\end{subfigure}\begin{subfigure}[t]{0.25\textwidth}\centering
            \includegraphics[width=\textwidth]{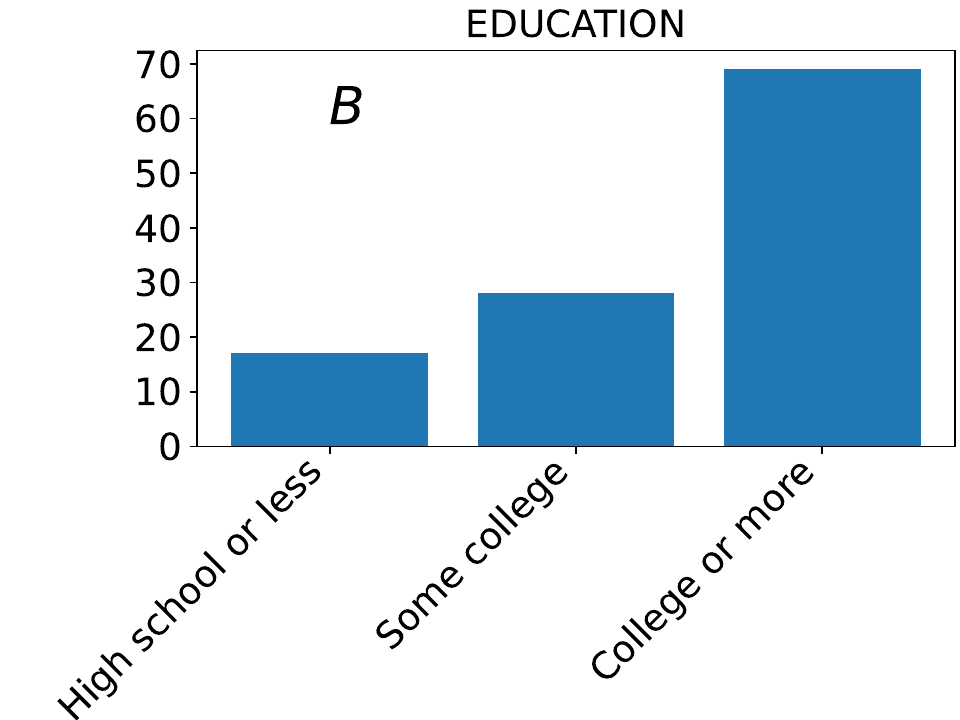}\end{subfigure}
	\begin{subfigure}[t]{0.25\textwidth}\centering
            \includegraphics[width=\textwidth]{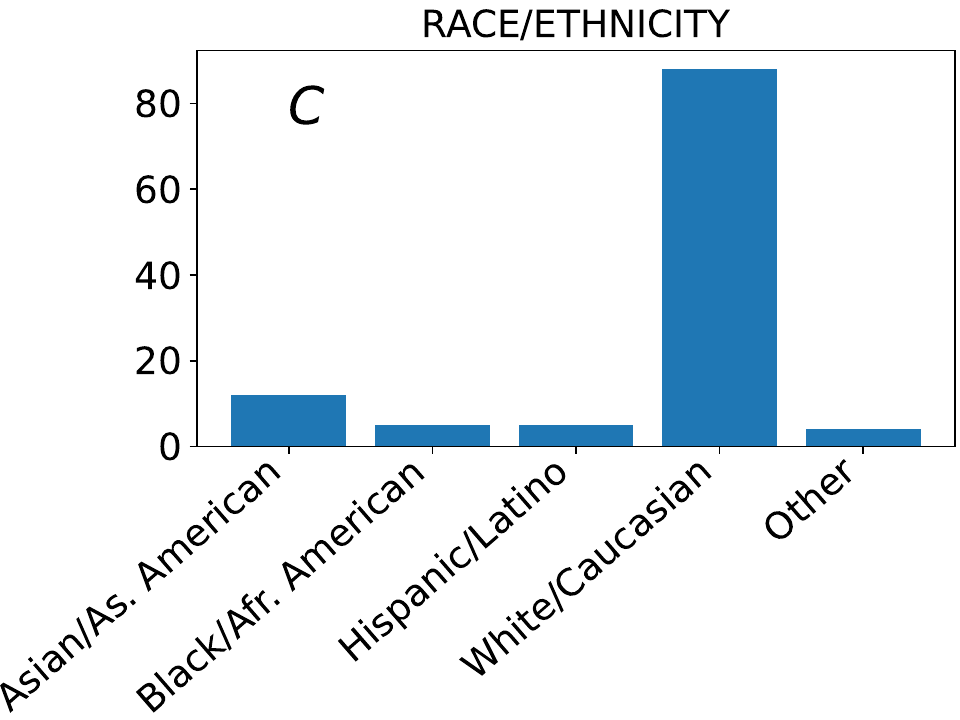}\end{subfigure}\begin{subfigure}[t]{0.25\textwidth}\centering
            \includegraphics[width=\textwidth]{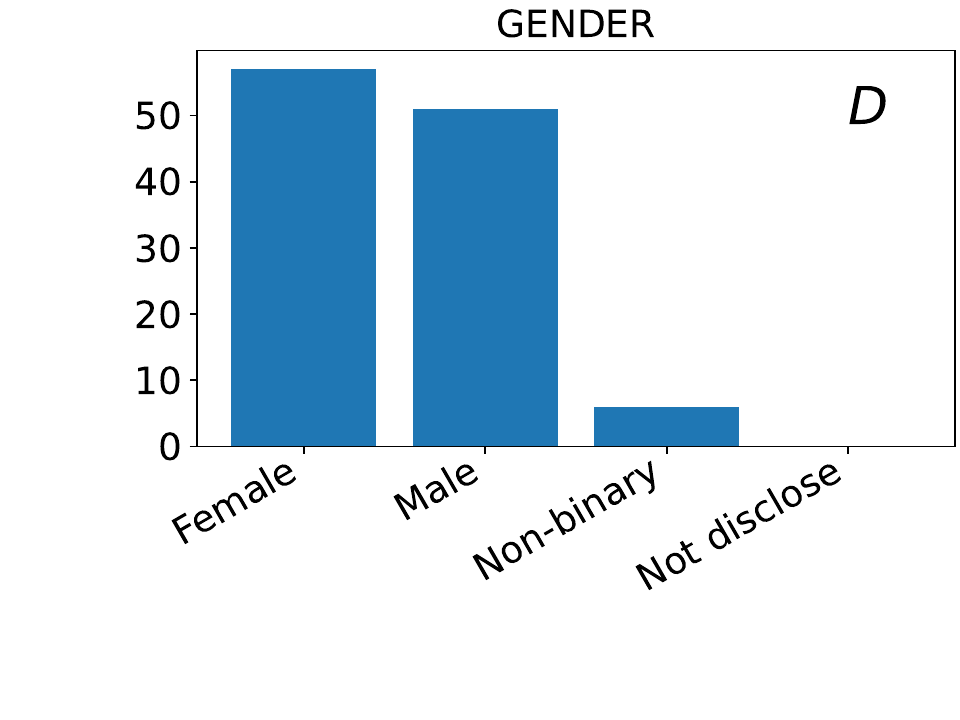}\end{subfigure} \\
	\begin{subfigure}[t]{0.25\textwidth}\centering
            \includegraphics[width=\textwidth]{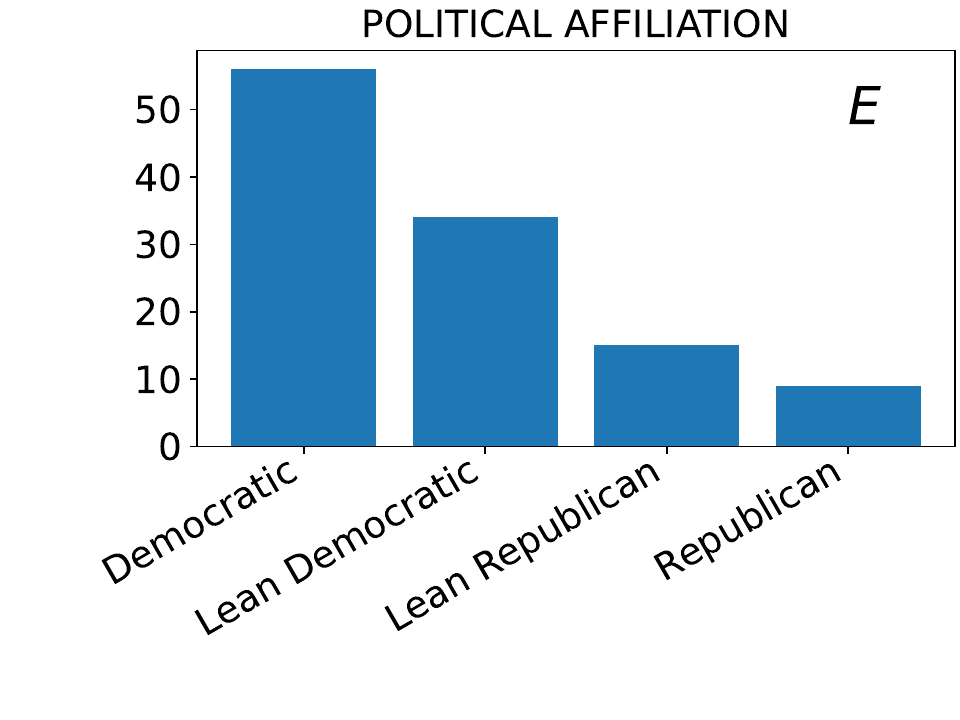}\end{subfigure}\begin{subfigure}[t]{0.25\textwidth}\centering
            \includegraphics[width=\textwidth]{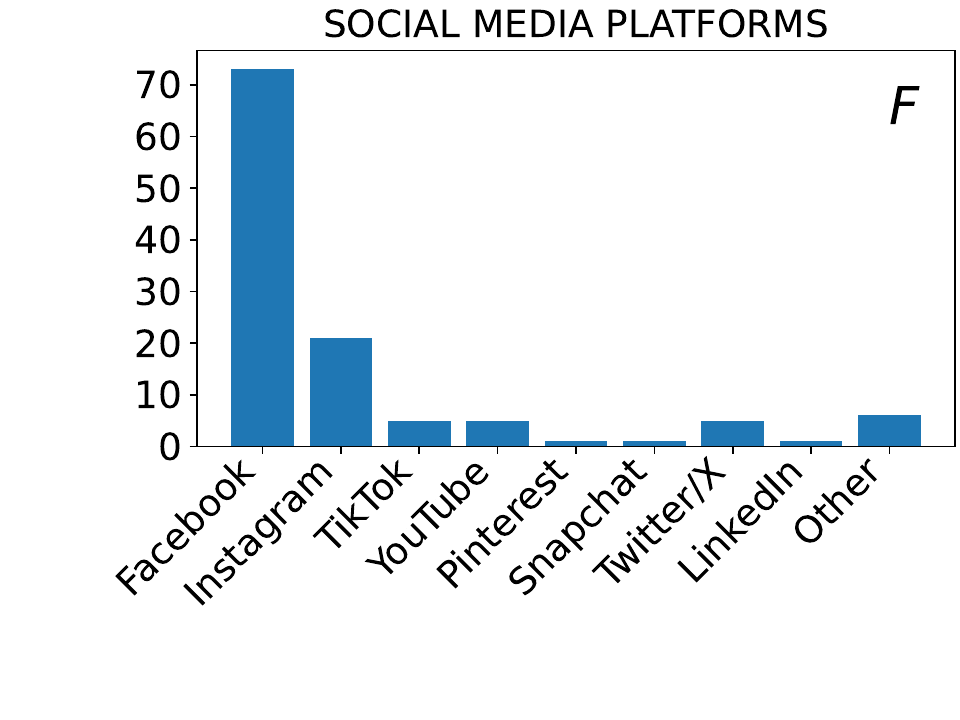}\end{subfigure}\begin{subfigure}[t]{0.25\textwidth}\centering
            \includegraphics[width=\textwidth]{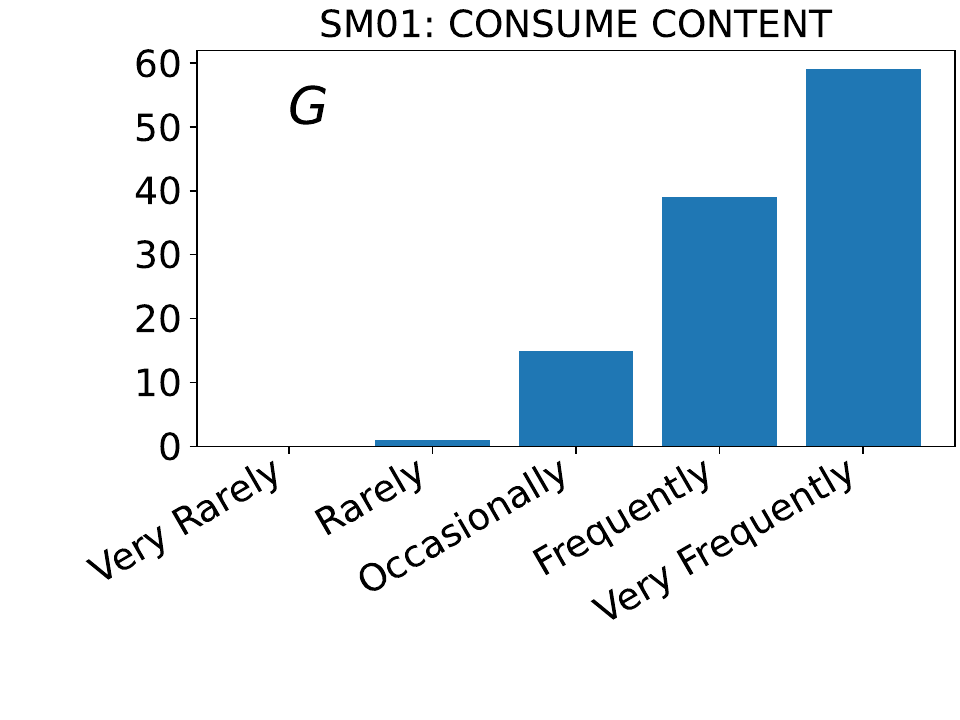}\end{subfigure}\begin{subfigure}[t]{0.25\textwidth}\centering
            \includegraphics[width=\textwidth]{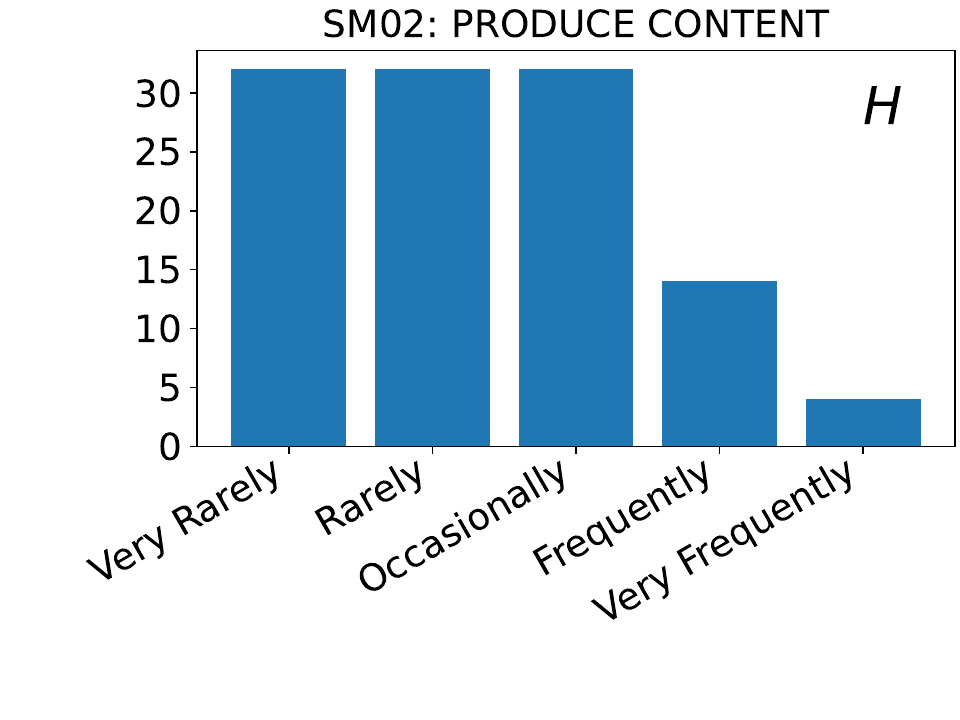}\end{subfigure} \\
    	\begin{subfigure}[t]{0.25\textwidth}\centering
            \includegraphics[width=\textwidth]{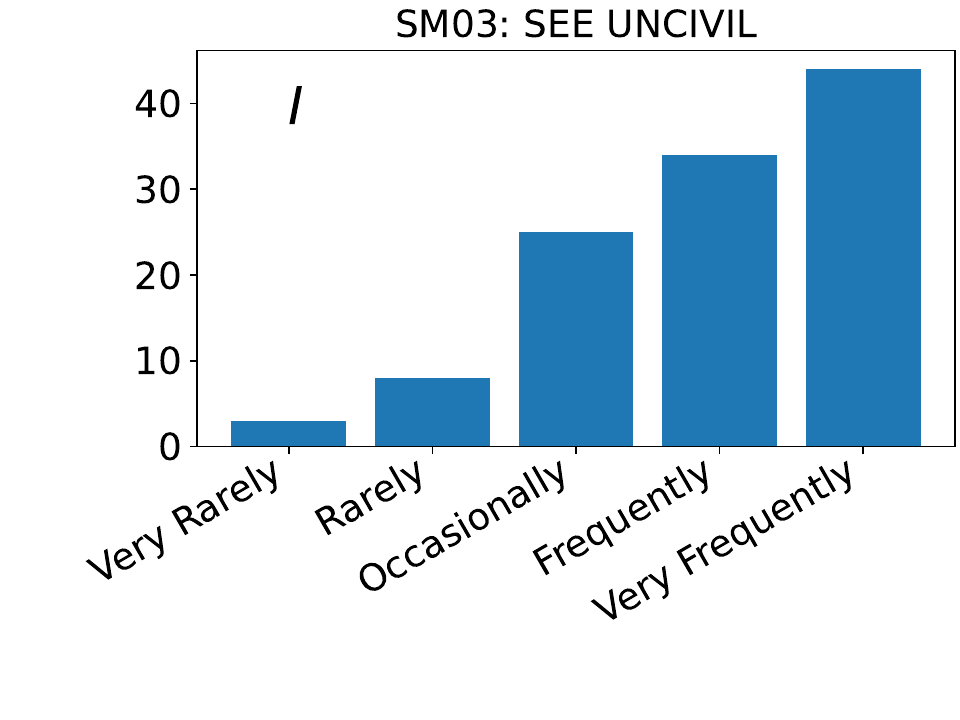}\end{subfigure}
    \begin{subfigure}[t]{0.25\textwidth}\centering
            \includegraphics[width=\textwidth]{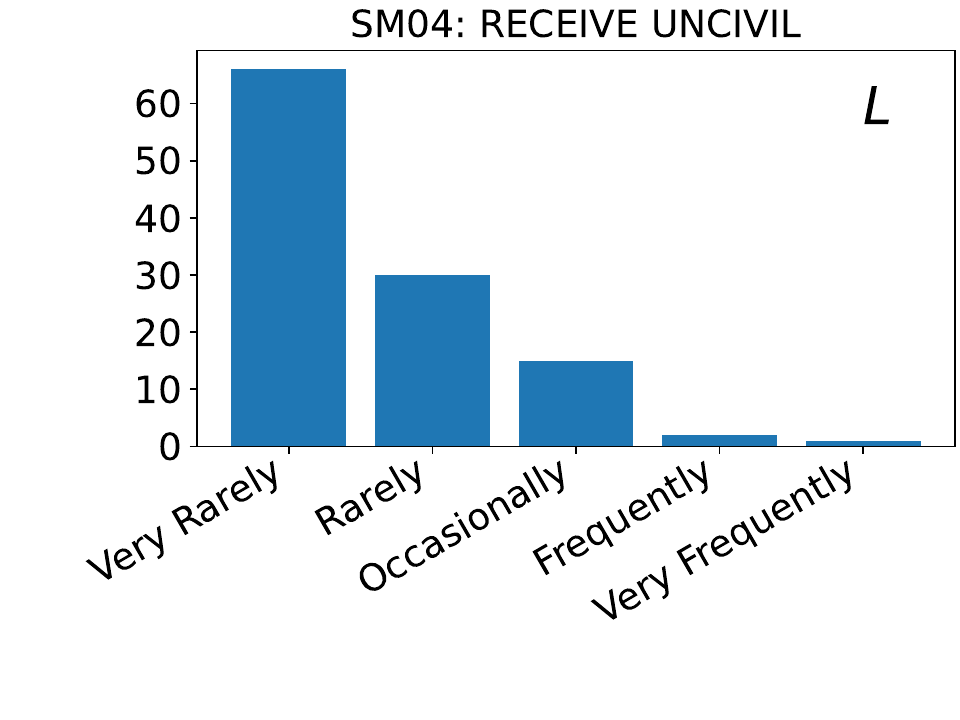}\end{subfigure}\begin{subfigure}[t]{0.25\textwidth}\centering
            \includegraphics[width=\textwidth]{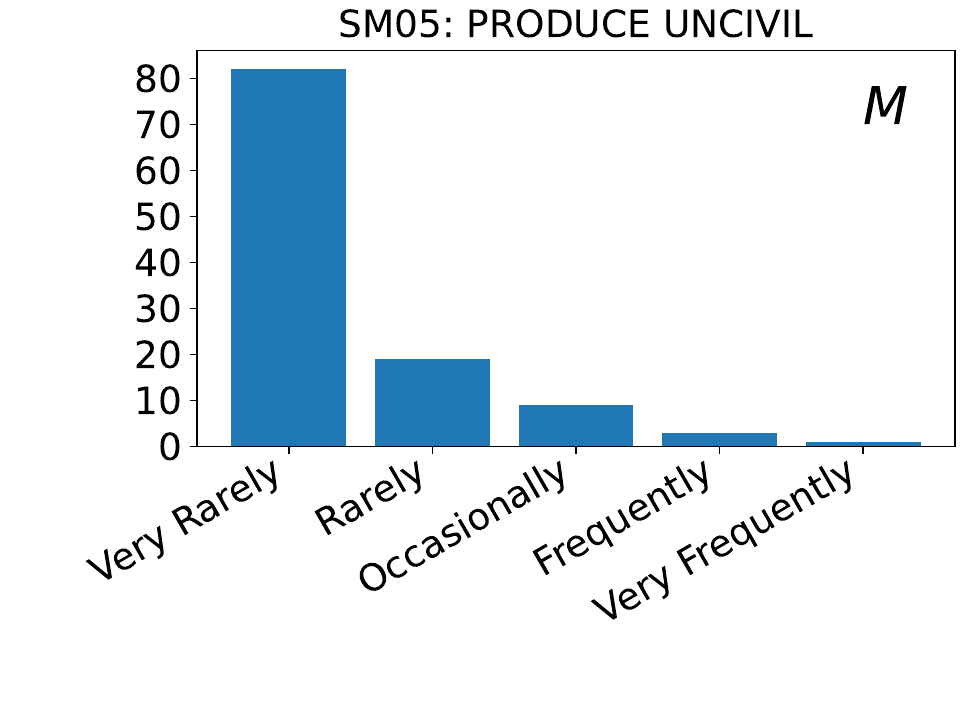}\end{subfigure}\caption{Socio-demographic characteristics and social media use habits of the participants in the crowdsourcing campaign.}
    \label{fig:socio-demo-distribution}
\end{figure*}

Figure \ref{fig:socio-demo-distribution} illustrates the distributions of participants’ socio-demographic characteristics and their responses to questions about social media usage, offering a clearer picture of the heterogeneous nature of the study sample.

\section{Hierarchical Clustering for Feature Selection}
\label{sec:app_feature_selection}

\begin{figure*}[h]
    \centering
    \begin{subfigure}{0.92\textwidth}
        \includegraphics[width=0.5\linewidth]{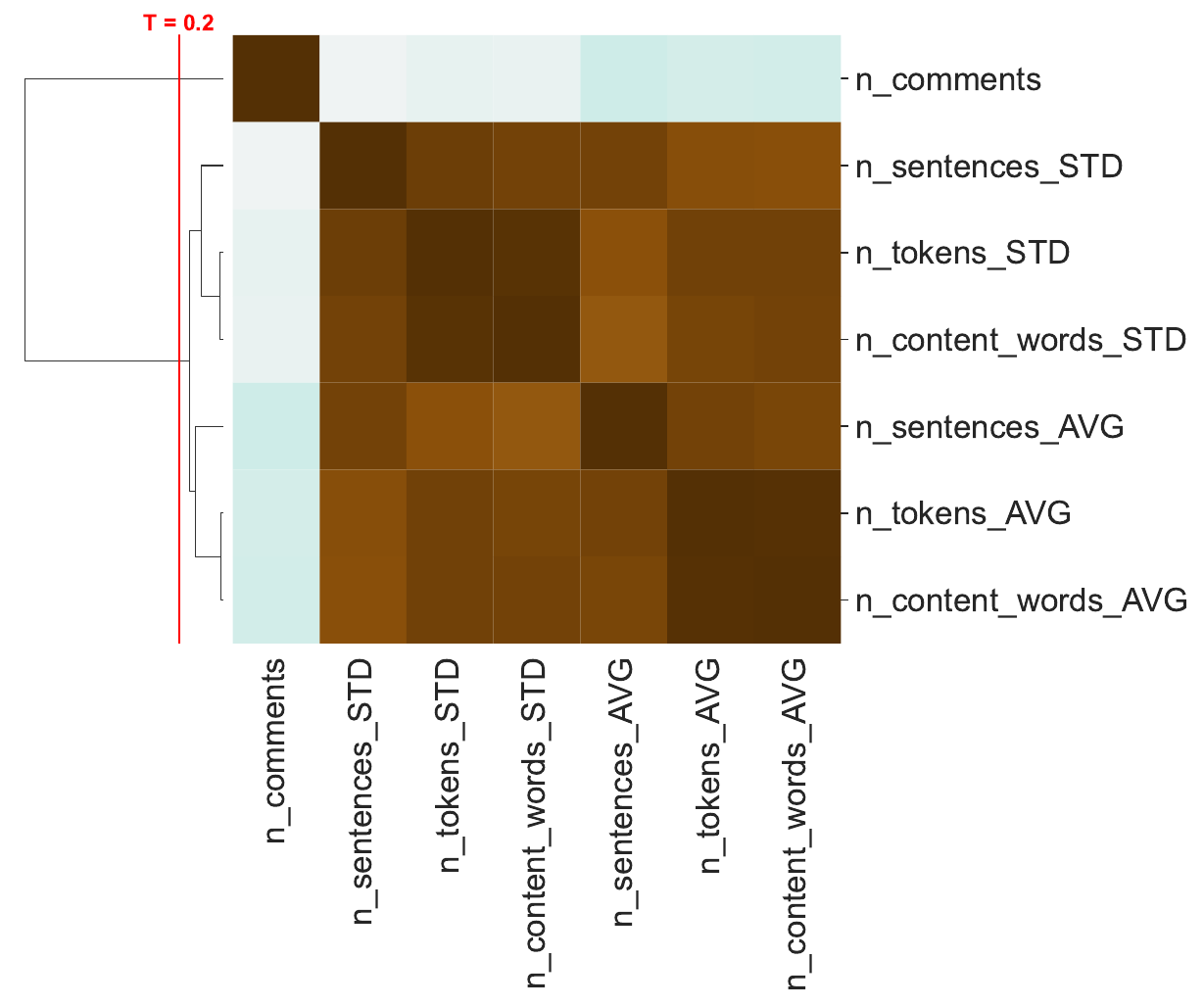}
        \centering
    \end{subfigure}
    \begin{subfigure}{0.05\textwidth}
        \includegraphics[width=0.95\linewidth]{pictures/colorbar_cropped.pdf}
        \vspace{1cm}
        \centering
    \end{subfigure}
    \caption{Pairwise Spearman correlation matrix for the features in the basic linguistic group, together with the hierarchical clustering dendrogram used for feature selection. The vertical red line indicates the threshold used to define clusters. For each cluster, the medoid was selected as the representative feature. The procedure removed 5 features (71\% reduction). The two remaining features, \texttt{n\_comments} and \texttt{n\_tokens\_STD}, exhibit a low residual correlation ($\rho = -0.061$).}
    \label{fig:dendrogram_basic_text}
\end{figure*}

\begin{figure*}[h]
    \centering
    \begin{subfigure}{0.92\textwidth}
        \includegraphics[width=1\linewidth]{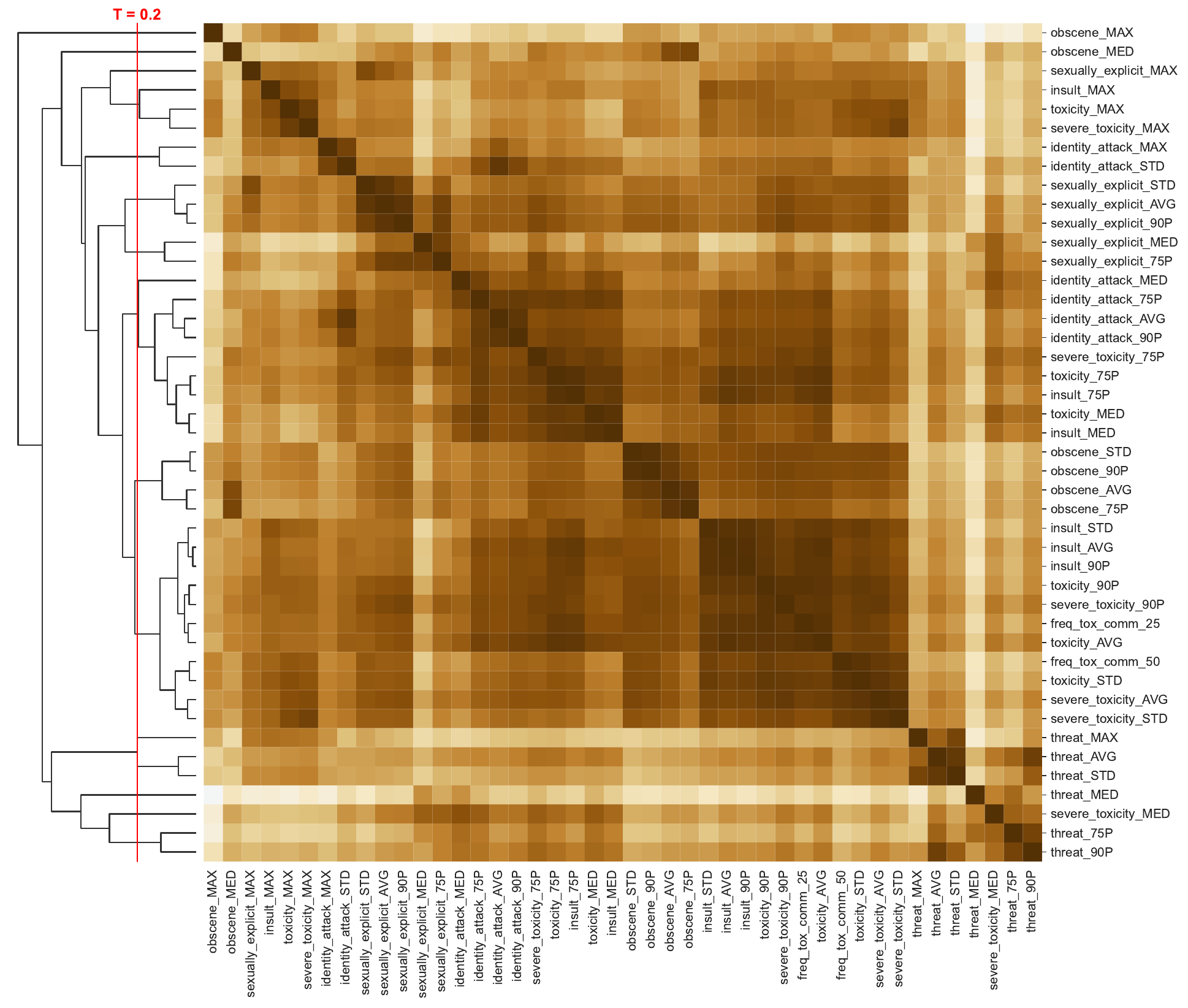}
        \centering
    \end{subfigure}
    \hspace{3pt}
    \begin{subfigure}{0.05\textwidth}
        \includegraphics[width=0.95\linewidth]{pictures/colorbar_cropped.pdf}
        \vspace{1.25cm}
        \centering
    \end{subfigure}
    \caption{Pairwise Spearman correlation matrix for the features in the toxicity group, together with the hierarchical clustering dendrogram used for feature selection. The vertical red line indicates the threshold used to define clusters. For each cluster, the medoid was selected as the representative feature.}
    \label{fig:dendrogram_toxicity}
\end{figure*}

\begin{figure*}[h]
    \centering
    \begin{subfigure}{0.45\textwidth}
        \includegraphics[width=1\linewidth]{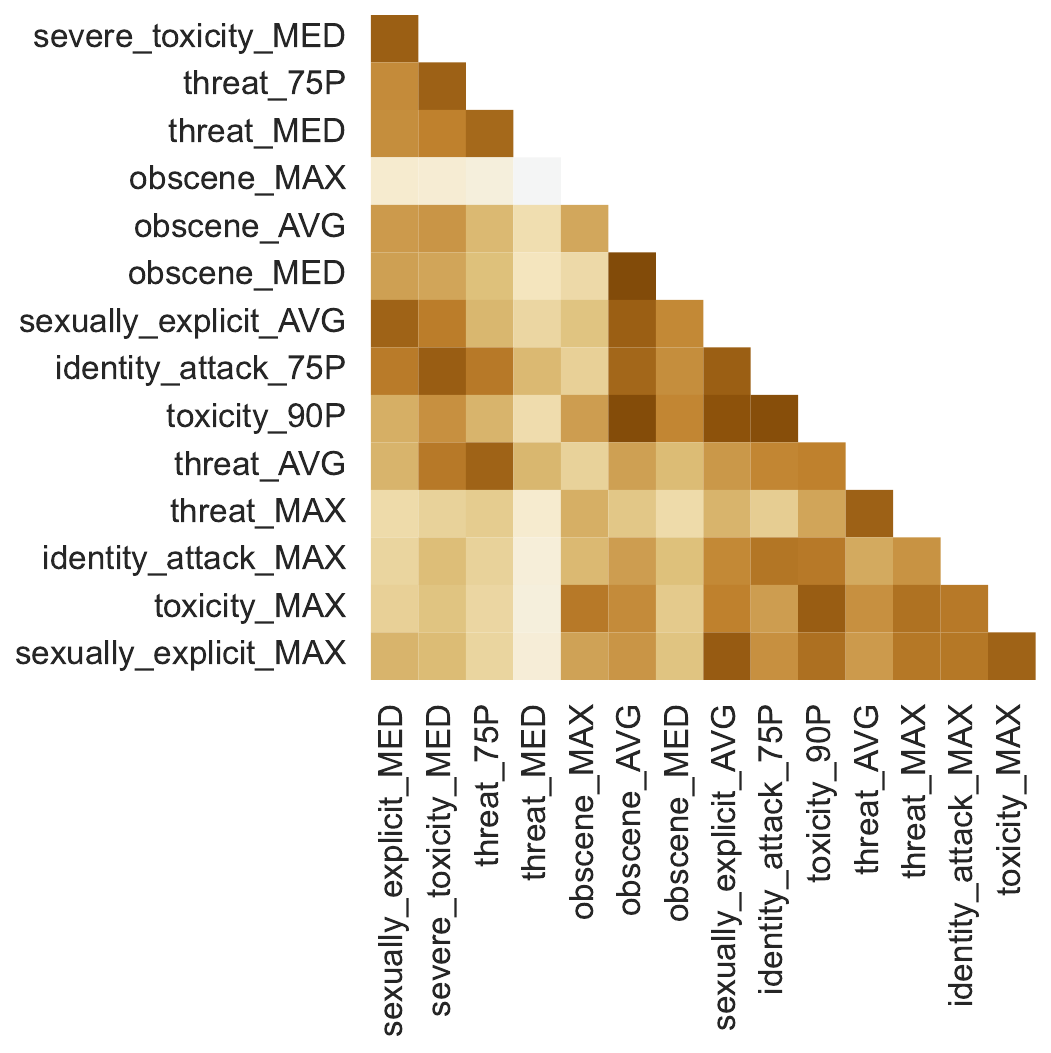}
        \centering
    \end{subfigure}
    \begin{subfigure}{0.1\textwidth}
        \includegraphics[width=0.5\linewidth]{pictures/colorbar_cropped.pdf}
        \vspace{2.52cm}
        \centering
    \end{subfigure}
    \caption{Pairwise Spearman correlation matrix for the selected features in the toxicity group after feature selection. The procedure removed 29 features (66\% reduction).}
    \label{fig:hierarchical_selected_toxicity}
\end{figure*}

\begin{figure*}[h!]
    \centering
    \begin{subfigure}{0.92\textwidth}
        \includegraphics[width=1\linewidth]{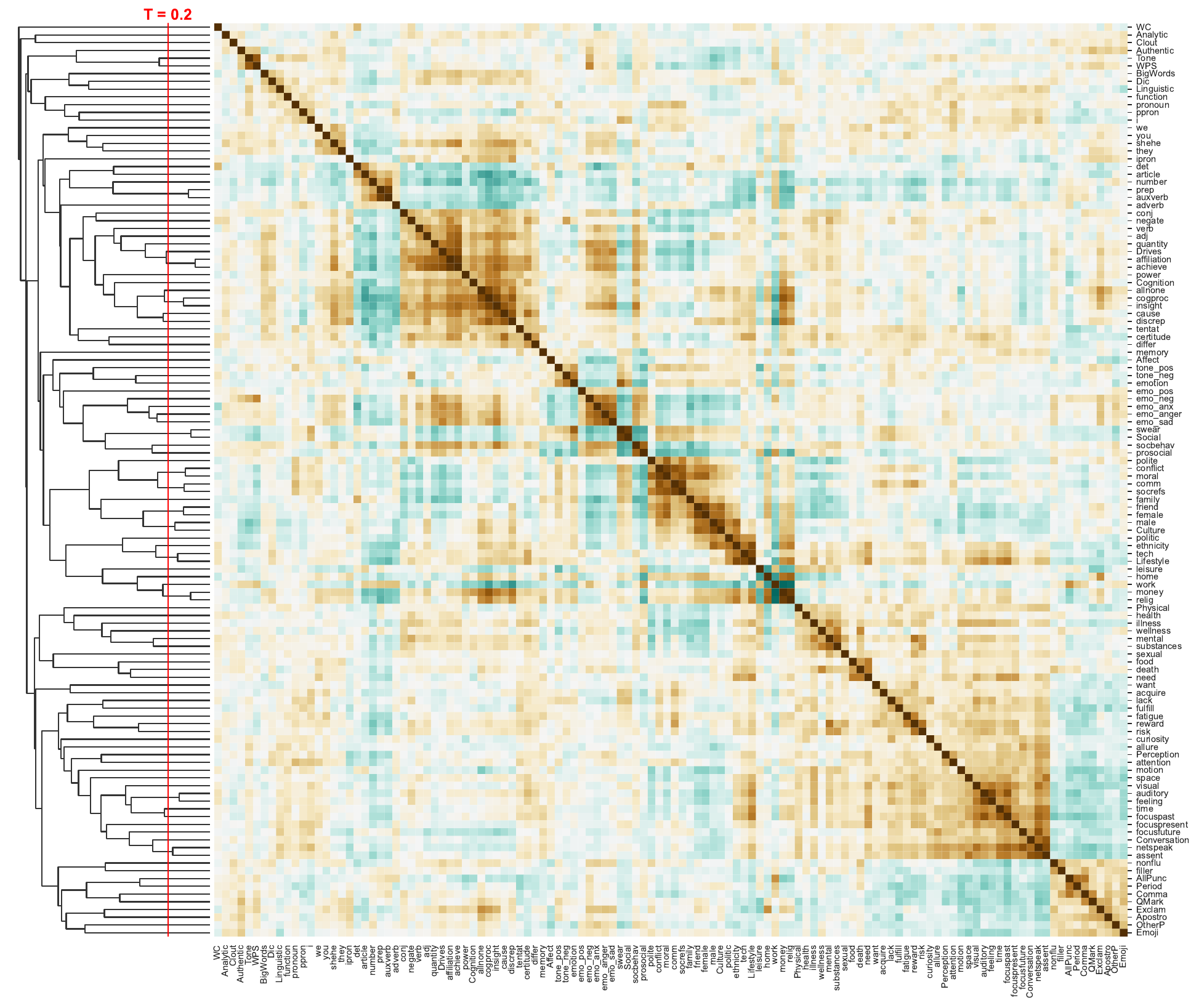}
        \centering
    \end{subfigure}
    \hspace{3pt}
    \begin{subfigure}{0.05\textwidth}
        \includegraphics[width=0.95\linewidth]{pictures/colorbar_cropped.pdf}
        \vspace{0.25cm}
        \centering
    \end{subfigure}
    \caption{Pairwise Spearman correlation matrix for the features in the psycholinguistic group, together with the hierarchical clustering dendrogram used for feature selection. The vertical red line indicates the threshold used to define clusters. For each cluster, the medoid was selected as the representative feature.}
    \label{fig:dendrogram_liwc}
\end{figure*}

\begin{figure*}[h!]
    \centering
    \hspace{-0.5cm}
    \begin{subfigure}{0.85\textwidth}
        \includegraphics[width=1\linewidth]{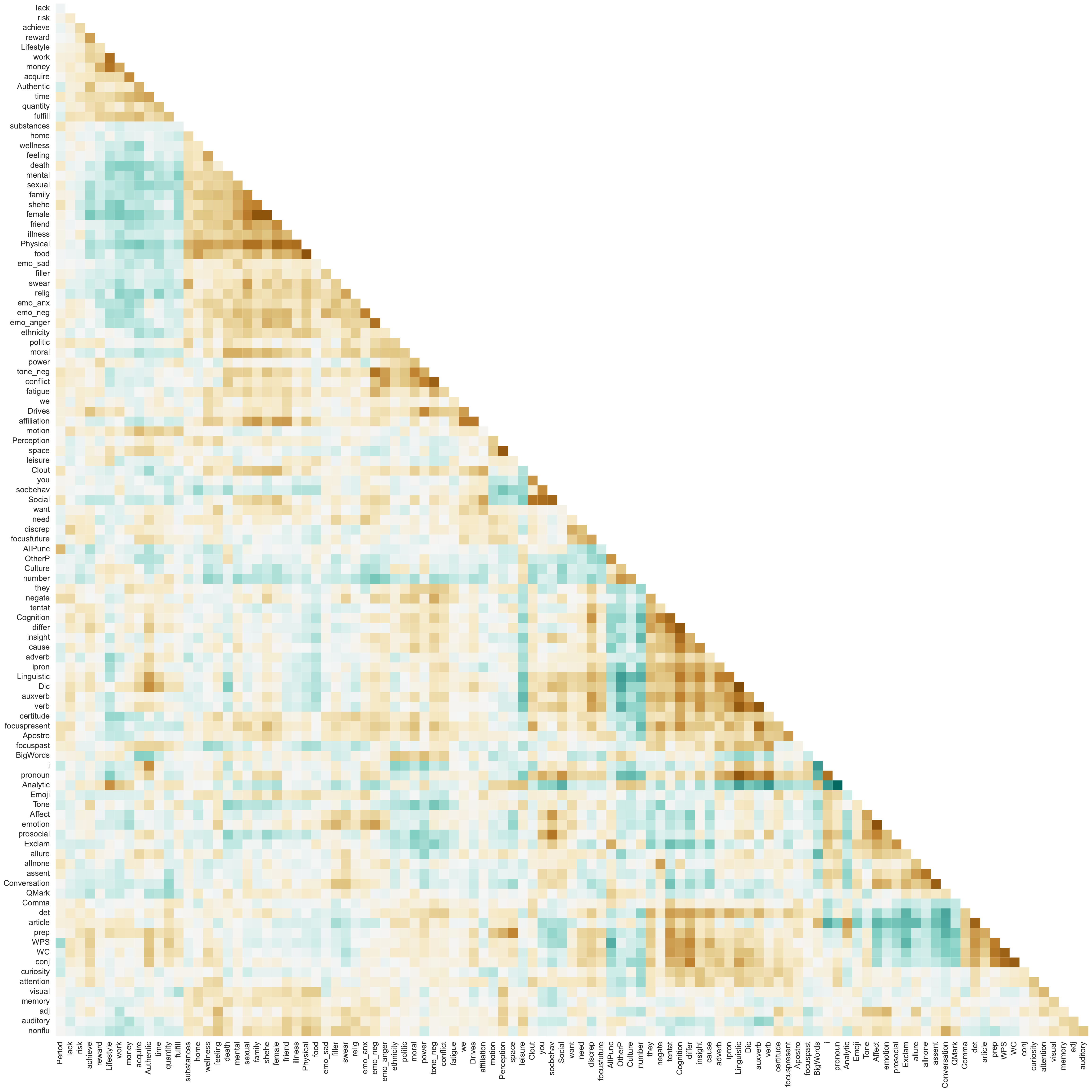}
        \centering
    \end{subfigure}
    \hspace{-2cm}
    \begin{subfigure}{0.1\textwidth}
        \includegraphics[width=0.5\linewidth]{pictures/colorbar_cropped.pdf}
        \vspace{5cm}
        \centering
    \end{subfigure}
    \caption{Pairwise Spearman correlation matrix for the selected features in the psycholinguistic group after feature selection. The procedure removed 12 features (10\% reduction).}
    \label{fig:selected_liwc}
\end{figure*}

This Appendix illustrates the dimensionality reduction process described in Section~\ref{sec:feature_selection} for the four feature macro-groups used in our study. Figures~\ref{fig:dendrogram_basic_text} through~\ref{fig:selected_liwc} visualize the pairwise correlations and the hierarchical clustering logic applied to select the final set of 154 features.

\section{Non-standardized Dimension Scores}
\label{sec:app_original_scores}
\begin{figure}[th!]
    \centering
    \begin{subfigure}{0.49\textwidth}
        \includegraphics[width=\linewidth]{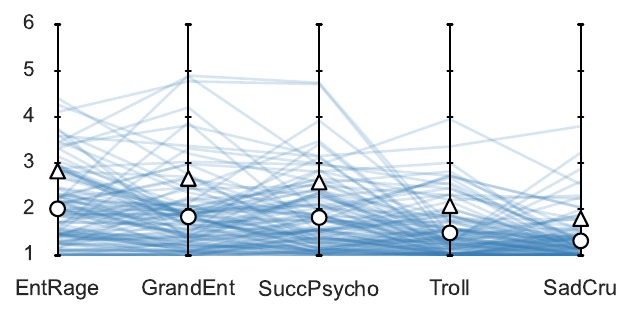}
        \caption{All participants ($N = 114$).}
        \label{fig:results-dist-all}
    \end{subfigure}
    \hfill
    \begin{subfigure}{0.49\textwidth}
        \includegraphics[width=\linewidth]{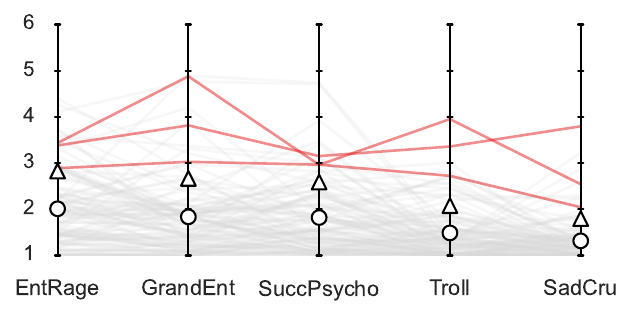}
        \caption{Participants exhibiting all traits ($N = 3$).}
        \label{fig:results-dist-all-dark}
    \end{subfigure}
    \begin{subfigure}{0.49\textwidth}
        \includegraphics[width=\linewidth]{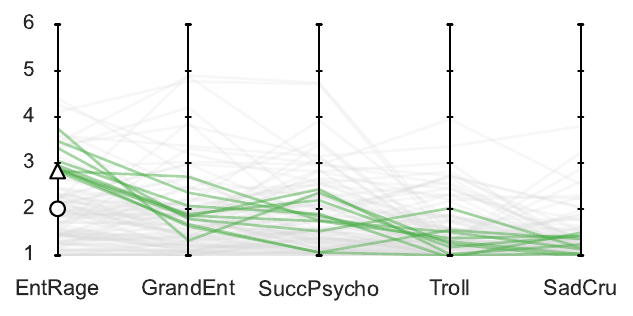}
        \caption{Participants exclusively exhibiting EntRage ($N = 10$).}
        \label{fig:results-dist-ent-rage}
    \end{subfigure}
    \hfill
    \begin{subfigure}{0.49\textwidth}
        \includegraphics[width=\linewidth]{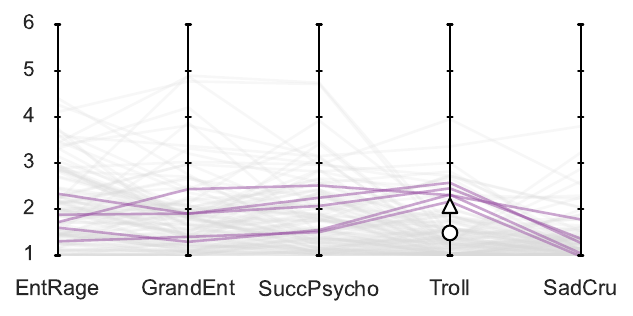}
        \caption{Participants exclusively exhibiting Trolling ($N = 5$).}
        \label{fig:results-dist-trolling}
    \end{subfigure}
    \begin{subfigure}{0.49\textwidth}
        \includegraphics[width=\linewidth]{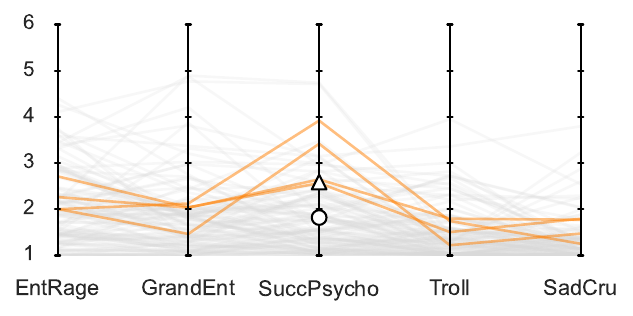}
        \caption{Participants exclusively exhibiting SuccPsycho ($N = 4$).}
        \label{fig:results-dist-succ-psycho}
    \end{subfigure}
    \hfill
    \begin{subfigure}{0.49\textwidth}
        \includegraphics[width=\linewidth]{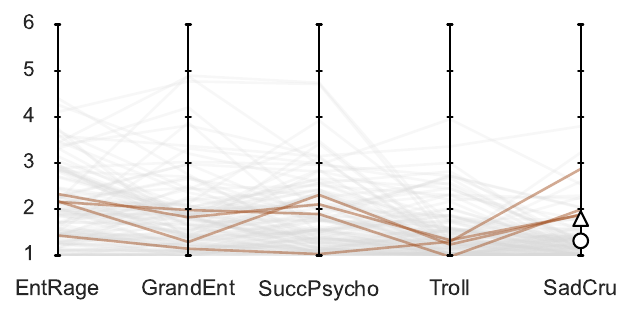}
        \caption{Participants exclusively exhibiting SadCru ($N = 4$).}
        \label{fig:results-dist-sad-cru}
    \end{subfigure}
    \caption{Distribution of raw self-reported trait and trolling behavior scores. Each vertical axis represents one of the measured dimensions, and each line shows a participant’s scores across all traits and trolling behavior. The white circle and triangle on an axis respectively display the mean and the threshold for the corresponding dimension. Dimensions are ordered from left to right by decreasing mean score. Figure~\ref{fig:results-dist-all} shows all participants, while the remaining ones highlight subsets of participants exhibiting certain dimensions.} \label{fig:parCoords}
\end{figure}

Figure~\ref{fig:parCoords} shows the distributions of the raw, non-standardized dimension scores assigned to participants. The white circle and triangle on an axis respectively indicate the mean $\mu$ and the threshold $\mu + \sigma$ for the corresponding dimension. Dimensions are ordered from left to right by decreasing mean score.

\begin{table}[t]
    \centering
    \small
    \begin{tabular}{lccccccccccccc}
        \toprule
        && \multicolumn{2}{c}{$\bar{X}_{\text{our}}$} & \multicolumn{2}{c}{$s_{\text{our}}$} && \multicolumn{2}{c}{$\bar{X}_{\text{Katz}}$} & \multicolumn{2}{c}{$s_{\text{Katz}}$} && \multicolumn{2}{c}{$d$} \\
        \cmidrule(lr){3-4} \cmidrule(lr){5-6} \cmidrule(lr){8-9} \cmidrule(lr){10-11} \cmidrule(lr){13-14}
        \textbf{dimension} && \textit{M} & \textit{F} & \textit{M} & \textit{F} && \textit{M} & \textit{F} & \textit{M} & \textit{F} && \textit{M} & \textit{F} \\
        \midrule
        SuccPsycho && 1.83 & 1.80 & 0.82 & 0.76 && 2.46 & 1.95 & 0.95 & 0.80 && $-$0.66 & $-$0.19 \\
        GrandEnt   && 1.80 & 1.89 & 0.83 & 0.87 && 2.30 & 2.04 & 1.00 & 0.95 && $-$0.50 & $-$0.16 \\
        SadCru     && 1.33 & 1.31 & 0.55 & 0.43 && 1.57 & 1.33 & 0.75 & 0.65 && $-$0.32 & $-$0.03 \\
        EntRage    && 1.94 & 2.07 & 0.78 & 0.86 && 2.50 & 2.37 & 1.05 & 1.05 && $-$0.53 & $-$0.29 \\
        \bottomrule
    \end{tabular}
    \caption{Comparison between our sample and the pooled reference values from Katz et al.~\cite{katz2022dark}, reported on the same item-level scale. Standardized distances ($d$) are computed as the difference between our sample mean ($\bar{X}_{\text{our}}$) and Katz's pooled mean ($\bar{X}_{\text{Katz}}$), divided by Katz's pooled standard deviation ($s_{\text{Katz}}$), separately for males (\textit{M}) and females (\textit{F}).}
    \label{tab:dshs-distances}
\end{table} 
\section{Comparison with the DSHS Validation Sample}
\label{sec:dshs-validation}
To contextualize the distribution of dark personality traits in our sample, we compared our results with those reported in the original validation study of the Dark Side of Humanity Scale (DSHS) by Katz et al.~\cite{katz2022dark}. Since Katz et al. report subscale scores as summed totals, whereas our study uses item-level averages, we first converted their means and standard deviations to the same scale by dividing by the number of items in each subscale. We then computed sex-specific pooled reference values across the four validation samples from~\cite{katz2022dark} using sample-size-weighted means,
\[
\bar{X}_{\text{Katz}} = \frac{\sum_{i} n_i \bar{X}_i}{\sum_{i} n_i},
\]
and corresponding pooled standard deviations that account for both within- and between-sample variability,
\[
s^2_{\text{Katz}} =
\frac{
\sum_{i} (n_i - 1)s_i^2 + \sum_{i} n_i(\bar{X}_i - \bar{X}_{\text{Katz}})^2
}{
\sum_{i} n_i - 1
}.
\]
Finally, to assess how typical our sample is relative to the validation data, we computed standardized distances between our sample means and the reference values from~\cite{katz2022dark},
\[
d = \frac{\bar{X}_{\text{our}} - \bar{X}_{\text{Katz}}}{s_{\text{Katz}}},
\]
separately for males and females. In the above formulas, $i$ indexes the validation samples ($i = 1, \dots, 4$) from Katz et al.~\cite{katz2022dark}; $n_i$ denotes the number of participants in sample $i$; $\bar{X}_i$ and $s_i$ are the mean and standard deviation of the given subscale in sample $i$, respectively; $\bar{X}_{\text{Katz}}$ and $s_{\text{Katz}}$ are the pooled reference mean and standard deviation across all samples; and $\bar{X}_{\text{our}}$ denotes the corresponding mean observed in our sample. The standardized distance $d$ thus expresses the difference between our sample and the reference distribution in units of the pooled standard deviation. This analysis provides a descriptive benchmark to determine whether our sample exhibits typical levels of dark trait endorsement or deviates meaningfully from the original validation population. Results are reported in Table~\ref{tab:dshs-distances}. As shown, our sample consistently exhibits lower mean scores than the Katz et al.~\cite{katz2022dark} references across all four dimensions, with standardized distances ranging from small to moderate in magnitude. Differences are more pronounced among males ($d \in [-0.66, -0.32]$) than females ($d \in [-0.29, -0.03]$), but in all cases remain well within one standard deviation of the reference distribution. These results indicate that, while our participants tend to report somewhat lower levels of dark trait endorsement, the sample is not markedly atypical relative to the original validation population. As such, the absence of strong associations between DSHS dimensions and behavioral or linguistic features observed in this study is unlikely to be driven by extreme or unrepresentative trait distributions, and instead reflects the relationships (or lack thereof) present within a reasonably typical range of variation.

\section{Bootstrap Procedure}
\label{sec:app:bootstrap}
As explained in Section~\ref{sec:analyses}, we applied bootstrap resampling to compute 95\% confidence intervals (CIs) for those correlations that remained statistically significant after \textit{p}-value correction. Estimating CIs allowed us to evaluate the reliability and precision of the correlation estimates, which is particularly crucial given the relatively small size of our participant sample. This section presents the results of the procedure. For each pair of correlated features, Table
\ref{tab:bootstrap} reports the \textit{p}-value and the corresponding level of statistical significance before and after correction, the correction method, and the obtained CI.

\begin{table}[h]
\setlength{\tabcolsep}{5pt}
\centering
\footnotesize
\begin{tabular}{llcrllrlc}
\toprule
\textbf{question} & \textbf{feature} & \textbf{$\boldsymbol{\rho}$} & \multicolumn{2}{c}{$\bm{p_{raw}}$} & \textbf{method} & \multicolumn{2}{c}{$\bm{p_{corr}}$} & \textbf{95\% CI}\\
\toprule
\multicolumn{5}{l}{\textit{dark personality traits and trolling behavior}}\\
 SM05 & Troll & 0.365 & 6.4$e$\textsuperscript{-5} & *** & Bonf & 1.6$e$\textsuperscript{-3} & *** & [0.191, 0.520]\\
 SM05 & SuccPsycho & 0.336 & 2.6$e$\textsuperscript{-4} & *** & Bonf & 6.5$e$\textsuperscript{-3} & *** & [0.175, 0.491]\\
 \hline
  \multicolumn{5}{l}{\textit{toxicity}}\\
  SM05 & toxicity\_median & 0.337 & 2.5$e$\textsuperscript{-4} & *** & Bonf & 1.5$e$\textsuperscript{-2} & ** & [0.169, 0.502] \\
  SM04 & toxicity\_median & 0.324 & 4.3$e$\textsuperscript{-4} & *** & Bonf & 2.6$e$\textsuperscript{-2} & ** & [0.150, 0.479] \\
  SM05 & toxicity\_75\textsuperscript{th} & 0.298 & 1.3$e$\textsuperscript{-3} & *** & Bonf & 7.8$e$\textsuperscript{-2} & * & [0.131, 0.459] \\
  SM04 & toxicity\_75\textsuperscript{th} & 0.295 & 1.4$e$\textsuperscript{-3} & *** & Bonf & 8.6$e$\textsuperscript{-2} & * & [0.113, 0.445] \\
  \hline
  \multicolumn{5}{l}{\textit{toxicity subtypes}}\\
SM04 & sev\_toxicity\_median & 0.308 & 8.5$e$\textsuperscript{-4} & *** & BH & 4.8$e$\textsuperscript{-2} & ** & [0.139, 0.456] \\
 SM05 & identity\_attack\_75\textsuperscript{th} & 0.317 & 5.9$e$\textsuperscript{-4} & *** & BH & 5.0$e$\textsuperscript{-2} & * & [0.124, 0.492] \\
 SM04 & identity\_attack\_75\textsuperscript{th} & 0.325 & 4.2$e$\textsuperscript{-4} & *** & BH & 7.2$e$\textsuperscript{-2} & * & [0.153, 0.471] \\
\hline
 \multicolumn{5}{l}{\textit{LIWC}}\\
 SM05 & tone\_neg & 0.354 & 1.1$e$\textsuperscript{-4} & *** & BH & 3.6$e$\textsuperscript{-2} & ** & [0.199, 0.519] \\
 SM04 & certitude & 0.330 & 3.4$e$\textsuperscript{-4} & *** & BH & 5.6$e$\textsuperscript{-2} & * & [0.155, 0.490] \\
 SM05 & moral & 0.315 & 6.5$e$\textsuperscript{-4} & *** & BH & 7.2$e$\textsuperscript{-2} & * & [0.143, 0.489] \\
 \hline
 \multicolumn{5}{l}{\textit{NRC-EIL}}\\
SM04 & sadness & 0.304 & 1.0$e$\textsuperscript{-3} & *** & BH & 7.8$e$\textsuperscript{-2} & * & [0.122, 0.478] \\
\hline
 \multicolumn{5}{l}{\textit{Moral Foundations}} \\
 SM05 & fairness.vice & 0.338 & 2.4$e$\textsuperscript{-4} & *** & BH & 3.7$e$\textsuperscript{-2} & ** & [0.167, 0.498] \\
SM05 & bias\_fairness & $-$0.339 & 2.2$e$\textsuperscript{-4} & *** & BH & 6.8$e$\textsuperscript{-2} & * & [$-$0.510, $-$0.167] \\
 SM05 & intensity\_authority & 0.302 & 1.1$e$\textsuperscript{-3} & *** & BH & 6.9$e$\textsuperscript{-2} & * & [0.137, 0.466] \\
SM05 & intensity\_general\_morality & 0.313 & 7.1$e$\textsuperscript{-4} & *** & BH & 7.3$e$\textsuperscript{-2} & * & [0.158, 0.481] \\
 SM05 & authority.vice & 0.290 & 1.8$e$\textsuperscript{-3} & *** & BH & 9.0$e$\textsuperscript{-2} & * & [0.095, 0.457] \\
\bottomrule
\end{tabular}
\caption{Correlations between social usages and all feature groups that remained significant after \textit{p}-value correction by means of Bonferroni (Bonf) or Benjamini-Hochberg (BH) methods, with 95\% confidence intervals estimated using the described bootstrapping procedure. Within each group of features, rows are ordered by increasing ${p_{corr}}$ values.}
\label{tab:bootstrap}
\end{table} 
\end{appendix} 
\end{document}